\documentclass{aa} 
\usepackage{graphicx}
\usepackage{epsfig}
\begin{document}

%
%

\title{The COMPTEL instrumental line background} 


\author{G. Weidenspointner\inst{1}\thanks{\emph{Present address:}
        {\sl i}. NASA Goddard Space Flight Center, Code 661,
        Greenbelt, MD 20771, USA \hspace{2ex}{\sl ii}. Universities
        Space Research Association, 7501 Forbes Blvd. \#206,
        Seabrook, MD 20706-2253, USA} 
        \and 
        M. Varendorff\inst{1}
          \and 
          U. Oberlack\inst{3} 
          \and 
          D. Morris\inst{2}
          \and
          S. Pl{\"u}schke\inst{1} 
          \and
          R. Diehl\inst{1}
          \and
          S. C. Kappadath\inst{4}
          \and
          M. McConnell\inst{2}
          \and
          J. Ryan\inst{2}
          \and
          V. Sch{\"o}nfelder\inst{1}
          \and
          H. Steinle\inst{1}
          }

\offprints{G. Weidenspointner, e-mail ggw@tgrosf.gsfc.nasa.gov}

\institute{Max-Planck-Institut f{\"u}r extraterrestrische Physik,
              D-85740 Garching, Germany
              \and
              Space Science Center, University of New Hampshire, Durham
              NH 03824, USA 
              \and
              Astrophysics Laboratory, Columbia University, New York,
              NY 10027, USA 
              \and
              Louisiana State University, Baton Rouge, Louisiana, USA
             }

\date{Received ???, 2000; accepted ???}




\abstract{
The instrumental line background of the Compton telescope COMPTEL
onboard the Compton Gamma-Ray Observatory is due to the activation
and/or decay of many isotopes. The major components
of this background can be attributed to eight
individual isotopes, namely $^{2}$D, $^{22}$Na, $^{24}$Na,
$^{28}$Al, $^{40}$K, $^{52}$Mn, $^{57}$Ni, and $^{208}$Tl.
The identification of instrumental lines with specific isotopes
is based on the line energies as well as on the variation of the
event rate with time, cosmic-ray intensity, and deposited radiation dose
during passages through the South-Atlantic Anomaly. The
characteristic variation of the event rate due to a specific isotope
depends on its life-time, orbital parameters such as the altitude of the
satellite above Earth, and the solar cycle.\\
A detailed understanding of the background contributions from 
instrumental lines is crucial at MeV energies for measuring the
cosmic diffuse gamma-ray background and for observing $\gamma$-ray line
emission in the interstellar medium or from supernovae and their
remnants. Procedures to determine the event rate from each background
isotope are described, and their average activity in spacecraft
materials over the first seven years of the mission is estimated.
}

\maketitle

%

\section{\label{introduction} Introduction}

Gamma-ray experiments in low-Earth orbit, such as the Compton
telescope COMPTEL onboard the Compton Gamma-Ray Observatory (CGRO),
operate in an intense and variable radiation environment. The main
constituents of the ambient radiation fields are primary cosmic-ray
particles, geomagnetically trapped radiation-belt particles, as well
as albedo neutrons and $\gamma$-ray photons.  The different particle
species interact with the spacecraft and detector materials, resulting
in the emission of instrumental background photons (for a review, see
e.g. Dean et al.\ \cite{dean_bgd-review}). COMPTEL data, dominated by
instrumental background, have a typical signal-to-noise ratio of a few
percent. Hence, a qualitative and quantitative understanding of the
instrumental background is crucial for conducting astrophysical
measurements, in particular of the cosmic diffuse gamma-ray background
(hereafter CDG), and of the $\gamma$-ray line emission in the
interstellar medium or from supernovae and their remnants.

The instrumental background experienced by COMPTEL is subdivided into
two major components according to their signature in energy space:
first, a continuum background discussed by Ryan et al.\
(\cite{ryan_ieee}); second, the instrumental line background, the
focus of this paper. 
The latter arises from a number of different radioactive isotopes
generated in the instrument material. This primarily occurs from
activation by trapped protons during passages through the
South-Atlantic Anomaly (SAA), from neutron absorption, and from
primordial radioactivity\footnote{In the following, ``primordial
radioactivity'' denotes unstable isotopes that already existed when
the Earth was formed.}.
An earlier report on activation in the COMPTEL
telescope was given by Morris et al.\ (\cite{morris_ieee}).

The discussion is structured as follows. After a brief description of
the COMPTEL instrument in Sect.~\ref{instrument}, general
characteristics of the instrumental (line) background are summarized
in Sect.~\ref{line_characteristics}. In
Sect.~\ref{identified_isotopes}, identifications of specific isotopes
are discussed. In Sect.~\ref{variations}, the variations of the
activity of individual isotopes are described. In
Sect.~\ref{comparisons_implications}, a comparison of instrumental
line backgrounds in different low-energy $\gamma$-ray experiments is
given. Also, average values for the activity of spacecraft materials
are presented. The results of this work are summarized and discussed
in Sect.~\ref{discussion}. Finally, appendices give the event
selections used in these line studies, and provide detailed
descriptions of the procedures employed for determining the background
contributions of individual isotopes in the CDG analysis, and also --
with slight modifications -- in the analysis of the galactic 1.8~MeV
line emission from $^{26}$Al.

%

\section{\label{instrument} Instrument description}

\begin{figure}
\epsfig{figure=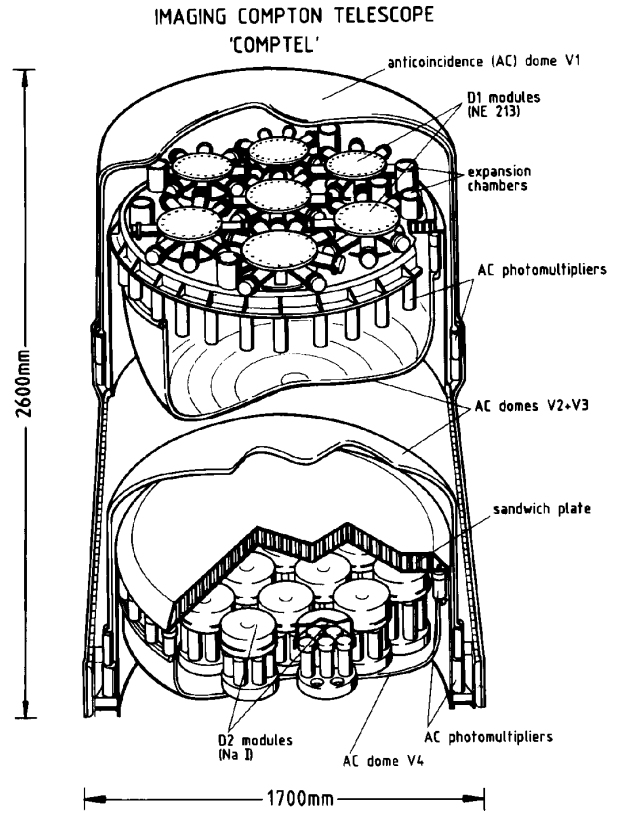,width=8.8cm,clip=}
\caption{A schematic view of the Compton telescope COMPTEL (from
Sch{\"o}nfelder et al.\ \cite{schoenfelder_comptel}).}
\label{comptel_scheme} 
\end{figure}

COMPTEL is the first double-scattering Compton telescope designed for
$\gamma$-ray astronomy to operate on a satellite platform. A detailed
description of the COMPTEL instrument, which is sensitive to
$\gamma$-rays in the 0.8--30~MeV range, can be found in
Sch{\"o}nfelder et al.\ (\cite{schoenfelder_comptel}). Briefly, the
instrument consists of two planes of detector arrays, D1 and D2,
separated by 1.58~m (see Fig.~\ref{comptel_scheme}). The D1 detector
consists of seven cylindrical modules filled with NE~213A organic
liquid scintillator. The D2 detector consists of 14 cylindrical
NaI(Tl) crystals. The D1 scintillator material has a low average
atomic number to optimize the occurrence of a single Compton scatter,
while the D2 scintillator crystals have a high density and average
atomic number to maximize their photon absorption properties. Each
detector array is surrounded by a pair of overlapping anti-coincidence
domes, manufactured of NE~110 plastic scintillator, to reject charged
particle triggers of the telescope. The in-flight performance of the
instrument is monitored with two calibration (CAL) units, each
composed of a $^{60}$Co-doped scintillator viewed by two 1/2 inch
photomultiplier tubes (PMTs), that provide tagged photons for
in-flight energy calibration (Snelling et al.\ \cite{snelling}). The
COMPTEL instrument accepts and registers coincident triggers in a
single D1-D2 module pair within the coincidence time window of $\sim
\!  40$~ns in the absence of a veto signal from the four charged
particle shields as valid events.  These interactions can be caused by
a single photon or by multiple photons and/or particles.  Among other
parameters, a time-of-flight (ToF) value and a so-called pulse shape
discriminator (PSD) value in D1 are recorded for each event. The ToF
is a measure of the time difference between the triggers in the D1 and
D2 detectors and is used to discriminate forward scattered
(D1$\rightarrow$D2) events, such as celestial photons with a ToF value
of about 5~ns, from backward scattered (D2$\rightarrow$D1) background
events which cluster around a ToF value of about $-5$~ns.  The PSD is
a measure of the shape of the scintillation light pulse in the D1
detector. The energy loss characteristics of recoil electrons
resulting from Compton scattering and of recoil protons resulting from
neutron scattering are different, allowing one to reject many neutron
induced events. The summed energy deposits in the two detectors,
$\mathrm{E}_\mathrm{1} + \mathrm{E}_\mathrm{2}$, are a measure of the
total energy of the incident photon, E$_\mathrm{tot}$, while the
photon scatter angle $\bar{\varphi}$ is determined from E$_\mathrm{1}$
and E$_\mathrm{2}$ through the Compton-scatter formula:
\begin{equation}\label{phibar_definition}
\cos \bar{\varphi} = 1 - \frac{m_o c^2}{\mathrm{E}_\mathrm{2}} +
\frac{m_o c^2}{\mathrm{E}_\mathrm{1} + \mathrm{E}_\mathrm{2}}, \;
\mathrm{with} \; m_o c^2 = 511~\mathrm{keV}
\end{equation}

%

\section{\label{line_characteristics} Characteristics of
instrumental line background}

\begin{figure}
\epsfig{figure=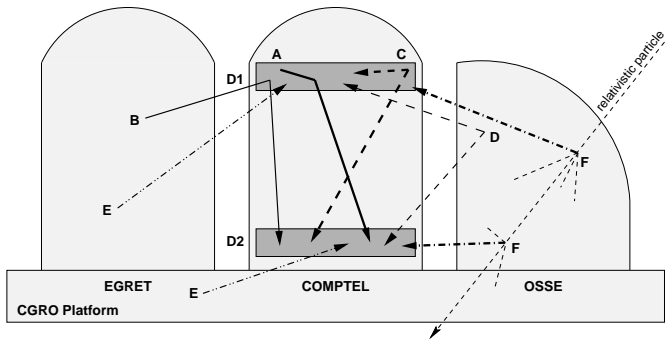,width=8.8cm}
\caption[]{An illustration of the main channels
for the triggering of valid events due to instrumental background
(adopted from van~Dijk \cite{vandijk_phd}). For
simplicity, BATSE was omitted in this schematic view of CGRO. The various
event types are explained in the text.}
\label{cgro_bgd-scheme}
\end{figure}

\begin{figure}
\resizebox{\hsize}{!}{\includegraphics{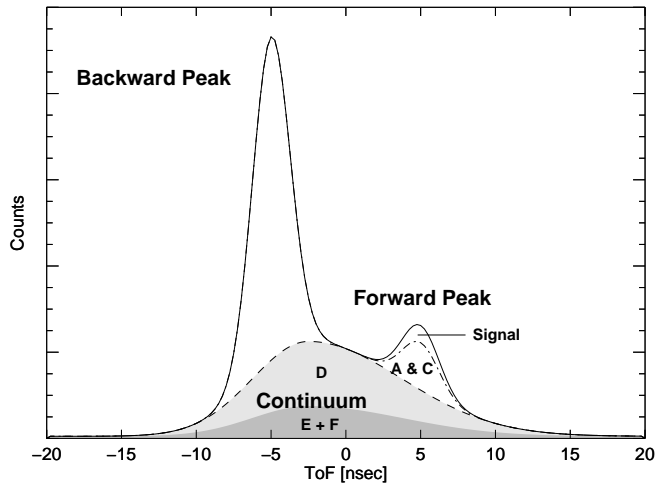}}
\caption[]{A schematic representation of the ToF distribution of valid
events. Three major components can be discerned: the ToF backward peak
and forward peak, centered at ToF values of about $-5$~ns and $+5$~ns,
respectively, and an underlying continuum distribution. The backward
peak is composed of all types of background events originating in and
around the D2 detector (these are not specified individually in this
illustration). The ToF forward peak contains the celestial signal as
well as background events originating in and around the D1 detector,
mostly of types {\bf A} and {\bf C}. The ToF continuum is dominated by
background events of types {\bf D}, {\bf E}, and {\bf F} originating
in the instrument structure between the two detectors and the
spacecraft structure in general. The relative magnitudes of the
different components, which depend on E$_\mathrm{tot}$, are only
represented approximately.}
\label{tof_scheme}
\end{figure}

The particle and photon environment of COMPTEL produces coincident
interactions in the D1 and D2 detectors that pass all the logic and
electronic criteria for a valid event.
In the following, an overview is given of different types of
instrumental background events in terms of their interaction process
and location, as illustrated in Fig.~\ref{cgro_bgd-scheme}. This
classification of background events (van~Dijk \cite{vandijk_phd})
provides a simple and versatile framework for discussing the COMPTEL
instrumental background.
ToF is of prime importance in identifying and rejecting instrumental
background events. A schematic representation of the ToF distribution
of valid events is depicted in Fig.~\ref{tof_scheme}.

$\bullet$ Event types {\bf A} and {\bf B}: {\sl events caused by the
double scattering of a single photon}. Any photon created in CGRO may
produce this type of background event, which, if the scattering is
from D1 to D2, is identical to a proper celestial event. It follows
that the ToF distribution of forward scattered single photon events is
identical to that of celestial photons and peaks around 5~ns in the
ToF forward peak.  Depending on where they originate, forward
scattered single photon events may be rejected. In particular, many of
the photons that originate from below the D1 detector (such as type
{\bf B} events) can be eliminated by a selection on the scatter angle
$\bar{\varphi}$ (see e.g.\ the event selections in
App.~\ref{event_selections}). The remaining instrumental background
from forward scattered single photons is therefore mostly due to
photons originating in, around and above the D1 detector (such as type
{\bf A} events); type {\bf B} events are negligible.  Line emissions
appear in energy space, particularly in E$_\mathrm{tot}$ and
E$_\mathrm{1}$-E$_\mathrm{2}$ space (see below).

Single photons may also scatter in D2 before interacting in D1. These
backward scattered single photon events, however, are identifiable by
their ToF distribution, which is confined in the backward peak at ToF
values of about $- 5$~ns. Finally, high-energy neutrons can undergo a
double scattering process analogously to photons, however, many of
these neutron events can be rejected by their PSD value.

$\bullet$ Event types {\bf C} and {\bf D:} {\sl events caused by two
or more photons both spatially and temporally correlated (so-called
multiple photon or cascade events)}.  In general, multiple photon
events are more efficient in generating a background event than single
photon events, since the probability for coincident interactions in
both detectors increases with the number of emitted photons.  Emission
of two or more photons can occur from a small region on a time-scale
shorter than the coincidence window of 40~ns. Nucleons that have been
excited above the first nuclear level, e.g.\ by proton or neutron
interactions, may promptly emit a cascade of photons. The multitude of
nuclear excitation levels often results in a rather featureless
continuum distribution in energy space. If, however, only a few
transition levels are involved, characteristic features appear in
E$_\mathrm{2}$ and E$_\mathrm{1}$-E$_\mathrm{2}$ space (see
below). Multiple photons may also arise from $\beta$-particle
bremsstrahlung, or the annihilation of a positron. Also, high-energy
neutrons may induce the emission of one or more $\gamma$-ray photons
outside of the D1 scintillator (thereby eluding rejection by PSD) and
some may interact in the D2 detector. More complicated nuclear
reactions such as the spallation (break-up) of a nucleus or the
initiation of a shower of secondaries by an incident cosmic-ray
particle or neutron may also produce multiple photons.

Frequently, the emission of photons is nearly simultaneous, i.e.\ on
time-scales much shorter than the coincidence window. The ToF value of
these event types is determined by the location of the emitting
nucleus relative to the D1 and D2 detectors. Multiple photons
originating in the vicinity of the D1 detector (such as type {\bf C}
events) will peak slightly below a ToF value of 5~ns (see
Sect.~\ref{24_na}) and therefore contribute to the ToF forward peak.
Photons emitted by nuclei in the spacecraft material between the D1
and D2 detectors (such as type {\bf D} events) will interact in the
two detectors near-simultaneously and have ToF values that are broadly
distributed around zero, while photons originating in and around the
D2 detector will contribute to the ToF backward peak. The full,
double-peaked ToF distribution of all multiple photon events anot only
reflects the location of the emitting nucleus, but also the mass
distribution of the entire spacecraft relative to the D1 and D2
detectors.

$\bullet$ Event type {\bf E:} {\sl events that are both spatially and
temporally uncorrelated (the so-called random coincidences)}. The
COMPTEL detectors are continuously exposed to a large flux of
$\gamma$-ray photons. This inevitably leads to coincident interactions
that qualify as valid events. The photons producing these random
coincidences are mostly of local or atmospheric origin. Since the two
photons creating the event are unrelated to one another, and, in
particular, not correlated in time, these events are uniformly
distributed in ToF. Type {\bf E} events may also involve a neutron
instead of a photon triggering the D2 detector, which has no PSD
capability.

$\bullet$ Event type {\bf F:} {\sl events caused by two photons that
are temporally correlated, but spatially uncorrelated}. High-energy
cosmic-ray particles or atmospheric neutrons may interact at several
different locations along their path through CGRO. Individual
interactions include those generating type {\bf C} and {\bf D} events
(e.g.\ spallations or showers). The whole interaction chain creating
type {\bf F} events is similar to multiple photon events, the
main difference being that their ToF distribution depends on both the
location of the interactions as well as the (relativistic) velocity of
the primary particle. The ToF distribution of this type of event is
broad and covers the entire coincidence window (also reflecting
the spacecraft mass distribution).

$\bullet$ Other processes, such as the interaction of a neutron in the
D2 detector after producing a photon in the D1 system, or direct
ionization losses of $\beta$-particles created in the housings of the
D1 scintillators, may also play a role. The event signatures of these
and other, more complicated processes, however, will be similar to the
event types described above.

Event types {\bf A} -- {\bf D} may arise from activation by primary
cosmic-ray protons or secondary particles as well as atmospheric
neutrons. The time between the interaction of the proton or neutron
within the telescope material and the actual triggering of a
background event varies, since it depends on the decay time of the
radio-isotopes produced. These processes and the resulting background
events can be crudely separated into ``prompt'' and ``delayed''
components. For prompt background events the time delay between the
primary particle interaction in the instrument and the resulting
emission of background photons is shorter than the coincidence window
of 40~ns for the triggering of a valid event. Thus the intensity of
prompt background components instantaneously follows the
(time-variable) incident local cosmic-ray flux (see
Sect.~\ref{variations}). For delayed events the time delay is longer
than the typical length of the fast-logic veto signal of $\sim
200$~ns. In contrast to protons, neutrons can travel to any location
in the spacecraft to produce $\gamma$-ray photons without triggering
the veto system. Inside the veto domes, particularly in the D1
detector, prompt events can therefore only be produced by neutrons.
Both protons and neutrons, however, can produce delayed background
events inside as well as outside the veto domes.

As described in Sect.~\ref{instrument}, events due to incident
celestial photons have ToF values around 5~ns. Only background
events with a similar ToF value will therefore interfere with
astrophysical analyses. As illustrated in Fig.~\ref{tof_scheme}, a
major portion of the instrumental background events, including those
in the backward peak, can be eliminated by a ToF selection. The most
important background event types in the ToF forward peak region around
ToF$ = 5$~ns are those originating in the D1 detector (types {\bf A} and
{\bf C}), and some of the background events produced in the satellite
structure (types {\bf D}, {\bf E}, and {\bf F}).

After ToF selection, the majority of the instrumental line
background is expected to arise from activation of the D1 detector
material because of the relatively high mass density and probability
for triggering a background event as compared to the general
spacecraft structure. The material composition of the D1 detector
system therefore provides important clues as to which radioactive
isotopes can effectively be produced and ultimately contribute to the
instrumental line background in the low-Earth orbit of CGRO. The D1
support structure and the D1 module and PMT housings are mostly
aluminium, the most abundant element in the instrument. The liquid
scintillator NE~213A is composed of hydrogen and carbon. The quartz
windows in the module housings contain silicon and oxygen. The PMTs
and electronics boxes contain, among other elements, copper, nickel,
and iron.

\begin{figure*}
\begin{minipage}{8.8cm}\makebox[0cm]{}\\
\centerline{\epsfig{figure=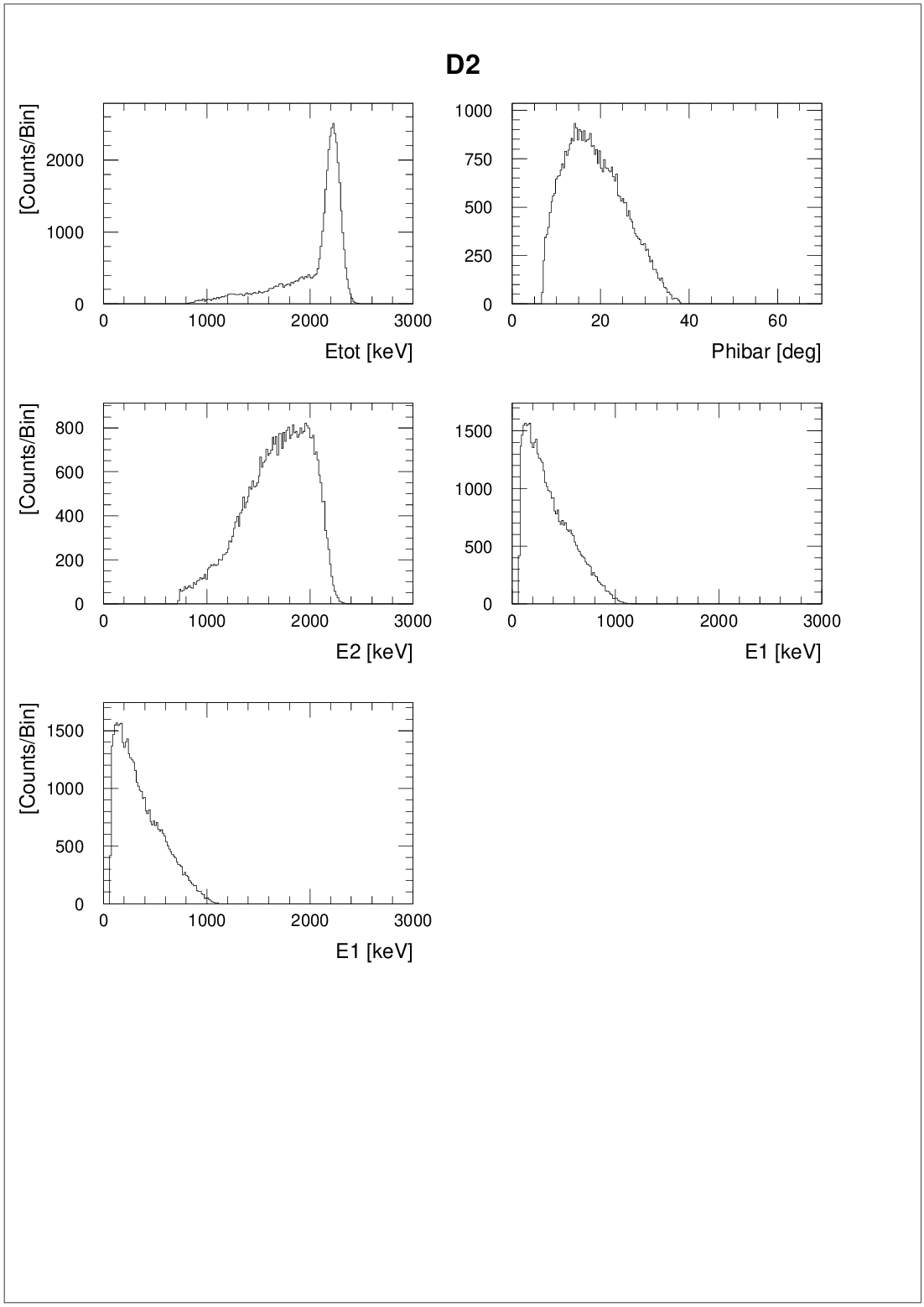,%
bbllx=38pt,bblly=584pt,bburx=282pt,bbury=744pt,width=6.75cm,clip=}}
\centerline{\epsfig{figure=H2362F4.ps,%
bbllx=38pt,bblly=412pt,bburx=282pt,bbury=572pt,width=6.75cm,clip=}}
\centerline{\epsfig{figure=H2362F4.ps,%
bbllx=38pt,bblly=240pt,bburx=282pt,bbury=400pt,width=6.75cm,clip=}}
\centerline{\epsfig{figure=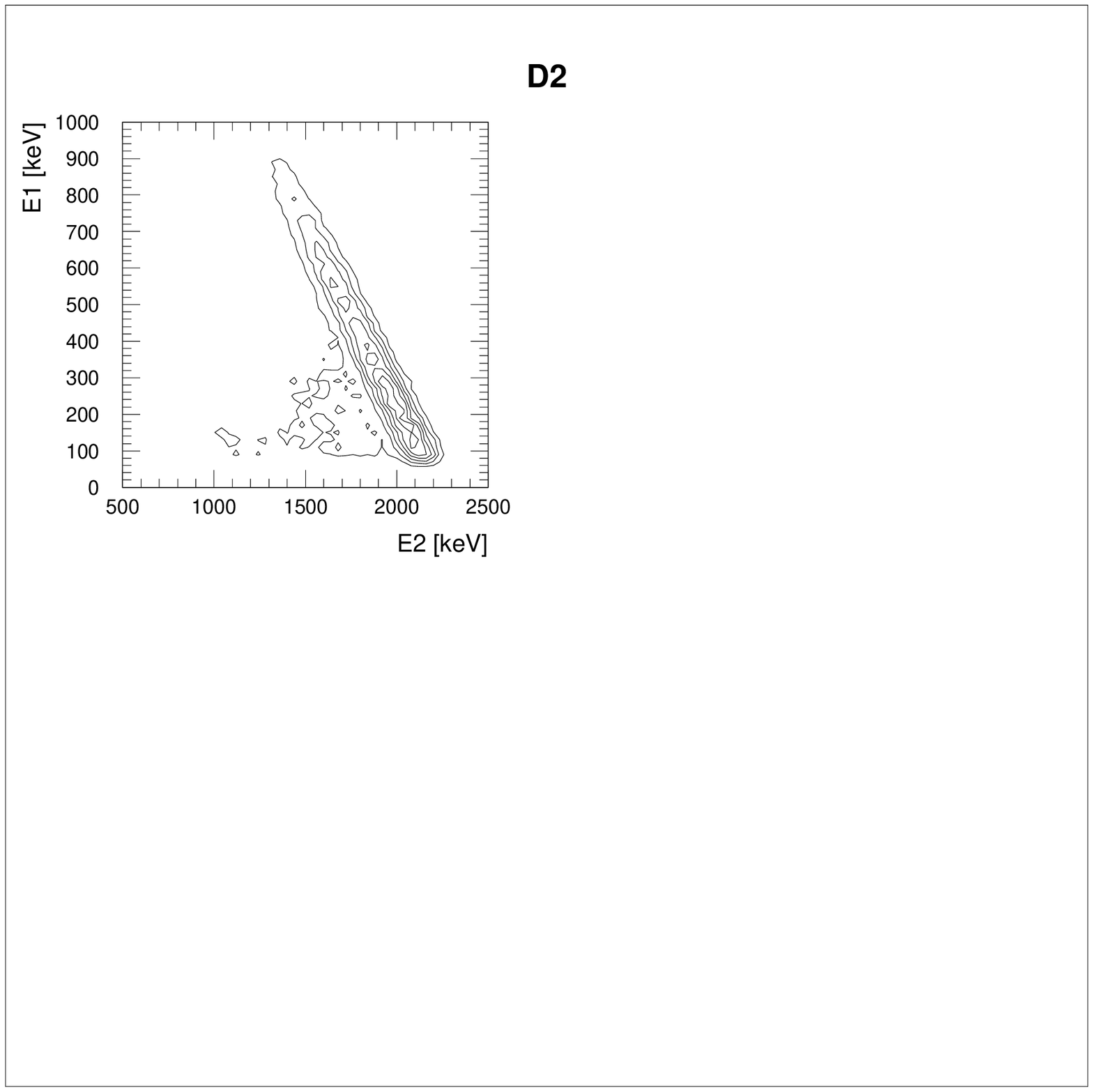,%
bbllx=38pt,bblly=412pt,bburx=282pt,bbury=636pt,width=6.75cm,clip=}}
\caption[]{The E$_\mathrm{tot}$, E$_\mathrm{2}$, E$_\mathrm{1}$, and
E$_\mathrm{1}$-E$_\mathrm{2}$ distribution of instrumental 2.22~MeV
photons from $^{2}$D production for CDG event selections as determined
from Monte Carlo simulations.  }
\label{2_d_dataspace}
\end{minipage}
\hfill
\begin{minipage}{8.8cm}
\centerline{\epsfig{figure=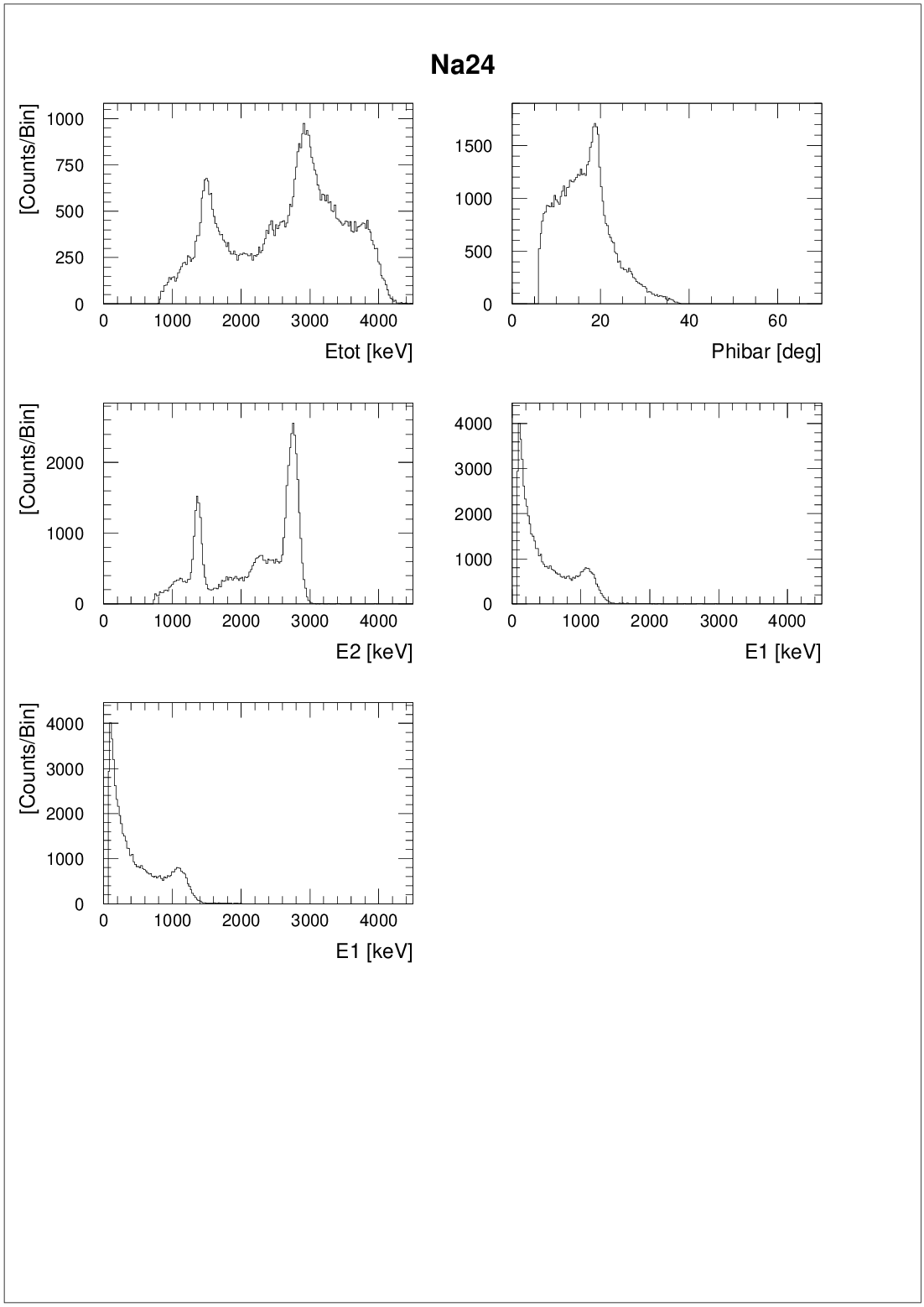,%
bbllx=38pt,bblly=584pt,bburx=282pt,bbury=744pt,width=6.75cm,clip=}}
\centerline{\epsfig{figure=H2362F6.ps,%
bbllx=38pt,bblly=412pt,bburx=282pt,bbury=572pt,width=6.75cm,clip=}}
\centerline{\epsfig{figure=H2362F6.ps,%
bbllx=38pt,bblly=240pt,bburx=282pt,bbury=400pt,width=6.75cm,clip=}}
\centerline{\epsfig{figure=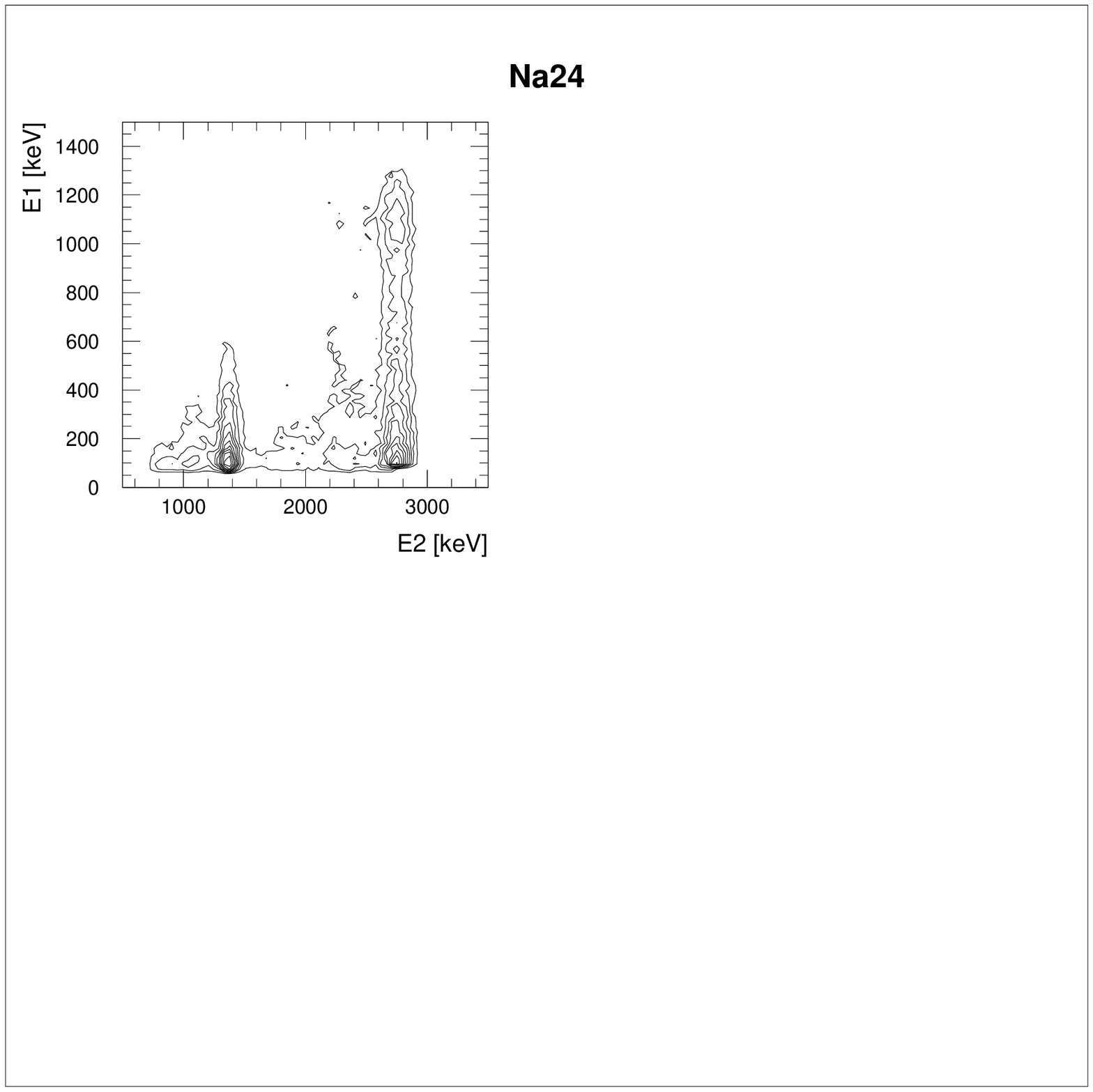,%
bbllx=38pt,bblly=412pt,bburx=282pt,bbury=636pt,width=6.75cm,clip=}}
\caption[]{The E$_\mathrm{tot}$, E$_\mathrm{2}$, E$_\mathrm{1}$, and
E$_\mathrm{1}$-E$_\mathrm{2}$ distribution of instrumental $^{24}$Na
events for CDG data selections as determined from Monte Carlo
simulations.  }
\label{24_na_dataspace}
\end{minipage}
\end{figure*}

Events from the instrumental line background produce conspicuous and
characteristic features in energy space that can be exploited to
distinguish them from the general (continuum) background. The ToF
distributions of the line background and the continuum background are
also different: the instrumental line background is concentrated in
the forward peak region, while the continuum background in energy is
throughout the ToF continuum as well as the ToF forward peak. In the
following, the characteristics of single photon (type {\bf A}) events
and multiple photon or cascade (type {\bf C}) events are illustrated
with examples of (background) events from $^{2}$D and $^{24}$Na,
respectively, both major contributors to the instrumental line
background (see Sect.~\ref{identified_isotopes}).

The instrumental 2.22~MeV photons emitted when $^{2}$D is produced in
the D1 scintillators (see Sect.~\ref{2_d}) are single photon (type
{\bf A}) background events. The E$_\mathrm{tot}$, E$_\mathrm{2}$,
E$_\mathrm{1}$, and E$_\mathrm{1}$-E$_\mathrm{2}$ distributions of
instrumental 2.22~MeV photons for CDG event selections (described in
App.~\ref{event_selections_cdg}) as determined from Monte Carlo
simulations are depicted in Fig.~\ref{2_d_dataspace}. The
E$_\mathrm{tot}$ distribution exhibits a peak at the energy of the
primary photon, while the distributions in E$_\mathrm{2}$ and
E$_\mathrm{1}$ are broad and relatively featureless. In
E$_\mathrm{1}$-E$_\mathrm{2}$ space the event distribution of type
{\bf A} events follows the diagonal $\mathrm{E}_1 + \mathrm{E}_2 =
\mathrm{E}_\gamma$, with $\mathrm{E}_\gamma = 2.22$~MeV for $^{2}$D.
The distributions in E$_\mathrm{tot}$ and
E$_\mathrm{1}$-E$_\mathrm{2}$ are the most important characteristics
of single photon background events.

The $\beta^-$-decay of $^{24}$Na results in the emission of two
photons with energies 1.37~MeV and 2.75~MeV causing type {\bf C} line
background events (see Sect.~\ref{24_na}). The $\beta^-$-particle is
of minor importance for the generation of a background event. Since it
rarely escapes the support structure, it contributes to the background
only through secondary bremsstrahlung photons. The $^{24}$Na
signature in E$_\mathrm{tot}$, E$_\mathrm{2}$, E$_\mathrm{1}$, and
E$_\mathrm{1}$-E$_\mathrm{2}$ as determined from Monte Carlo
simulations is shown, for CDG event selections (see
App.~\ref{event_selections_cdg}), in Fig.~\ref{24_na_dataspace}. The
most important characteristic of $^{24}$Na, and any other isotope
emitting multiple photons, is the E$_\mathrm{2}$ distribution, which
exhibits peaks at the energies of each of the primary photons. In
E$_\mathrm{1}$ the only conspicuous feature is the Compton edge of the
1.37~MeV photon (the Compton edge of the 2.75~MeV photon is suppressed
by the CDG data selections). The E$_\mathrm{tot}$ spectrum is more
complex and exhibits less pronounced, line-like features just above
the individual photon energies. These come from the absorption of one
photon in D2 with the other photon scattering in D1 with an energy
deposit near the D1 threshold. In addition, in E$_\mathrm{tot}$ a
shoulder is present at about 3.9~MeV. This is analogous to the sum
peak in standard spectroscopy employing a single detector. Under CDG
data selections, this shoulder is due to the absorption of the
2.75~MeV photon in D2, with the 1.37~MeV photon scattering in D1 with
an energy deposit at the Compton edge. The $^{24}$Na event
distribution in E$_\mathrm{1}$-E$_\mathrm{2}$ space is also very
characteristic: the events cluster along two bands parallel to the
E$_\mathrm{1}$ axis located in E$_\mathrm{2}$ at the energies of the
two decay photons. The band at 2.75~MeV in E$_\mathrm{2}$ extends in
E$_\mathrm{1}$ from the threshold up to the Compton edge of the
1.37~MeV photon. The band at 1.37~MeV in E$_\mathrm{2}$ is suppressed
by the CDG data selections.

For other isotopes, if the energy of one of a pair of cascade photons
is below the D2 threshold, then the two photons produce a
cascade (type {\bf C}) event only if the lower-energy photon scatters in
D1, and the higher-energy photon interacts in D2.
The E$_\mathrm{2}$ and E$_\mathrm{1}$-E$_\mathrm{2}$ distribution of
such a cascade photon pair is simple: there is only one photopeak in
E$_\mathrm{2}$ and only one band in E$_\mathrm{1}$-E$_\mathrm{2}$.

Radioactive decays that result in the (simultaneous) emission of two
or more photons can produce line background events other than of type
{\bf C}. 
For example, an individual photon of the emitted
photon multiple can produce a type {\bf A} event, provided all other
photons escape from the instrument without interacting in any of the
detectors. 
As apparent in the E$_\mathrm{1}$-E$_\mathrm{2}$ event distribution of
$^{24}$Na, however, the E$_\mathrm{tot}$ signatures corresponding to
the photon energies 1.37~MeV and 2.75~MeV are much weaker than the
cascade structure.
Monte Carlo simulations show that less than 10\% of the
$^{24}$Na background events are due to type {\bf A} events.
This illustrates the fact that multiple photon decays in the D1
detector material are more efficient in generating a background event
than decays that result in the emission of a single photon only.

\begin{figure}
\epsfig{figure=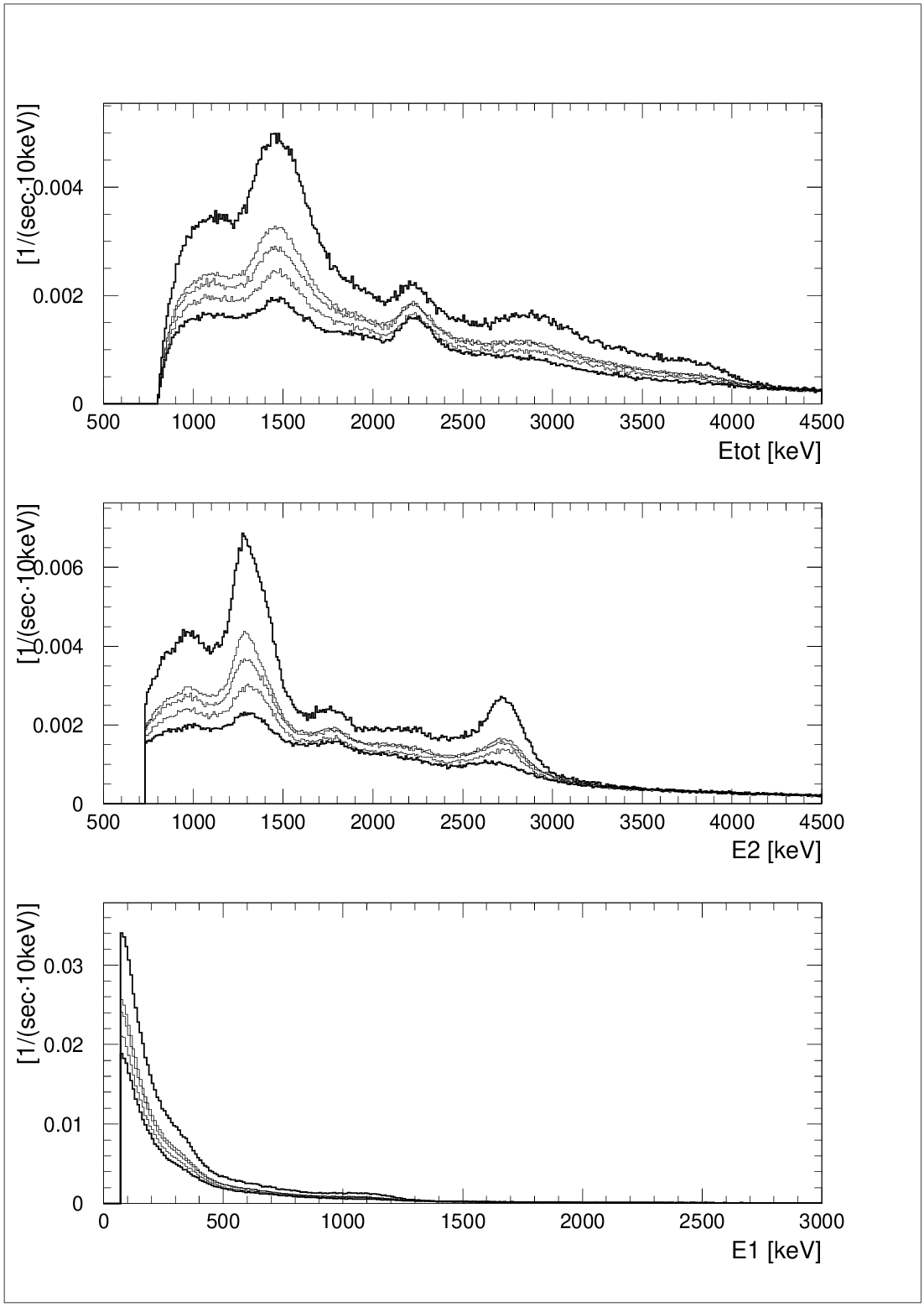,%
bbllx=38pt,bblly=298pt,bburx=544pt,bbury=744pt,width=8.8cm,clip=}
\caption[]{An illustration of the spectral distributions in E$_\mathrm{tot}$
(top) and E$_\mathrm{2}$ (bottom) as a function of time. The two spectra
plotted with thick lines represent the times of highest (after the
second reboost in May 1997) and lowest (from the beginning of the
mission until the first reboost, May 1991 -- Nov.\ 1993) contributions
from long-lived background isotopes. The three intermediate spectra,
plotted with thin lines, cover the time periods Nov.\ 1993 -- Oct.\
1994 (lowest), Oct.\ 1994 -- Oct.\ 1995 (middle), and Oct.\ 1995 --
May 1997 (highest).}
\label{etot_e2_spectra}
\end{figure}


\section{\label{identified_isotopes} Identified radioactive isotopes}

\begin{table*}
\resizebox{12cm}{!}{
\begin{tabular}{|cclc|}   \hline
\rule[0ex]{0cm}{3.5ex}\makebox[10ex]{Isotope} & 
Half-Life & \makebox[35ex]{Decay Modes and} &
        \makebox[25ex]{Main} \\ 
        &          & \makebox[35ex]{Photon Energies [MeV]} &
        Production Channels \\[1ex] \hline 
\rule[0pt]{0ex}{3ex}
$^2$D   & prompt & \makebox[35ex]{2.224} &
$^{1}$H(n$_\mathrm{ther}$,$\gamma$) \\[1.5ex] 
$^{22}$Na & 2.6~y & \makebox[35ex]{$\beta^+$ (91\%): 0.511, 1.275} &
$^{27}$Al(p,3p3n), \\ 
        &       & \hspace{4.7ex} EC (9\%) \hspace{0.4ex}: 1.275 &
Si(p,4p$x$n)  \\[1.5ex] 
$^{24}$Na & 14.96~h & \makebox[35ex]{$\beta^-$: 1.37, 2.75} &
$^{27}$Al(n,$\alpha$), \\  
        &        &                    & $^{27}$Al(p,3pn)  \\[1.5ex]
$^{28}$Al & 2.2~min & \makebox[35ex]{$\beta^-$: 1.779} &
$^{27}$Al(n$_\mathrm{ther}$,$\gamma$) \\[1.5ex]
$^{40}$K & $1.28 \times 10^9$~y & \makebox[35ex]{EC (10.7\%): 1.461} &
primordial \\[1.5ex]
$^{52}$Mn & 5.6~d & \hspace{0ex} EC (64\%): 0.744, 0.935, 1.434 &
Fe(p,x), Cr(p,x),   \\ 
          &       & \makebox[35ex]{$\beta^+$ (27\%): 0.511, 0.744,
0.935, 1.434}  & Ni(p,x) \\[1.5ex]  
$^{57}$Ni & 35.6~h & \makebox[35ex]{$\beta^+$ (35\%): 0.511, 1.377}  & 
Ni(p,x), Cu(p,x) \\  
          &        & \hspace{4.6ex} EC (30\%): 1.377  &  \\[1.5ex]  
$^{208}$Tl  & $1.4 \times 10^{10}$~y & \hspace{1.5ex} $\beta^-$ (50\%):
0.583, 2.614  & primordial  \\ 
           &    \rule[-1.2ex]{0cm}{3.6ex}($^{232}$Th)        &
\makebox[35ex]{$\beta^-$ (25\%): 0.511, 0.583, 2.614} &    \\[1.5ex] \hline  
\end{tabular}
}
\hfill
\parbox[t]{5.5cm}{\vspace*{-25ex}
\caption[]{A summary of the isotopes identified in the COMPTEL
instrumental line background. For simplicity, only the photon energies
of the most frequent decay modes are listed. If $\beta$-decays are
involved, the $\beta$-particles have been included in the response
simulations. The identification of $^{208}$Tl has to be considered
tentative. The label ``prompt'' for the half-life of the stable
isotope $^{2}$D refers to the time-scale for the emission of the
2.22~MeV photon.}
\label{identified_isotopes_table}
}
\end{table*}

The major components of the COMPTEL instrumental line background can
be attributed to eight individual isotopes, namely $^{2}$D, $^{22}$Na,
$^{24}$Na, $^{28}$Al, $^{40}$K, $^{52}$Mn, $^{57}$Ni, and $^{208}$Tl
(Weidenspointner \cite{weidenspointner_phd}). Identification of these
isotopes was achieved in an iterative process, starting from the most
prominent lines. The E$_\mathrm{tot}$ and E$_\mathrm{2}$ spectra (see
Fig.~\ref{etot_e2_spectra}) are particularly useful for the
identification of isotopes that give rise to single photon (type {\bf
A}) or multiple photon (type {\bf C}) events, respectively. The
diagnostic power of E$_\mathrm{1}$ spectra is limited since they do
not exhibit line features at the energy of the incident photons, but
only rather broad features at the corresponding Compton edges.  Due to
the correlated signatures in E$_\mathrm{1}$-E$_\mathrm{2}$ space,
individual spectral features can be accentuated by applying suitable
event selections.  Viable isotope identifications are required to
account self-consistently for spectral features in selected regions of
the E$_\mathrm{1}$-E$_\mathrm{2}$ dataspace (see the detailed
explanations in App.~\ref{line-fitting_procedure}), as well as for
their variation with time and/or incident cosmic-ray intensity
(discussed in Sect.~\ref{variations}). The telescope response to
individual isotopes was modelled through Monte Carlo simulations.
The isotopes, their half-lifes, most important decay channels, and
main production channels are summarized in
Table~\ref{identified_isotopes_table}. Below, a review of the isotope
identifications is given (a more detailed account on the COMPTEL
instrumental line background can be found in, e.g., Weidenspointner
\cite{weidenspointner_phd}).

\subsection{\label{2_d} $^{\mathsf 2}$D}

Thermal-neutron capture on hydrogen results in the production of a
stable $^{2}$D nucleus and the emission of a single 2.22~MeV
photon, as seen in E$_\mathrm{tot}$ (see top panel in
Fig.~\ref{etot_e2_spectra}). The liquid scintillator NE~213A in the D1
detector modules consists of 9.2\% H and 90.8\% C by mass, making it
an efficient moderator for neutrons. The instrumental 2.22~MeV
photons are isotropically emitted from the D1 scintillators and
constitute single photon (type {\bf A}) background events. The average
moderation and absorption times in the D1 scintillator for an incident
neutron are a few $\mu$s and about $3 \times 10^{-4}$~s, respectively,
hence the instrumental 2.22~MeV line is a quasi-prompt background
component that follows the local, instantaneous cosmic-ray intensity
(Weidenspointner et al.\ \cite{weidenspointner_3c}). This quasi-prompt
nature of the production/emission of the 2.22~MeV line is the reason
for treating, in the following, the stable isotope $^{2}$D as if it
were unstable, and for referring to it as ``short-lived''. In
principle, the initial scatterings of the incident neutrons could be
identified by their PSD values. However, the time-scale for the
moderation and absorption of an incident high-energy neutron is too
long to associate the initial scatterings with the delayed neutron
capture producing the $\gamma$-ray. The instrumental 2.22~MeV events
therefore cannot be rejected by a selection on PSD.

The bulk of the neutrons producing $^{2}$D are expected to be of
atmospheric origin, with secondary neutrons produced in the spacecraft
being of minor importance. This follows from measurements of the
fast-neutron flux in the D1 detector and Monte Carlo simulations of the
production of secondary neutrons in cosmic-ray interactions (Morris et
al.\ \cite{morris_fast_neutrons}), as well as from an estimate for the
neutron absorption efficiency of the D1 detector.

\subsection{\label{24_na} $^{\mathsf{24}}$Na}

The main production channels for $^{24}$Na are neutron-capture
reactions such as $^{27}$Al(n,$\alpha$)$^{24}$Na and proton
reactions such as $^{27}$Al(p,3pn)$^{24}$Na. The Al structure of
the D1 detector system is the primary source of $^{24}$Na
activation.
As described in Sect.~\ref{line_characteristics}, spectral features of
the $^{24}$Na background in both the E$_\mathrm{tot}$ and
E$_\mathrm{2}$ spectra can be understood by considering its cascade
nature (type {\bf C} events). In particular, the two photons with
energies of 1.37~MeV and 2.75~MeV produce line features at about
1.3~MeV and 2.7~MeV in E$_\mathrm{2}$ (see bottom panel in
Fig.~\ref{etot_e2_spectra}).

\begin{figure}
\resizebox{\hsize}{!}{\includegraphics{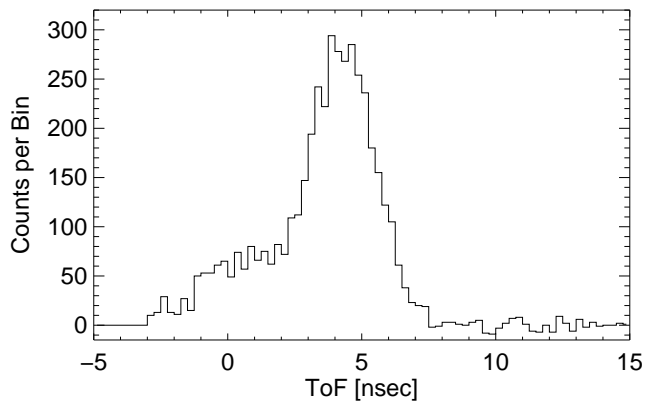}}
\caption[]{The measured ToF distribution of $^{24}$Na events from
the vicinity of the D1 detector.}
\label{na24_tof-spc}
\end{figure}

In addition, the time-variation of the $^{24}$Na event rate is
consistent with expectations for an isotope with a half-life of 15~h
that is produced during SAA passages (see
Sect.~\ref{time_variation}). Also, ToF distributions summed using data
selections that favour $^{24}$Na events have the characteristics of
type {\bf C} background events, i.e.\ they peak slightly below a ToF
value of 5~ns (see Fig.~\ref{na24_tof-spc}), corresponding to an
average distance of the location of the photon emission from the D1
module of about 20--30~cm.

\subsection{\label{40_k} $^{\mathsf{40}}$K}

$^{40}$K is one of the contributors to the spectral line feature at
about 1.4~MeV in E$_\mathrm{tot}$ (see top panel of
Fig.~\ref{etot_e2_spectra}). A prominent line at this energy was
present before launch in some of the calibration data, and at
that time was tentatively identified with primordial $^{40}$K
radioactivity, contained e.g.\ in the concrete of the buildings. 
The 1.4~MeV line was still present after launch, however, and did
not vary with any orbital parameter, indicating an instrumental
origin as well.

A viable explanation for the instrumental 1.4~MeV line was the
potassium in the glass of the D1 PMTs (van~Dijk \cite{vandijk_phd}).
Electron capture by $^{40}$K results in the emission of a single
1.46~MeV photon (type {\bf A} event). This origin is supported by the
fact that a disproportionate fraction of the photons at 1.46~MeV
interact in the outer sections of the D1 scintillator, indicating that
the photons enter the D1 detectors from the sides, consistent with the
location of the D1 PMTs. Also, the observed $\bar{\varphi}$
distribution at 1.46~MeV peaks at high $\bar{\varphi}$ values, again
indicating that the photons enter from the sides. These
characteristics were reproduced in Monte Carlo
simulations. Furthermore, the potassium content in the D1 PMTs derived
from the observed 1.4~MeV line rate is consistent with the
manufacturer's specifications, as was confirmed by radiological
measurements of a PMT in the laboratory. The mass of the PMT front
window (made of Corning France 801.51) is about 24.9~g, with a
potassium mass fraction of 6\%; the mass of the PMT side and back
glass (both made of Schott 8245) is about 40.7~g and 9.1~g,
respectively, both with a potassium mass fraction of 0.14\%. The front
window contains more than 95\% of the potassium in the PMTs and is
closest to the detectors. The $^{40}$K background therefore is
dominated by the front window. The $^{40}$K activity of the front
window, normalized to its mass, is about 0.2~decays~s$^{-1}$~g$^{-1}$.

\subsection{\label{22_na} $^{\mathsf{22}}$Na}

Routine data processing and the investigation of galactic 1.8~MeV line
emission from radioactive $^{26}$Al indicated the build-up of a broad
background component at about 1.5~MeV in E$_\mathrm{tot}$ (see top
panel of Fig.~\ref{etot_e2_spectra}). Investigations of the energy
distribution of this background feature, as well as of its ToF
dependence, led to its identification with the $\beta^+$-decay of
$^{22}$Na produced in the D1 detector material (Oberlack
\cite{oberlack_phd}). This background component is produced by a two
photon cascade (type {\bf C} event) with energies of $\sim \!
500$~keV and $\sim \!  1.3$~MeV, consistent with the absorption of the
1.275~MeV photon in D2, and the scattering of an annihilation 511~keV
photon in D1 (the reverse process, involving absorption of the 511~keV
photon in D2, is suppressed because 511~keV is below the D2
threshold). The ToF distribution of the 1.275~MeV $^{22}$Na line in
E$_\mathrm{2}$ is consistent with the ToF signature of the $^{24}$Na
cascade. The build-up of the feature is consistent with an isotope of
half-life of 2.6~y (see Sect.~\ref{time_variation}).

The most important process for the production of $^{22}$Na in the
D1 detector are proton interactions in the aluminium structure, the
main channel being $^{27}$Al(p,3p3n)$^{22}$Na. Based on the
production cross-sections, some $^{22}$Na should also be produced
in proton interactions with silicon, found in the glass of the D1
PMTs, such as Si(p,4p$x$n). It is expected that most of the
$^{22}$Na production occurs during SAA passages (see
Sect.~\ref{time_variation}).

\subsection{\label{28_al} $^{\mathsf{28}}$Al}

The isotope $^{28}$Al is produced by thermal-neutron capture on
aluminium, the most abundant element in the D1 detector. The existence
of the strong $^{2}$D line due to thermal-neutron capture on
hydrogen indicates the presence of a thermal-neutron flux in the D1
scintillators. The cross-section for thermal-neutron capture on
aluminium has a value of 0.28~mb (c.f.\ 0.33~mb for hydrogen). Upon
its $\beta^-$-decay, $^{28}$Al can generate a type {\bf C}
background event: the 1.78~MeV photon is absorbed in D2, with a
bremsstrahlung photon (from the $\beta^-$) scattering in D1. A
small fraction of the $^{28}$Al decays result in type {\bf A}
background events from the double scattering of the 1.78~MeV
photon.

From the beginning of the mission a weak, albeit significant, line
feature at about 1.8~MeV has consistently been present in the
E$_\mathrm{2}$ distribution (see bottom panel of
Fig.~\ref{etot_e2_spectra}). The position of this feature is
independent of E$_\mathrm{1}$, as expected for a type {\bf C} event.
Additional support for the identification of $^{28}$Al comes from the
fact that, as expected, the 1.8~MeV feature is short-lived (see
Sect.~\ref{veto_variation}).

\subsection{\label{52_mn} $^{\mathsf{52}}$Mn and $^{\mathsf{57}}$Ni}

The above mentioned isotopes were not sufficient to account entirely
for the pronounced and broad $\sim \! 1.4$~MeV feature in
E$_\mathrm{2}$ that built up over the mission. After subtracting the
effects of the other isotopes, there was clear evidence for a build-up
of the 1.4~MeV residual, whose half-life was estimated to be between
that of $^{24}$Na and that of $^{22}$Na (assuming that it is due to a
single isotope). The E$_\mathrm{1}$ dependence of the 1.4~MeV residual
in E$_\mathrm{2}$ suggested that the line is due to a $\beta^+$-decay
and therefore arises from type {\bf C} background events.
After detailed modelling of the E$_\mathrm{2}$ and E$_\mathrm{tot}$
spectra, taking into account the material composition of the D1
detector, it was concluded that the 1.4~MeV residual is due to a blend
of two different isotopes, namely $^{52}$Mn (1.43~MeV) and $^{57}$Ni
(1.38~MeV).

The isotope $^{52}$Mn is produced in SAA-proton interactions
with the Fe, Cr, and Ni in the D1
detector\footnote{The D1 detector mass fraction of Fe, Cr, and Ni
relative to Al is about 20\%, 5\%, and 6\%, respectively.}, found mainly
in the electronics. Matters are complicated by the
fact that $^{52}$Mn can be produced in either its ground state or
an isomeric state. These two states have different half-lifes and
decay schemes. The time variation
of the $^{52}$Mn event rate suggests that this isotope is
more likely produced in the ground state ($T_{1/2} = 5.6$~d) than in
the isomeric state ($T_{1/2} = 21.1$~min), as described in
Sect.~\ref{time_variation}. In this case, most $^{52}$Mn events are
due to the absorption of a 1.43~MeV photon in D2, accompanied by the
scattering of another photon, from the radioactive decay or positron
annihilation, in D1.

The isotope $^{57}$Ni is expected to be mostly produced by proton
interactions with Ni and Cu in the D1 detector during SAA
passages\footnote{The D1 detector mass fraction of Cu relative to Al
is about 13\%.}. The $^{57}$Ni background is mainly produced by the
absorption of the 1.38~MeV photon in D2, with one of the 511~keV
annihilation photons scattering in D1 (type {\bf C} event). Similar to
$^{52}$Mn, the identification of $^{57}$Ni, which has a
half-life of about 36~h, is supported
by modelling the observed long-term variation of its
background contribution (see Sect.~\ref{time_variation}).

\subsection{\label{208_tl} $^{\mathsf{208}}$Tl}

Similar to the 1.4~MeV feature, the 2.7~MeV feature in E$_\mathrm{2}$
represents a blend of lines from more than a single isotope. In
addition to $^{24}$Na, at least one component with a very long
half-life (exceeding that of $^{22}$Na) is present. The position of
the residual is about 2.6~MeV. $^{208}$Tl, which is part of the
natural $^{232}$Th decay chain, is the most viable candidate.

The isotope $^{208}$Tl undergoes $\beta^-$-decay through several
channels, all of which involve the emission of at least two photons,
implying that the $^{208}$Tl multiple photon (type {\bf C}) events are
efficient in triggering the telescope. The half-life of $^{208}$Tl is
only 3.1~min. If the isotope is part of a natural decay chain,
however, then its effective half-life is equal to the longest
half-life of any of its parent isotopes, which in this case is the
isotope $^{232}$Th with a half-life of $1.4 \times 10^{10}$~y. The
main channel for the production of background events by $^{208}$Tl is
the absorption of the 2.61~MeV photon in D2, with another photon
scattering in D1.

It was assumed that $^{208}$Tl is distributed like $^{40}$K and is
only present in the front glass of the D1 PMTs. This was tested by
recording $\gamma$-ray spectra of a spare D1 PMT using a Ge
spectrometer. Taking the COMPTEL results for the $^{40}$K and
$^{208}$Tl lines at face value, and assuming that both isotopes are
equally distributed in the D1 PMTs, one expects that the $^{208}$Tl
activity is about 10$^{-2}$~decays~s$^{-1}$~g$^{-1}$ in the front
window or about 6\% of that of $^{40}$K, based on the simulated
efficiencies, corresponding to a $^{232}$Th mass fraction of a few
$10^{-8}$.  Unfortunately, the laboratory measurement of $^{208}$Tl
was inconclusive. The 1.46~MeV $^{40}$K line is detected at the
12$\sigma$ level above the general background ($1700 \pm
140$~counts). The strongest $^{208}$Tl lines at 2.61~MeV and 0.58~MeV,
however, with intensities of 100\% and 85.5\%, were only detected at
the 0.9$\sigma$ ($50 \pm 56$~counts) and 1.4$\sigma$ ($128 \pm
92$~counts) level, respectively. Although consistent with
expectations, these numbers are insufficient to test the
assumptions. Much longer integration times are needed.

\section{\label{variations} Variability of instrumental lines}

The activity of isotopes contributing to the COMPTEL background
(except for the primordial radio-isotopes $^{40}$K and $^{208}$Tl) in
general varies in time. They are produced by interactions of primary
and secondary cosmic-ray particles and radiation-belt (SAA) particles
within the instrument material. The intensity of each of these
particle species, and hence the level of activation, changes in time
due to orbit details (e.g.\ the satellite's altitude, the geomagnetic
cut-off rigidity, the SAA radiation dose) and the solar cycle.
The activity of a specific radioactive isotope then results from the
competing processes of activation (production) and decay. 

The variation of the activity of a specific isotope is complex in
general. 
If the isotope's half-life $T_{1/2}$ and the typical time-scale(s)
$\tau$ for the intensity variation of the particle population(s)
producing it are very different, however, the activity variations can
be described rather simply.
For $T_{1/2} \ll \tau$, the isotope's activity approximatly follows
the incident cosmic-ray particle intensity and its orbital variation
(see Sect.~\ref{veto_variation}). This is, e.g., the case for
short-lived isotopes ($T_{1/2} \stackrel{\textstyle _<}{_{\sim}}$ a
few minutes) produced by primary cosmic-ray particles or neutrons,
such as $^{2}$D and $^{28}$Al. For $T_{1/2} \gg \tau$, a long-term
build-up occurs since the activated nuclei do not decay away between
consecutive SAA transits (see Sect.~\ref{time_variation}). This is,
e.g., the case for long-lived isotopes ($T_{1/2} \gg 90$~min, i.e.\
the orbital period) produced during SAA passages (which occur 6--8
times each day), such as $^{22}$Na and $^{24}$Na. Intermediate
half-lifes produce more complicated variability.

The long-term and orbital variations of spectral features are
useful in identifying their physical origin.
Spectral features arising from primordial radioactivity will show
no variation. Prompt and short-lived components will vary with
incident cosmic-ray intensity, i.e.\ on time-scales shorter than an
orbital period. In addition, prompt and short-lived components may
also vary over long time-scales if their production is sensitive to
changes in the cosmic-ray intensity due to the solar cycle or the
orbit altitude.  Spectral features due to long-lived isotopes will
exhibit variations on time-scales that reflect their half-life as well
as the changes in SAA dosage. With regard to a spectral feature not
yet identified, these long-term variations provide a crude estimate of its
half-life. With regard to a tentatively identified isotope, comparison
of its time-dependent activity with an empirical model serves as an
important cross-check for the correctness of its identification.

\subsection{\label{time_variation} Long-term variation}

The long-term variation of the activity of
long-lived radio-nuclides arises from the combined effects of
the isotopes' decay and the time history of the activation episodes
during SAA transits. Activation outside the SAA by cosmic-ray
particles is negligible (see e.g.\ Kurfess et al.\
\cite{kurfess_smm_lines}), since the average daily fluence due to SAA
particles by far exceeds that of cosmic-ray particles. 
The count rate in the CAL-units can be used to model the
production of long-lived radio-nuclides (Varendorff et al.\
\cite{varendorff_4c}, hereafter referred to as ``activity model'').
The small plastic scintillators (thickness 3~mm, diameter 12~mm)
of the CAL-units are sensitive to protons and other charged particles,
but insensitive to secondary $\gamma$-ray photons. They therefore provide a
measure of the charged-particle flux at the instrument (Snelling et
al.\ \cite{snelling}).
The trigger rates of the two CAL-units above both a low and a high
threshold value are available for all times, including SAA transits.
Most other instrument systems are switched off during SAA passages,
including the veto shields which therefore cannot be used as charged
particle monitors for the activity model.  The high-threshold trigger
rate of CAL-unit B ($\mathrm{r}_\mathrm{BH}$) is most useful because
it is the least affected by noise and long-term efficiency
degradation. Only telemetry data gaps interrupt this measure of the
cosmic-ray intensity.

To achieve continuous information on the charged-particle flux during
every SAA passage, the activity model employs a neural net to describe
$\mathrm{r}_\mathrm{BH}$ as a function of orbit
altitude, geographic longitude and latitude, time since launch (to
include variations due to the solar cycle), and orientation (azimuth
and zenith) of the satellite relative to its velocity vector (to
account for asymmetries in the incident SAA-proton flux, see e.g.\
Watts et al.\ \cite{watts_saa_anisotropy}). The number of nuclei of a
specific isotope as a function of time, $N(t)$, is then given by:
\begin{eqnarray}\label{nuclei_from_activation_model}
N(t + \delta t) & = & N(t) \cdot e^{-\lambda \cdot \delta t} +
N_\mathit{new}(t) \nonumber \\ & \mathrm{and} & \\ N_\mathit{new}(t) &
= & \alpha \cdot \delta t \cdot (\mathrm{r}_\mathrm{BH}(t) -
\mathrm{r}_\mathrm{BH,0}) \nonumber
\end{eqnarray}
where the time interval $\delta t$ usually is 16.384~s, $\lambda = \ln
2 / T_{1/2}$ is the decay constant of the isotope with half-life
$T_{1/2}$, $\alpha$ is the proportionality factor between the CAL-unit
trigger rate and the isotope activation, and
$\mathrm{r}_\mathrm{BH,0}$ is the value of $\mathrm{r}_\mathrm{BH}$
outside the SAA. The isotope activity $A(t)$ then is
\begin{equation}\label{isotope_activity}
A(t) = \lambda \cdot N(t)
\end{equation} 
The value of $\alpha$ is not well known as it depends on the
cross-sections for the production of the isotope in the telescope
material, particularly in the D1 detector.  For each isotope, a
scaling factor, which is the product of $\alpha$ and the efficiency
for triggering a background event (see
Sect.~\ref{comparisons_implications}), must be determined from a
measurement of the event rate from the COMPTEL data.

\begin{figure*}
\centerline{\hbox{
\epsfig{figure=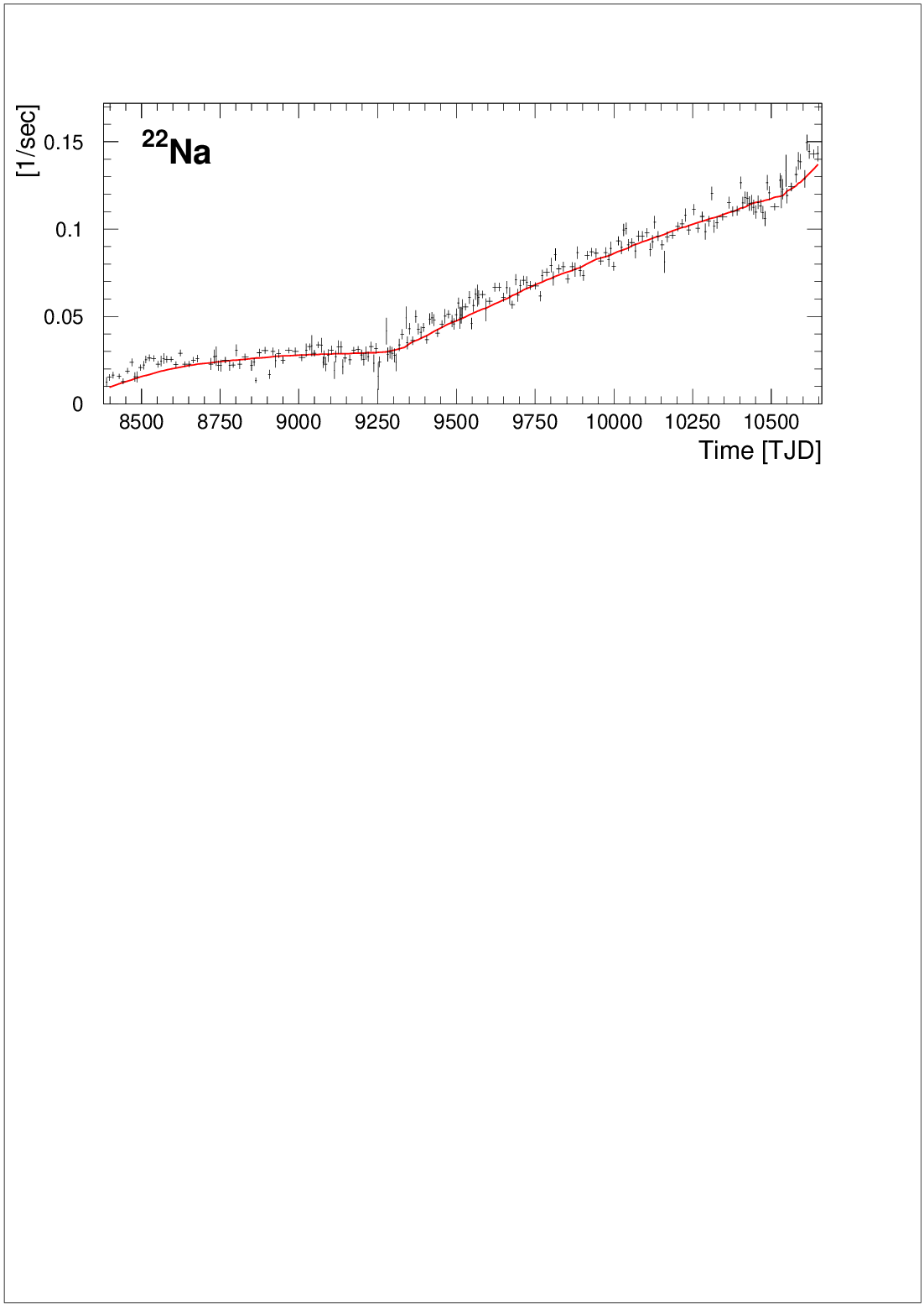,%
bbllx=38pt,bblly=526pt,bburx=548pt,bbury=744pt,width=8.8cm,clip=}
\epsfig{figure=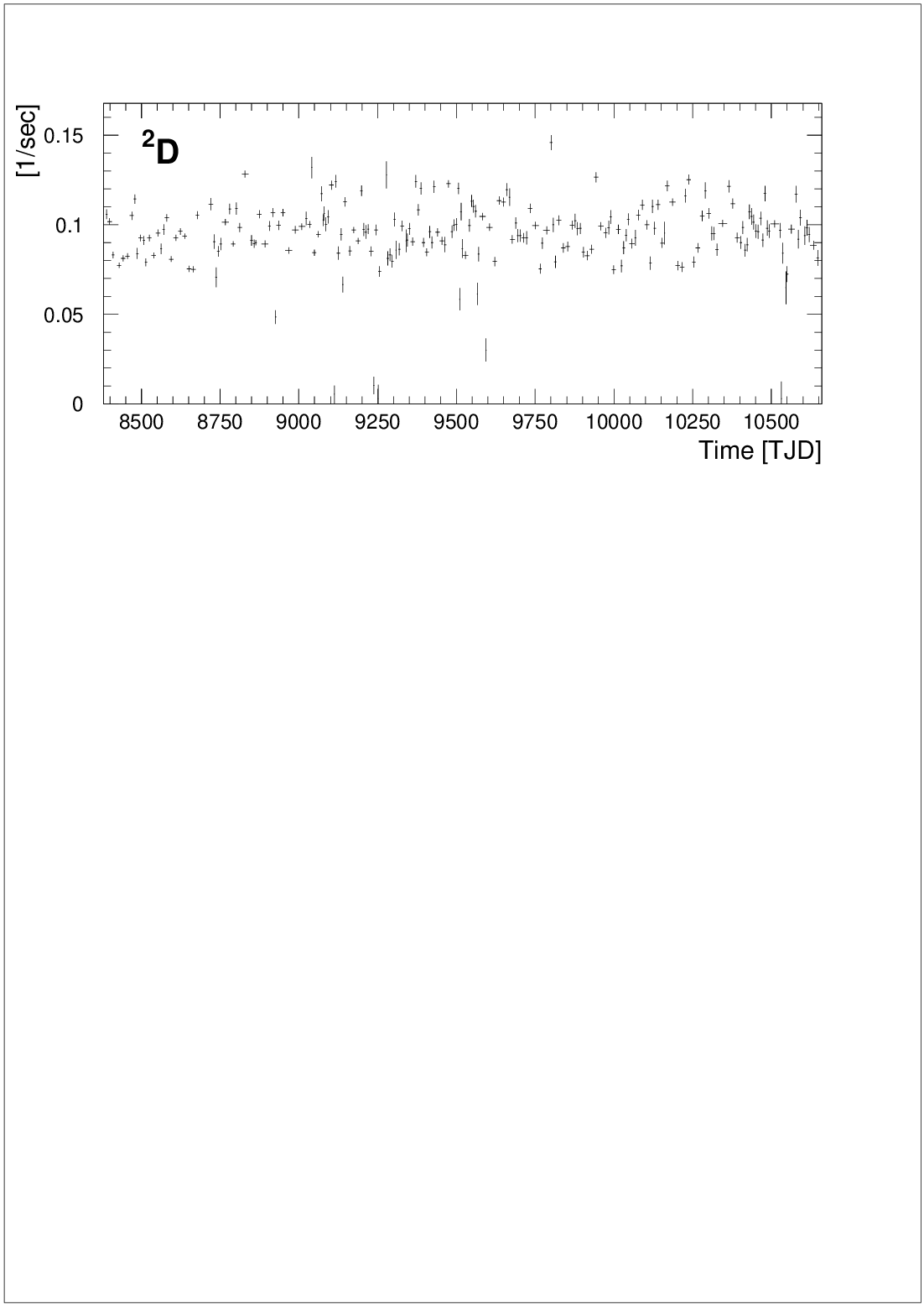,%
bbllx=38pt,bblly=526pt,bburx=548pt,bbury=744pt,width=8.8cm,clip=}
}}
\centerline{\hbox{
\epsfig{figure=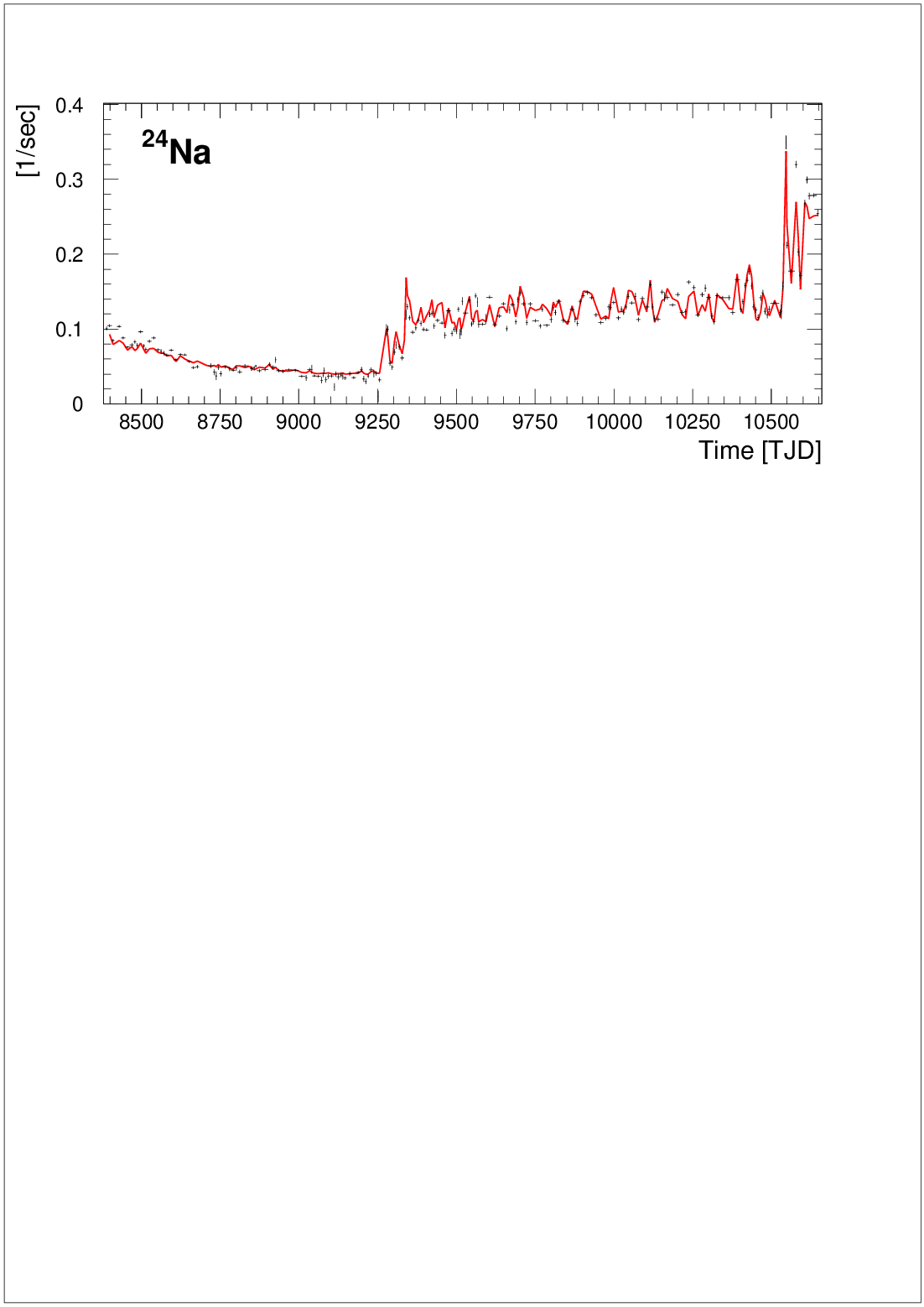,%
bbllx=38pt,bblly=526pt,bburx=548pt,bbury=744pt,width=8.8cm,clip=}
\epsfig{figure=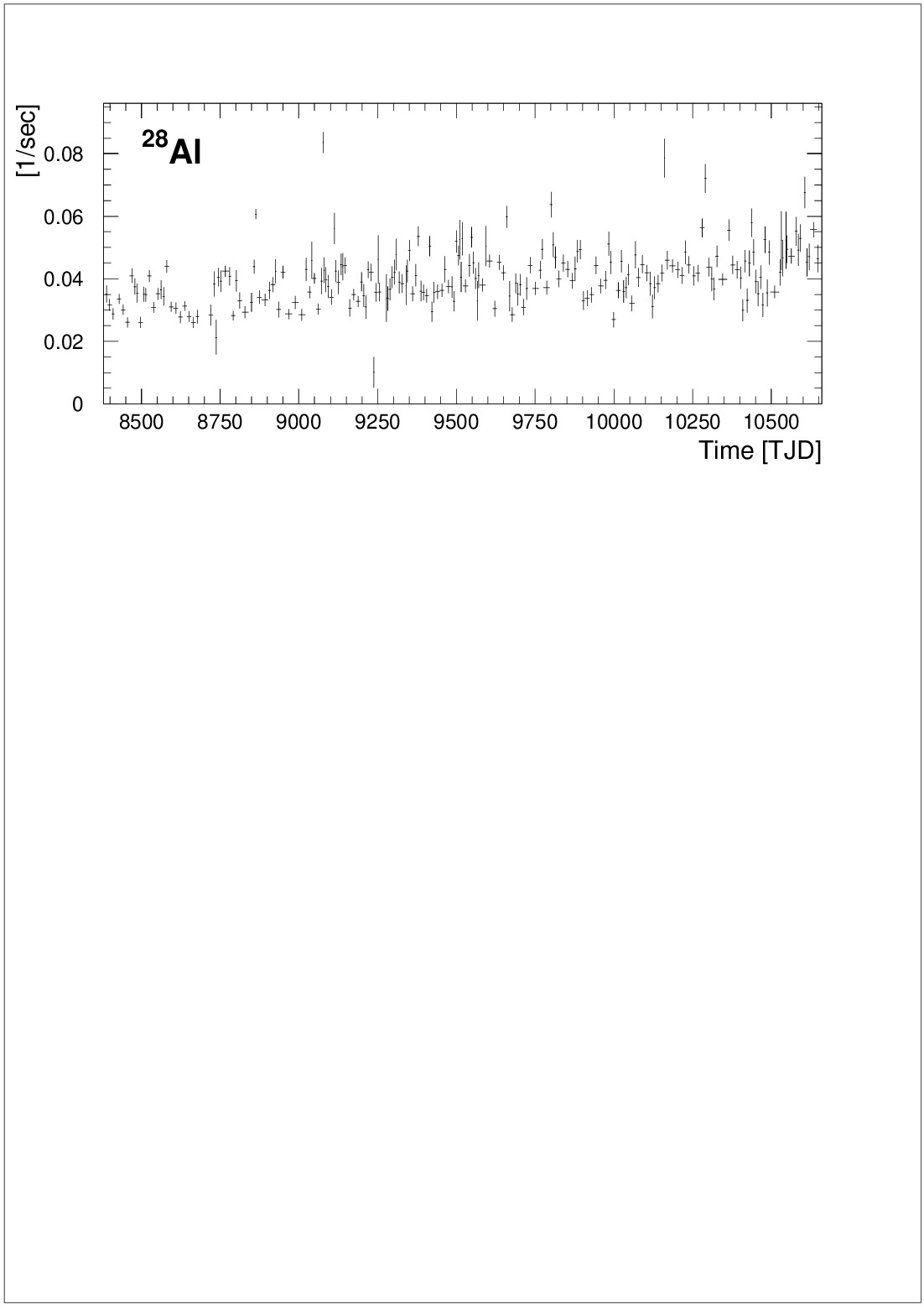,%
bbllx=38pt,bblly=526pt,bburx=548pt,bbury=744pt,width=8.8cm,clip=}
}}
\centerline{\hbox{
\epsfig{figure=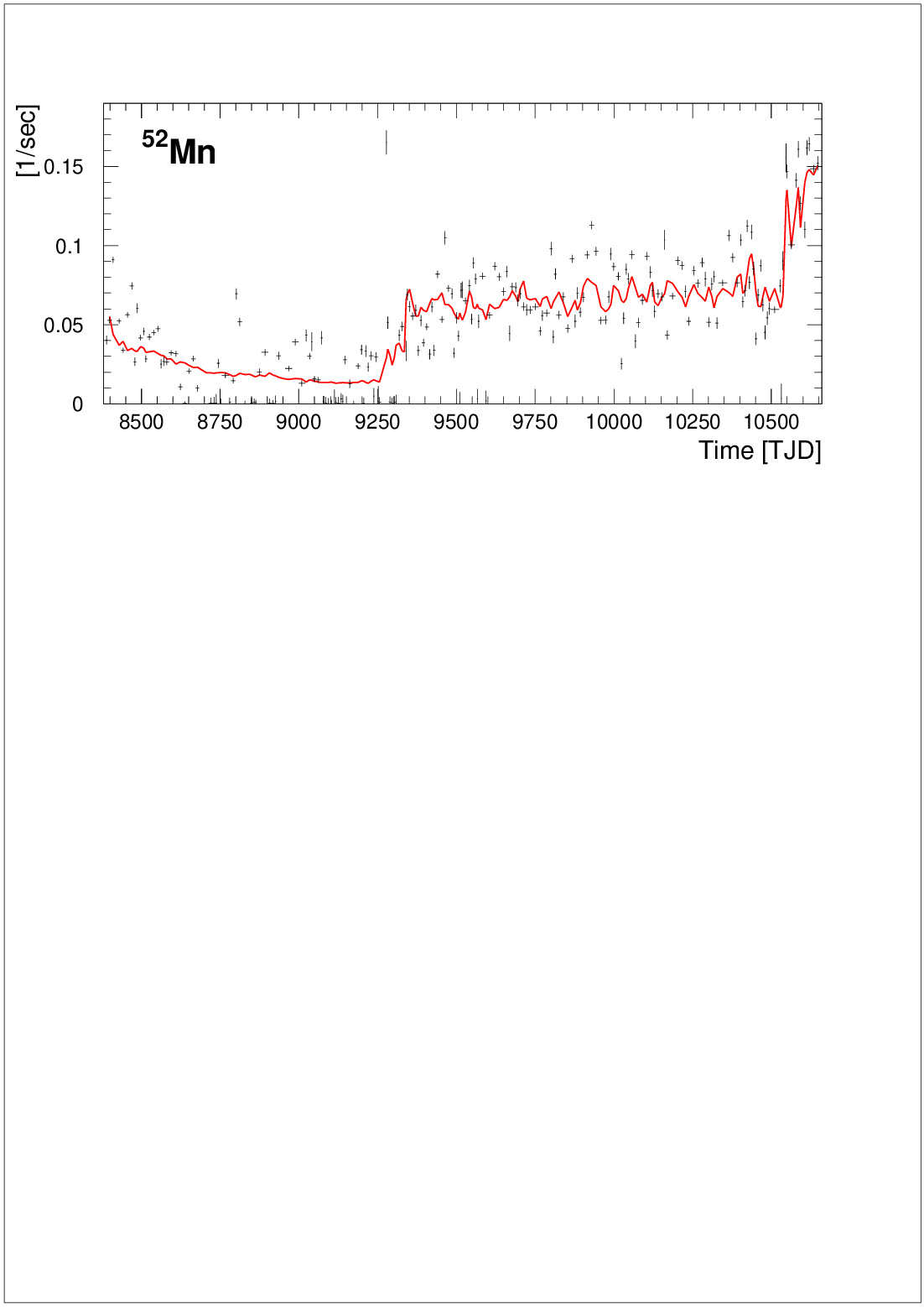,%
bbllx=38pt,bblly=526pt,bburx=548pt,bbury=744pt,width=8.8cm,clip=}
\epsfig{figure=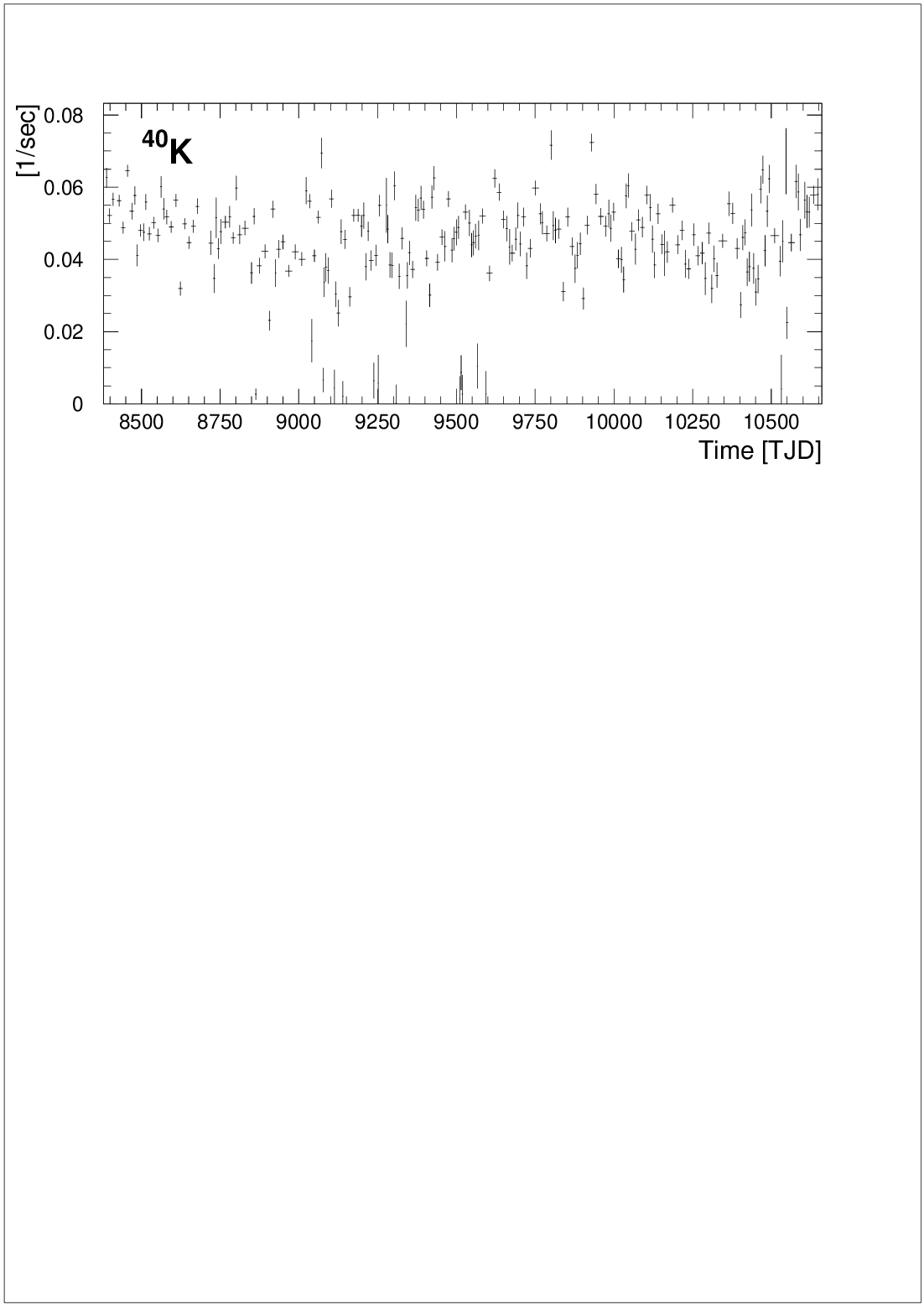,%
bbllx=38pt,bblly=526pt,bburx=548pt,bbury=744pt,width=8.8cm,clip=}
}}
\centerline{\hbox{
\epsfig{figure=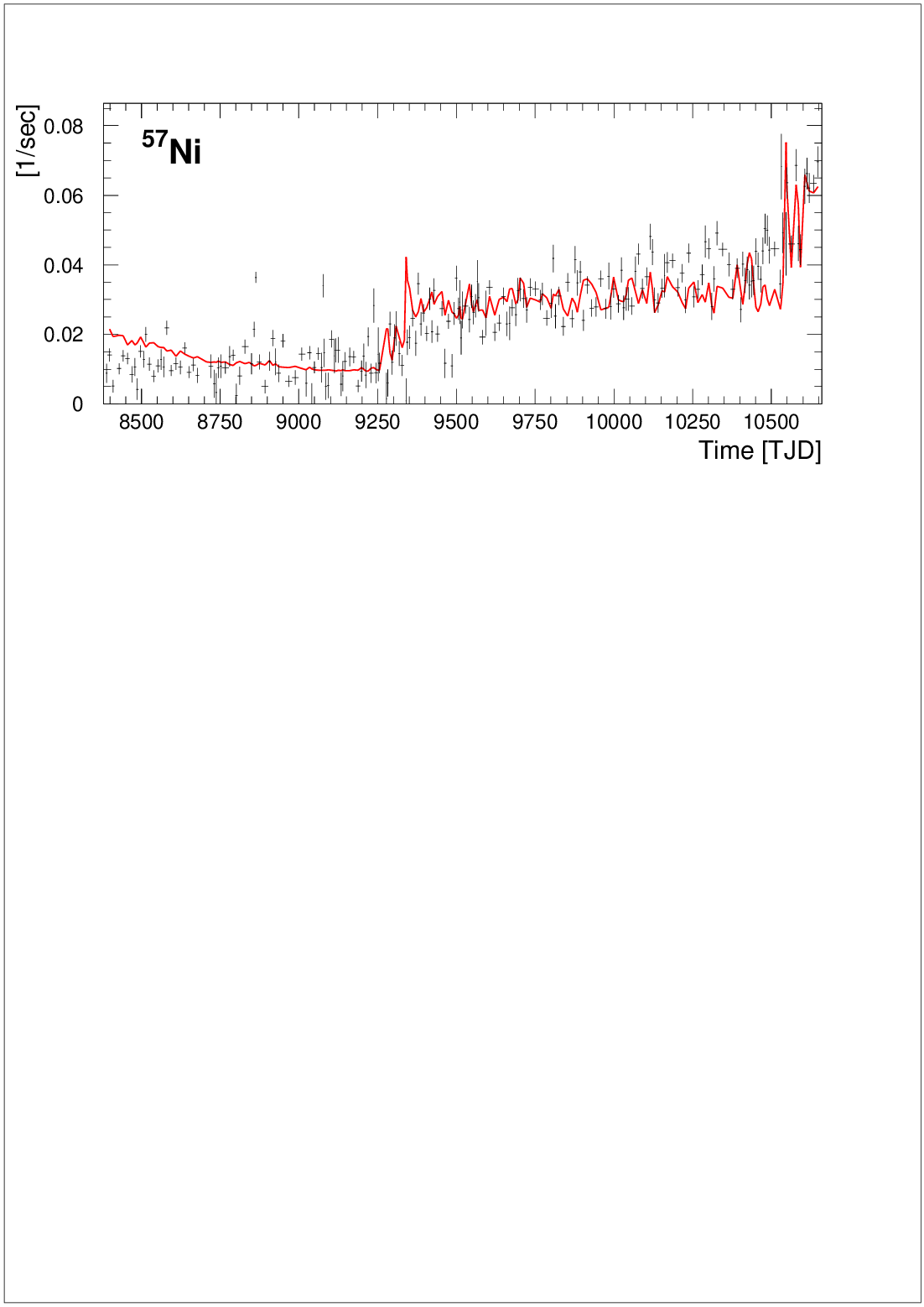,%
bbllx=38pt,bblly=526pt,bburx=548pt,bbury=744pt,width=8.8cm,clip=}
\epsfig{figure=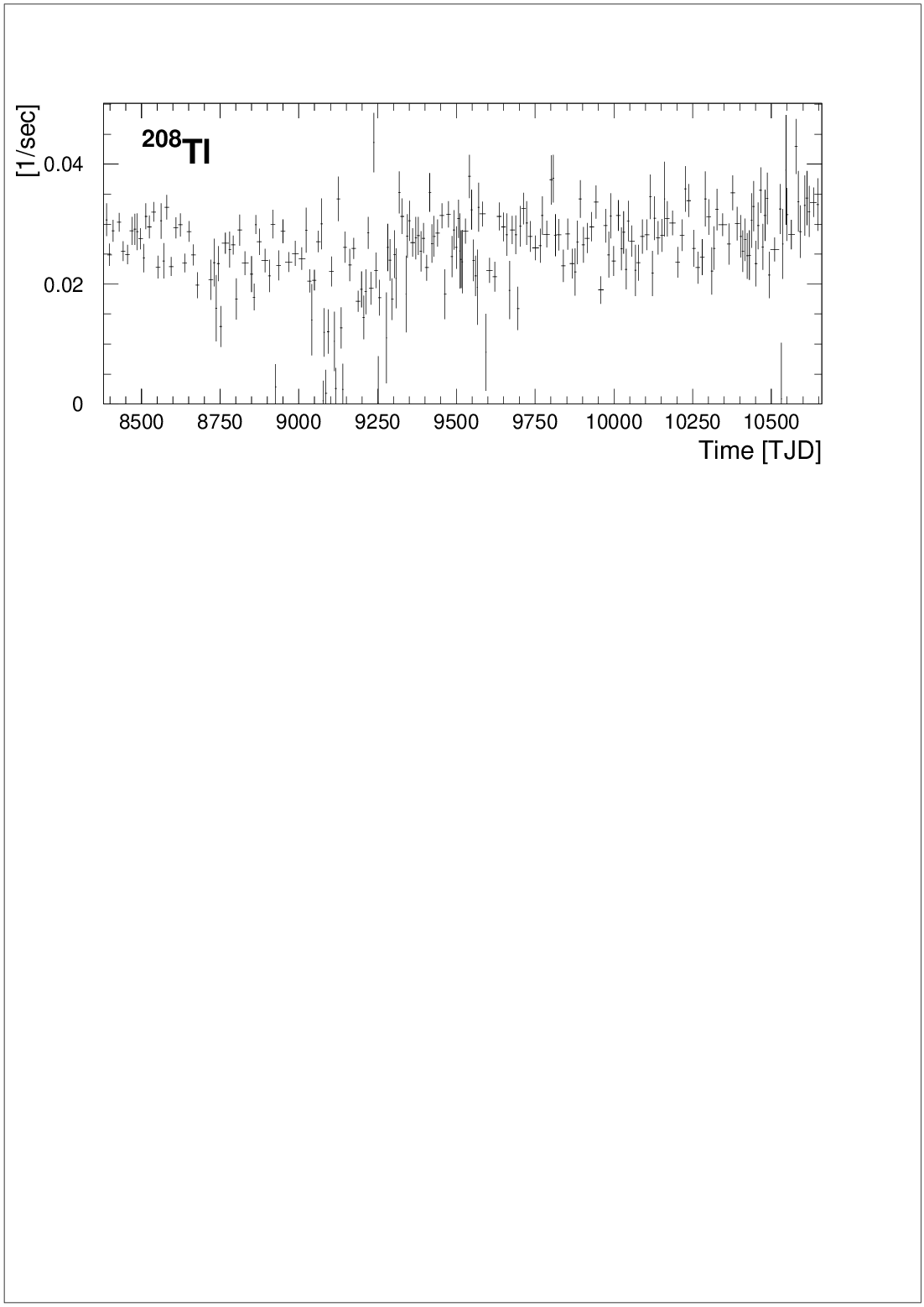,%
bbllx=38pt,bblly=526pt,bburx=548pt,bbury=744pt,width=8.8cm,clip=}
}}
\caption[]{The measured event rates due to the long-lived isotopes
$^{22}$Na, $^{24}$Na, $^{52}$Mn, and $^{57}$Ni, the
short-lived isotopes $^{2}$D and $^{28}$Al, and the primordial
isotopes $^{40}$K and $^{208}$Tl as a function of time for the
first 6~years (May~1991 through July~1997) of the CGRO mission for
imaging event selections. In addition, the predictions of the
normalized activity model for the long-lived isotopes is depicted with
solid grey lines.}
\label{lines_long-term_plot} 
\vspace{2ex}
\centerline{
\begin{minipage}{8.8cm}\makebox[0cm]{}\\
\epsfig{figure=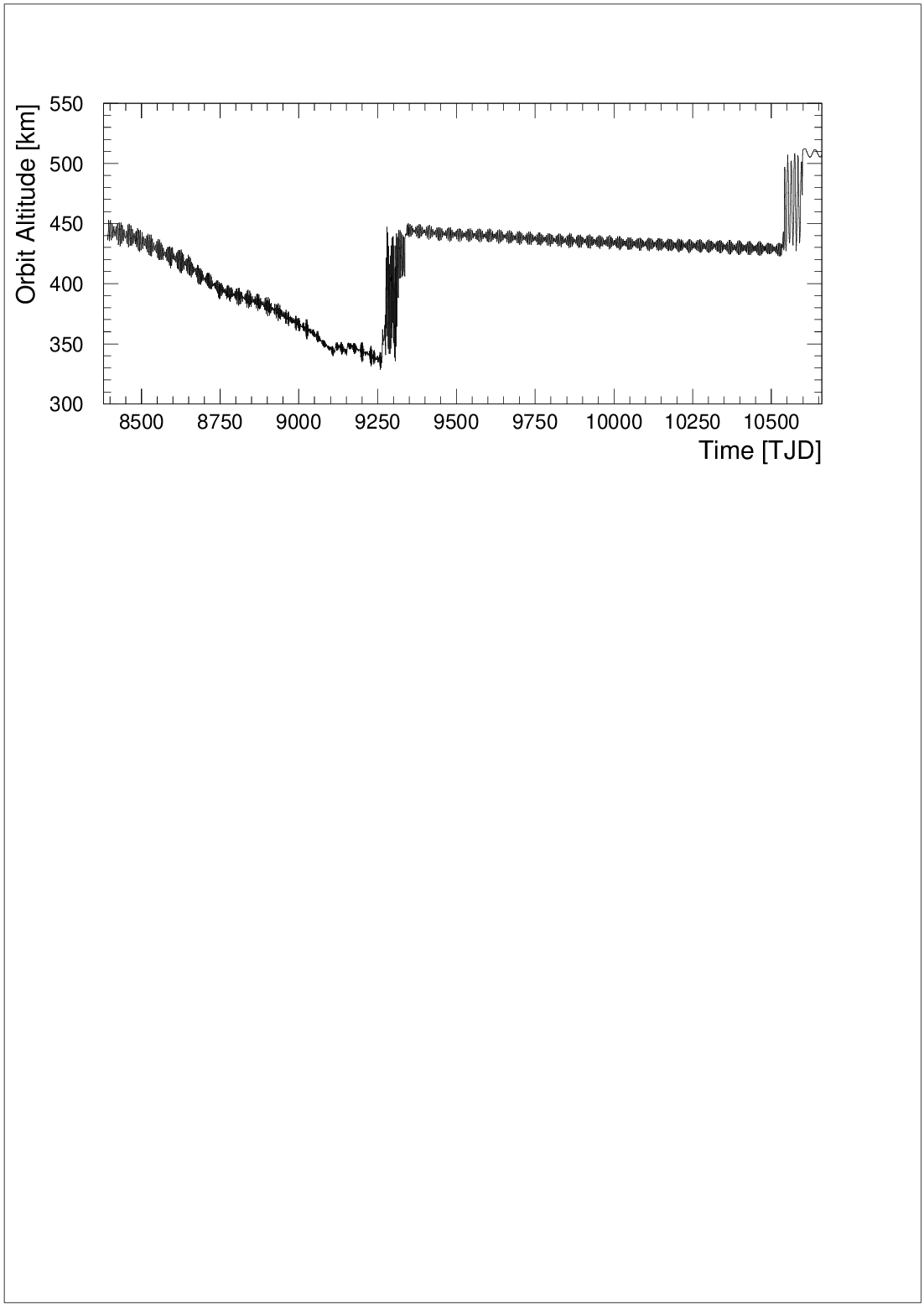,%
bbllx=38pt,bblly=525pt,bburx=544pt,bbury=744pt,width=8.8cm,clip=}
\end{minipage}
\begin{minipage}{8.8cm}
\caption[]{The altitude of the CGRO orbit as a function of time from
May~1991 through July~1997. The two reboosts of the orbit around
TJD~9280 (Oct.--Dec.\ 1993) and TJD~10560 (Apr.--May 1997) are clearly
seen. The rapid decay of the orbit at the beginning of the mission
results from the increased drag of the expanded atmosphere at solar
maximum.}
\label{cgro_orbit-alt}
\end{minipage}
}
\end{figure*}

A comparison of the measured event rates due to the long-lived
isotopes $^{22}$Na, $^{24}$Na, $^{52}$Mn, and $^{57}$Ni with the
predictions of the normalized activity model from May~1991 through
July~1997 is depicted in the left panels of
Fig.~\ref{lines_long-term_plot}. In addition, the long-term variation
determined for the event rates arising from the short-lived isotopes
$^{2}$D and $^{28}$Al, as well as from the primordial radio-nuclides
$^{40}$K and $^{208}$Tl are shown in the right panels of
Fig.~\ref{lines_long-term_plot}. These rates were obtained from the
study of the galactic $^{26}$Al 1.8~MeV line emission, which requires
a measure of the isotopes' background contributions for each
observation period (see App.~\ref{line_fitting_al26}) under imaging
event selections (see App.~\ref{event_selections_imaging}). The
scatter of the eight sets of event rates, particularly those for the
primordial radio-isotopes $^{40}$K and $^{208}$Tl, is larger than what
is expected from statistics alone. Part of the additional scatter is
expected from intrinsic inadequacies of the fitting procedure (such as
the correction for the Earth-horizon angle selection, see
App.~\ref{line_fitting_al26}) and of the fit models employed (these
cannot account for all structures in the E$_\mathrm{2}$ and
E$_\mathrm{tot}$ spectra, which particularly affects weak components,
see below and App.~\ref{line-fitting_procedure}). Partly, however, the
additional scatter may be attributed to known physical effects. Due to
the precession of the satellite orbit the radiation dose from SAA
transits may vary considerably between observation periods. In
addition, the orientation of the satellite relative to the anisotropic
SAA-particle flux affects the radiation dose in the D1 detector, which
is the major source of instrumental line background. The resulting
differences in SAA radiation dose between observation periods
contribute to the scatter in the event rates of the long-lived
isotopes. Since the CAL-units are located half-way between the D1 and
D2 detectors, shielding effects due to surrounding spacecraft
materials may result in less variation in $r_\mathrm{BH}$ than in the
D1 radiation dose, so that the activity model would underpredict this
variation.  Orientation effects may also result in variations of the
effective cosmic-ray particle flux between observation periods and add
some scatter to the event rates of short-lived isotopes.

The long-term variation of the long-lived isotopes reflects the
variation of the SAA radiation dose, which depends, among other
parameters, on orbit altitude and solar cycle. $^{24}$Na is the
best tracer of the changing intensity of the encountered SAA-particle
fluxes, as it is the strongest component of the line background and
has a half-live of about 15~h. From the beginning of the mission until
the first reboost (around TJD~9280), the decay of the orbital altitude
(see Fig.~\ref{cgro_orbit-alt}) results in a monotonic decrease of the
incident SAA-particle fluxes. The SAA radiation dose increased after
the first reboost; the effect of the slow decrease of the orbital
altitude is probably more than compensated by the solar cycle, which
proceeded towards solar minimum, resulting in the observed net
increase of SAA radiation dose over time. The second reboost (around
TJD~10560) again resulted in a significant increase of activation
during SAA passages. The long-term variation of the $^{52}$Mn and
$^{57}$Ni event rates is similar to that of $^{24}$Na, since the
half-lifes of these isotopes are of the same order. The
long-term behaviour of the $^{22}$Na event rate is quite
different. Its 2.6~y half-life is significantly
longer than the typical time-scale of the changes in SAA
radiation dose. Therefore $^{22}$Na never reaches an equilibrium
between activation and decay, but is continuously built up, with the
two reboosts resulting in increases in the slope of the $^{22}$Na
activity trend.

The fact that the activity model reproduces the long-term variation of
the event rates from long-lived isotopes confirms the model
assumptions, in particular that activation predominantly occurs during
SAA transits. The model can also verify the isotope identifications,
as the activity of an isotope depends, among other
parameters, on its half-life (see
Eqs.~\ref{nuclei_from_activation_model} and \ref{isotope_activity}).

The background produced by the short-lived isotopes $^{2}$D and
$^{28}$Al is not expected to be influenced by SAA radiation dose,
rather, any long-term trends will arise from variations in the average
cosmic-ray intensity. The count rate in the 2.22~MeV line does not
exhibit any significant long-term variation. The orbital variation of
the event rate in the 2.22~MeV line (Weidenspointner et al.\
\cite{weidenspointner_3c}) is similar to that of the fast-neutron flux
($\mathrm{E}_n > 12.8$~MeV) in the D1 detector (Morris et al.\
\cite{morris_fast_neutrons}), as discussed in
Sect.~\ref{veto_variation}. In addition, Morris et al.\
(\cite{morris_fast_neutrons}) demonstrated that the solar cycle
variation of the fast-neutron flux is much less than its orbit
variation. It is therefore expected that the relatively weak solar
cycle dependence of the 2.22~MeV line is dominated by the much greater
orbital variations, which presumably accounts for much of the observed
scatter.  Similar to $^{2}$D, the isotope $^{28}$Al is predominantly
produced by thermal-neutron captures. Hence $^{28}$Al is expected to
exhibit the same long-term variation as $^{2}$D. However, the low
count rate from $^{28}$Al exhibits a slight, increasing trend, hinting
at deficiencies of the fit models and/or contributions from yet
unidentified isotopes.

The activity of the primordial radio-nuclides $^{40}$K and
$^{208}$Tl is constant over the duration of
the mission. Both data sets are consistent with this fact,
although a small long-term increase may be present for $^{208}$Tl.
Again, this apparent increase may be due to deficiencies of the
fitting procedure and/or due to yet unidentified
line background components.

The activity model can not only be used to predict the number of
nuclei activated during SAA passages, but also to estimate the average
daily SAA-proton fluence from the daily increase in the number of
proton-produced nuclei. For example, based on the measured
$^{22}$Na event rate a daily average SAA-proton fluence
($\mathrm{E}_p > 100$~MeV) of $2.3 \times 10^5$~protons~cm$^{-2}$ was
inferred by Varendorff et al.\ (\cite{varendorff_4c}) for the
beginning of the mission during solar maximum at an altitude of
440~km. Considering the uncertainties in this measurement, and the
large altitude gradient of the SAA-proton flux (Stassinopoulos
\cite{stassinopoulos}), this value is in good agreement with the
prediction of $5 \times 10^5$~protons~cm$^{-2}$ for an altitude of
462~km (Dyer et al.\ \cite{dyer_cgro-bgd}).

\subsection{\label{veto_variation} Variation with cosmic-ray intensity}

\begin{figure}
\epsfig{figure=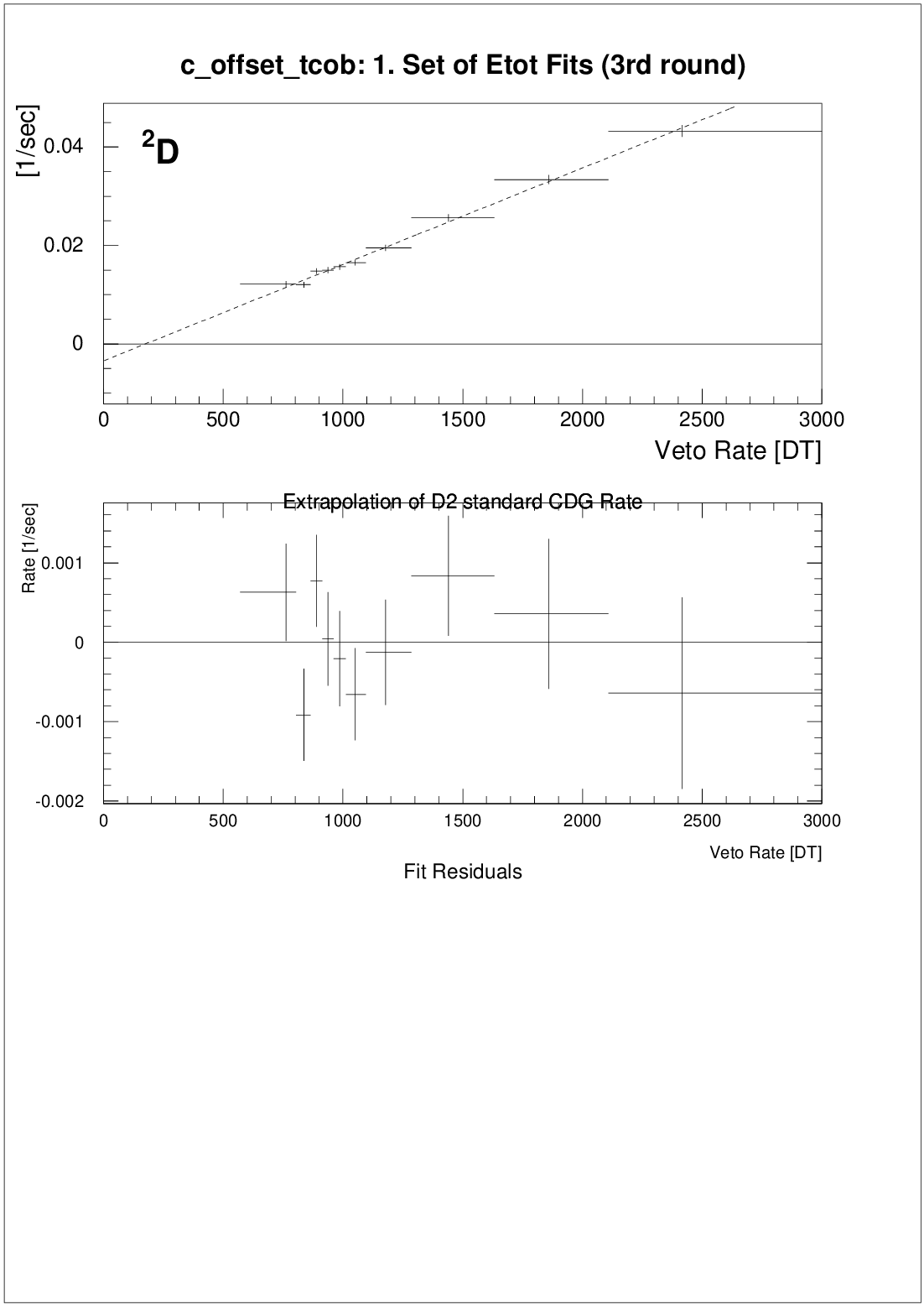,%
bbllx=38pt,bblly=525pt,bburx=518pt,bbury=742pt,width=8.8cm,clip=}
\\[2ex]
\epsfig{figure=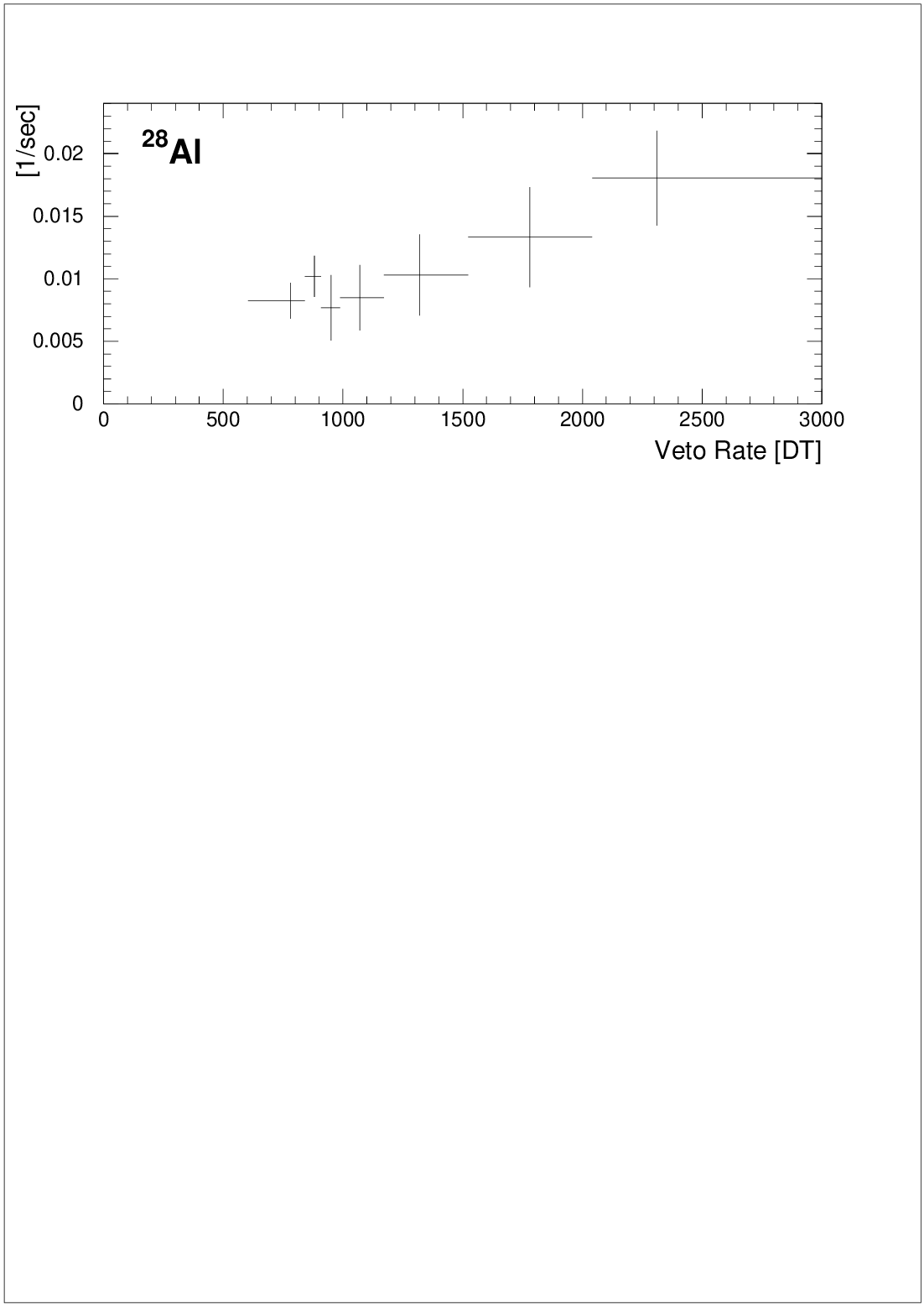,%
bbllx=38pt,bblly=525pt,bburx=518pt,bbury=742pt,width=8.8cm,clip=}
\caption[]{The veto rate variation of the count rate due the
background isotopes $^{2}$D and $^{28}$Al as measured for CDG
event selections. The dashed line is a linear fit to the measured
$^{2}$D event rate.}
\label{d2-al28_veto-var}
\end{figure}

The prompt instrumental background closely tracks the local,
instantaneous cosmic-ray intensity, which can, e.g., be parameterized
by a geomagnetic cut-off rigidity. Another way of parameterizing the
incident cosmic-ray intensity is to use the count rate of the
anti-coincidence domes of the COMPTEL instrument, referred to as
``veto rate'' in the following\footnote{This veto rate is the number
of deadtime clock counts that occur during the presence of a veto
signal. For each veto dome the deadtime clock counts are accumulated
over the length of a telemetry packet (2.048~s) once during each
superpacket (16.384~s) and recorded in the house-keeping datasets in
units of [deadtime clock counts/2.048~s~$\equiv$~DT]}. To a good
approximation, the prompt background components vary in proportion to
the incident cosmic-ray intensity as monitored by the veto rate.  The
best example of this are $^{2}$D and $^{28}$Al
(Fig.~\ref{d2-al28_veto-var}), determined for CDG event selections
(see App.~\ref{event_selections_cdg}).  In the following discussion,
``veto rate'' and ``incident cosmic-ray intensity'' are therefore used
interchangeably.
In contrast to prompt background components, the activity of the
primordial radio-isotopes $^{40}$K and $^{208}$Tl is independent
of incident cosmic-ray intensity and hence does not vary with veto
rate. The activity variation of long-lived isotopes such as
$^{22}$Na with veto rate is complex and depends on the isotopes'
half-life as well as on the geophysical environment. Similar to the
study of the long-term variation, the study of the variation of the
event rate of a background component with cosmic-ray intensity can be
helpful in identifying the responsible isotope.

In the following, the physical origin of the complex and -- in general
-- non-linear variation of the event rate of long-lived isotopes with
incident cosmic-ray intensity as monitored by veto rate is illustrated
with the isotope $^{22}$Na, which has a half-life of 2.6~y. To
investigate the variation of a line background component with veto
rate over a given time period the events are sorted in energy spectra,
one for each of a set of veto rate intervals, according to the veto
rate value at the time the events were recorded. The event rate due
to the background isotope under study is then determined for each veto
rate bin in an iterative line fitting procedure described in
App.~\ref{line_fitting_cdg}. Each of these rates represents the
``average'' event rate due to the isotope in a given veto rate
interval over a given time period.

\begin{figure}
\epsfig{figure=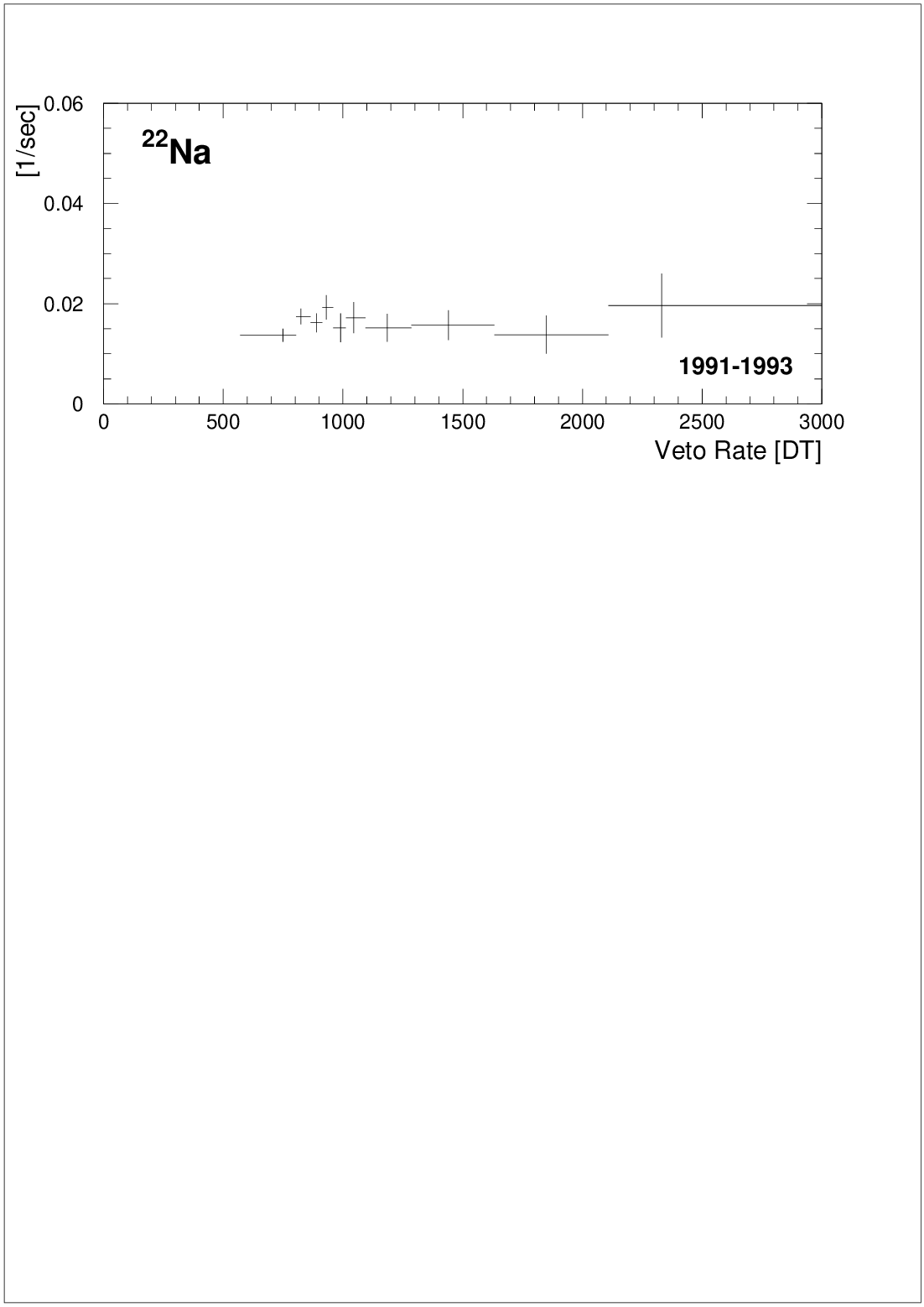,%
bbllx=38pt,bblly=525pt,bburx=520pt,bbury=742pt,width=8.8cm,clip=}
\\[2ex]
\epsfig{figure=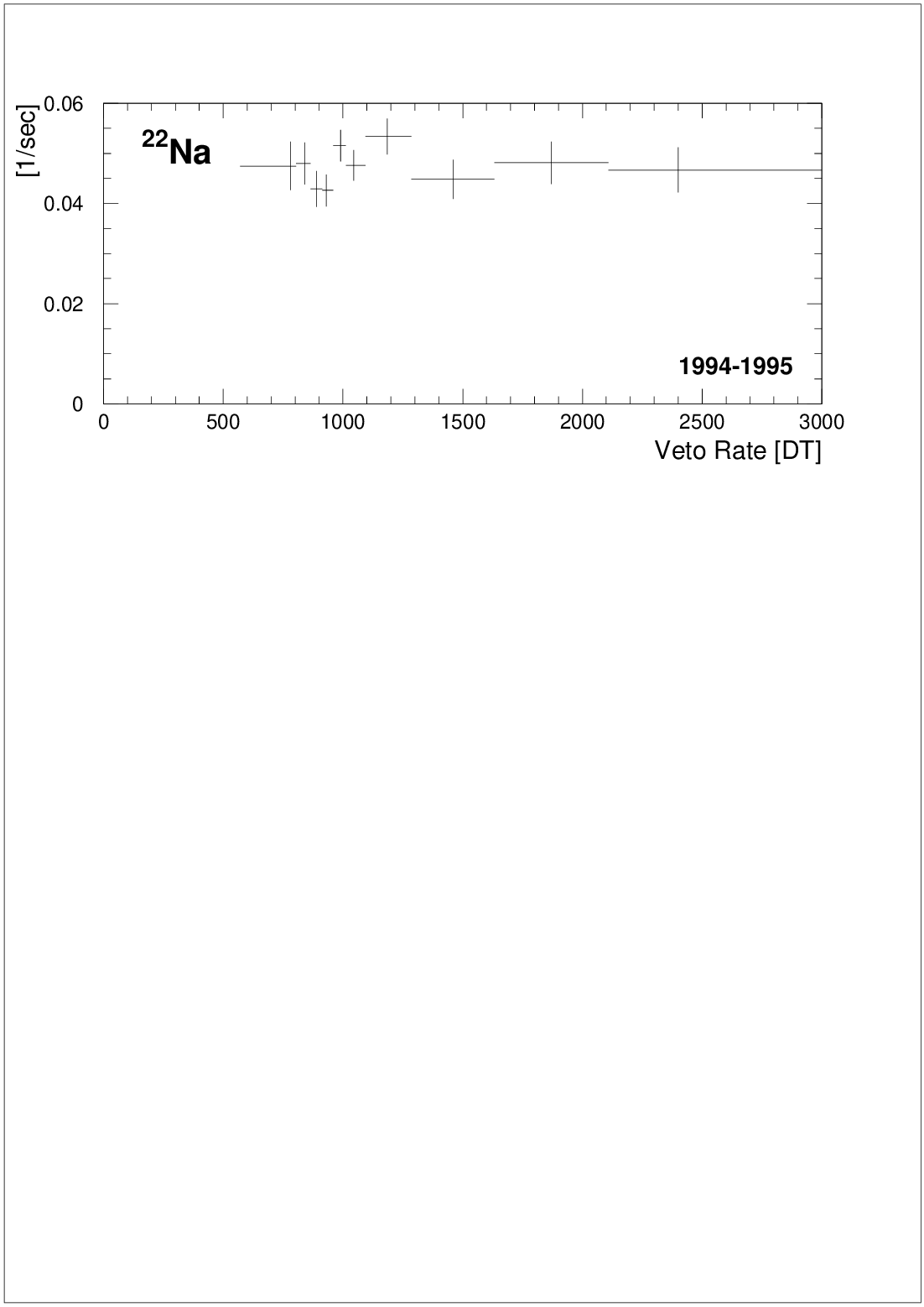,%
bbllx=38pt,bblly=525pt,bburx=520pt,bbury=742pt,width=8.8cm,clip=}
\caption{The count rate from $^{22}$Na, determined under CDG event
selections, as a function of veto rate for data from 1991--1993 and
from 1994--1995.}
\label{na22_ind_phases}
\vspace{5ex}
\epsfig{figure=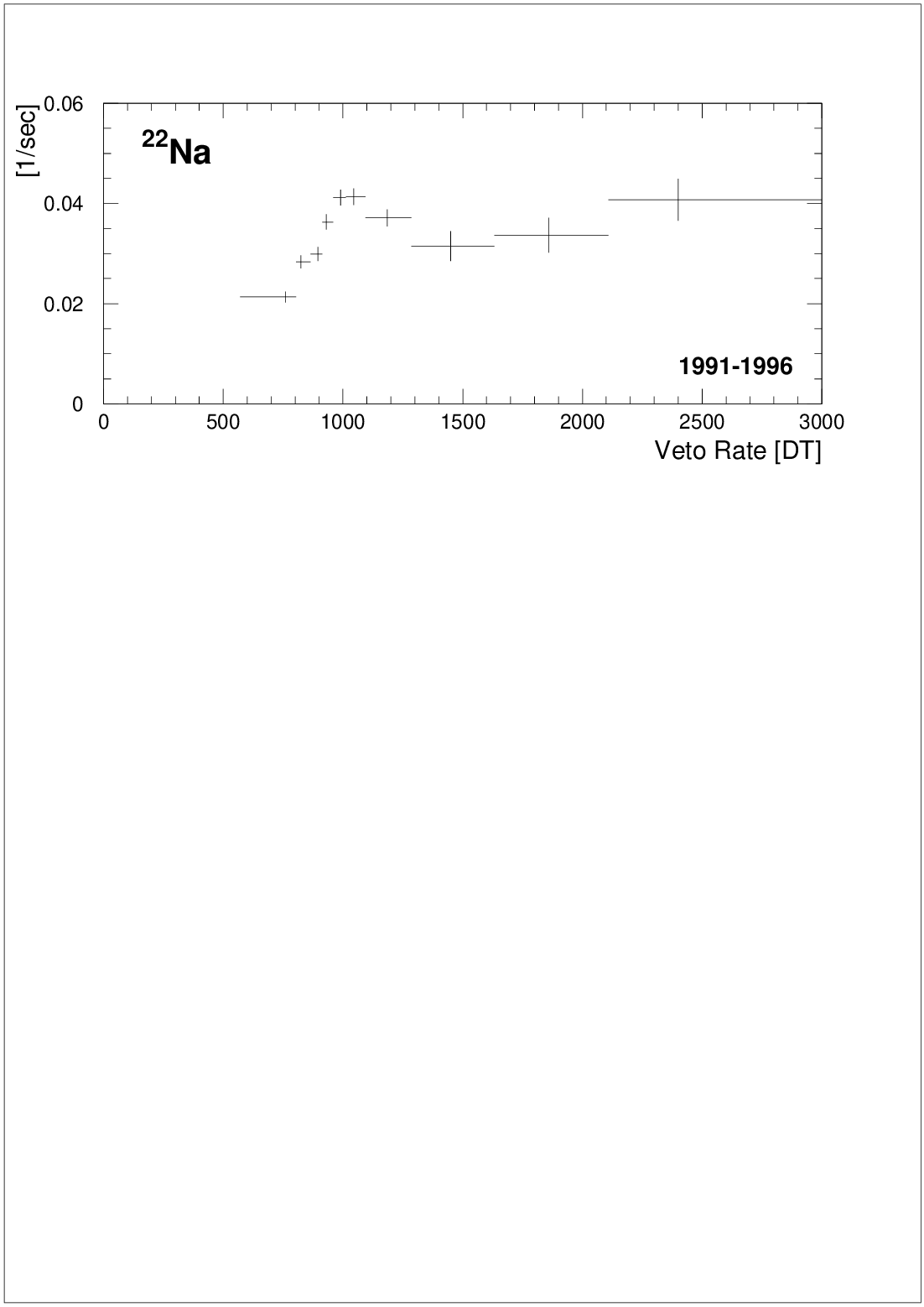,%
bbllx=38pt,bblly=525pt,bburx=520pt,bbury=742pt,width=8.8cm,clip=}
\caption{The count rate from $^{22}$Na, determined under CDG event
selections, as a function of veto rate for data from 1991--1996.}
\label{na22_comb_phases}
\end{figure}

The veto rate variation of the event rate due to $^{22}$Na over
relatively short periods of time, during which the geophysical
environment did not change significantly, is illustrated in
Fig.~\ref{na22_ind_phases} for data from 1991--1993 (CGRO Phases~I and
II) and from 1994--1995 (CGRO Cycle~4). To a good approximation, the
$^{22}$Na event rate is independent of veto rate, as expected for a
half-life of 2.6~years. The average $^{22}$Na event rate increased
with time due to the build-up of this isotope during successive SAA
passages (see Fig.~\ref{lines_long-term_plot}). An increasing
background contribution is also observed for long-lived isotopes with
shorter half-lifes, such as $^{24}$Na with $T_{1/2} =
15$~h. However, due to a correlation between the encountered
geomagnetic cut-off rigidity and the time since the last SAA transit,
there is some correlation between the event rate of these isotopes and
veto rate, which is generally complex even over relatively short
periods of time. Nevertheless, the following conclusions drawn from
the veto rate variation of the $^{22}$Na event rate apply to
long-lived isotopes with shorter half-lifes such as $^{24}$Na as
well.

The average variation of the event rate arising from $^{22}$Na over
the extended time period from 1991--1996 (CGRO Phase~I through
Cycle~5), during which the geophysical environment changed
significantly due to the first reboost (see Fig.~\ref{cgro_orbit-alt})
and the solar cycle, is depicted in
Fig.~\ref{na22_comb_phases}. Typically, a bump-like feature, hereafter
referred to as the veto rate bump, appears at low veto rate values
when studying the veto rate variation of the background contribution
of long-lived isotopes over extended periods of time that include at
least one reboost. This is even true for an isotope such as
$^{22}$Na, whose event rate is independent of veto rate over short
periods of time. It must be emphasized, however, that changes in the
geophysical environment do not always produce a bump-like feature, but
rather result in a complex, non-linear variation of the contribution
from long-lived background components with veto rate.

\begin{figure}
\epsfig{figure=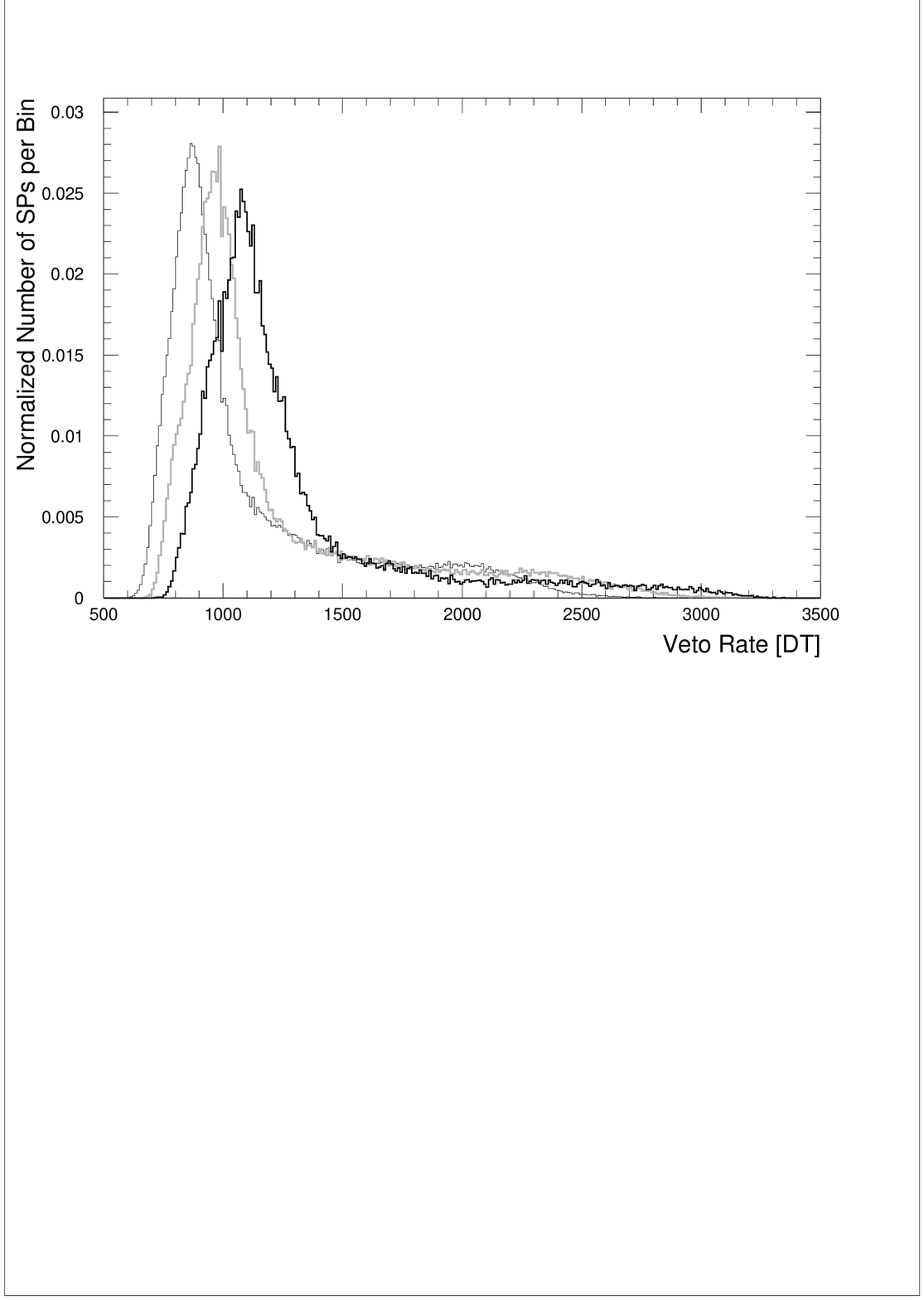,width=8.8cm,clip=}
\caption[]{The variation of the normalized veto rate frequency
distribution with time, exemplified for veto dome~2. Depicted are the
normalized veto rate frequency distributions, in units of 16.384~s
time intervals (superpackets, SP) per veto rate bin, for the time periods
1991/1992 (thin black line), 1994/1995 (grey line), and 1997/1998
(thick black line).}
\label{time-var_of_veto-distr}
\end{figure}

To understand the origin of the veto rate bump it is necessary to
realize that the changing geophysical environment not only affects the
activity of long-lived isotopes (see Fig.~\ref{lines_long-term_plot}),
but also the veto rate frequency distribution, mostly due to
variations of the incident cosmic-ray intensity with orbit altitude
and solar cycle. This is illustrated in
Fig.~\ref{time-var_of_veto-distr}, depicting the normalized veto rate
frequency distributions for the time periods 1991/1992, 1994/1995, and
1997/1998 (i.e.\ CGRO Phase~I, Cycle~4, and Cycle~7, respectively).
For clarity, the distributions for other Phases/Cycles were omitted as
they are fully consistent with the trend described in the
following. The generic veto rate distribution is characterized by a
pronounced peak at low veto rates, with a flat tail extending to
higher veto rate values. To a good approximation, the time variation
of this generic veto rate frequency distribution can be described by a
slow but steady shift of the peak to higher veto rate values, with
some slight changes of the shape of the distribution such as an
increasing extension of the tail.

\begin{figure}
\resizebox{\hsize}{!}{\includegraphics{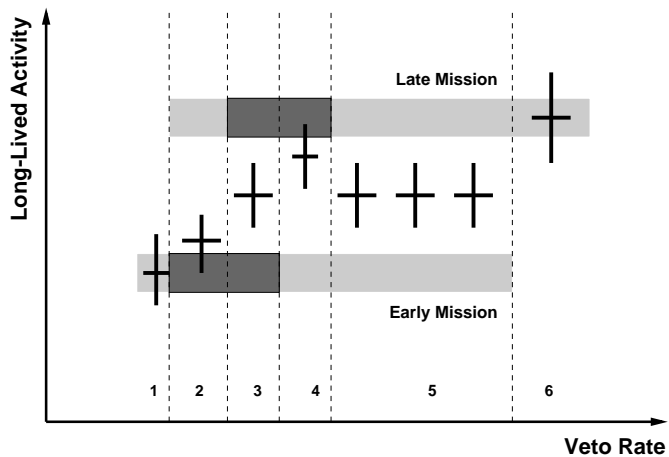}}
\caption[]{Illustration of the origin of the veto rate bump. Features
such as the veto rate bump result from the combined effects of the
time variation of the isotope activity (see
Fig.~\ref{lines_long-term_plot}) and of the veto rate frequency
distribution (see Fig.~\ref{time-var_of_veto-distr}), as explained in
detail in the text.}
\label{veto-rate-bump_illustration}
\end{figure}

As is illustrated in Fig.~\ref{veto-rate-bump_illustration}, the veto
rate bump is nothing more than an artifact of the combined effects of
the long-term variation in the activity of long-lived isotopes (see
Fig.~\ref{lines_long-term_plot}) and of the long-term shift of the
veto rate frequency distribution (see
Fig.~\ref{time-var_of_veto-distr}).  For simplicity, let us consider a
long-lived isotope whose activity is independent of veto rate over
periods of time during which the geophysical conditions do not change,
such as $^{22}$Na. Let us further consider a combination of two such
time intervals of equal duration from early and late in the
mission. Late in the mission the isotope activity is higher than early
in the mission, and the veto rate distribution is shifted to higher
veto rate values. The two sets of data are represented by the two bars
in Fig.~\ref{veto-rate-bump_illustration}, with the grey-shading
indicating the veto rate distribution in terms of the number of
16.384~s sampling intervals (superpackets, see Footnote~4) per veto
rate interval, i.e.\ dark grey indicates that the corresponding veto
rate values occurred more often (compare
Fig.~\ref{time-var_of_veto-distr}).
The average activity (determined as described above) then increases
from interval~1 through interval~4 as the late-mission data become
relatively more important. In interval~5 both sets of data contribute
equally and the average activity is the mean of the individual
activities. Finally, only data from the late mission contribute to
veto rate interval~6.
The net result for the average
activity exhibits a general tendency to increase with increasing veto
rate with a pronounced bump superimposed -- similar to what has been
measured for $^{22}$Na (compare to Fig.~\ref{na22_comb_phases}). It
can easily be demonstrated, following the above example, that the
veto rate bump is an almost universal indicator for the presence of
radio-isotopes with half-lifes ranging from several minutes to several
years when combining data covering extended periods of time during
which the geophysical environment inevitably changed. However, the
veto rate bump cannot be used to estimate the half-life of an isotope,
as can be done with the activity model. Also, it should be noted that
non-linear variations of the event rate due to a long-lived isotope with
veto rate other than the veto rate bump are to be expected once the
isotope activity decreases and/or the veto rate frequency distribution
shifts towards lower values as the mission continues.

Among other things, a detailed understanding of the variation of the
background contribution of long-lived isotopes with veto rate is
necessary for a reliable measurement of the cosmic diffuse gamma-ray
background (CDG) at MeV energies. The COMPTEL analyses make use of
the variable nature of the instrument background to isolate the CDG
signal, which is assumed to be constant (see e.g.\ Kappadath et al.\
\cite{kappadath_cdg}, Weidenspointner et al.\
\cite{weidenspointner_cdg}). In particular, above about 4~MeV the
instrumental background is exclusively due to prompt processes. These
are also present in the background at lower energies, but with large
contributions from long-lived radioactive isotopes. Because prompt
background components vary, to a good approximation, linearly with
incident cosmic-ray intensity as monitored by veto rate, they can be
eliminated from the data by an extrapolation technique (Kappadath et
al.\ \cite{kappadath_3c}). In contrast, the event rate of long-lived
isotopes does not vary linearly with veto rate. 
The contributions of long-lived isotopes must therefore be subtracted,
as a function of veto rate, before the veto rate extrapolation
(Weidenspointner \cite{weidenspointner_phd}).  
The validity of a veto rate extrapolation for removing prompt
instrumental backgrounds thus depends on the absence of any long-lived
background components, since only then can the event rate be assumed
to vary linearly with veto rate. Subtraction of non-prompt background
components after the veto rate extrapolation is only possible for
constant background components, such as primordial radioactivity.


\section{\label{comparisons_implications} Background line intensities:
comparisons and implications}

The instrumental line background of an instrument is determined by its
material composition, the radiation environment, and the employed
detection principle. In this respect it is interesting to compare the
COMPTEL instrumental lines with those identified in the data from
other low-energy $\gamma$-ray experiments on satellite platforms such
as the NaI Gamma-Ray Spectrometer (GRS) onboard the Solar Maximum
Mission (SMM) and the Ge Gamma-Ray Spectrometer (GRS) flown on
HEAO~3. In doing so, one must keep in mind that an instrument's energy
resolution is important for resolving lines. At 1~MeV, the energy
resolution (FWHM) of COMPTEL is 9.8\% and 8.8\% in E$_\mathrm{tot}$
and E$_\mathrm{2}$, respectively (Sch{\"o}nfelder et al.\
\cite{schoenfelder_comptel}). At the same energy, the resolution of
GRS-SMM was about 5.4\% (Forrest et al.\ \cite{forrest_smm}), and that
of GRS-HEAO~3 was $\stackrel{\textstyle _<}{_{\sim}}\!0.3$\% (Mahoney
et al.\ \cite{mahoney_heao3}). Therefore, the Ge spectrometer onboard
HEAO~3 is far superior for $\gamma$-ray line studies compared to
COMPTEL and GRS-SMM, which have similar resolution.  The CGRO
satellite is kept in a circular orbit of 28.5$^\circ$ inclination at
altitudes between about 330~km and 515~km (see
Fig.~\ref{cgro_orbit-alt}). The SMM and HEAO~3 missions were operated
in circular orbits of 28.5$^\circ$ inclination at about 490--570~km
altitude, and 43.6$^\circ$ inclination at about 500~km altitude,
respectively. Similar to COMPTEL, the most abundant element in the
material composition of both GRS-SMM (E.~Chupp, priv.\ comm., 1999)
and GRS-HEAO~3 (Wheaton et al.\ \cite{wheaton_heao3}) is Al in the
instrument structures. Also, all instruments contain significant
amounts of Fe, Ni, Cu, and Cr in, e.g., electronics components.

Five of the eight isotopes identified in the COMPTEL line background,
namely $^{22}$Na, $^{24}$Na, $^{28}$Al, $^{52}$Mn, and
$^{57}$Ni, are due to activation of Al and Fe, Ni, Cu, and Cr
nuclei in the instrument structure.
As expected, each of these isotopes has also been identified
in GRS-SMM (Share et al.\ \cite{share_smm_lines}) and GRS-HEAO~3
(Wheaton et al.\ \cite{wheaton_heao3}) data. For GRS-SMM, however, an
additional production channel for $^{22}$Na and $^{24}$Na were
proton and neutron reactions, respectively, on $^{23}$Na in the NaI
detectors. Other isotopes identified in GRS-SMM and GRS-HEAO~3 that
result from activation of the five elements and which have major lines
below about 1~MeV should in principle also be present in the COMPTEL
background, however, their events are suppressed by the D1 and
D2 detector thresholds used in COMPTEL analyses. On the other hand,
instrumental lines due to activation of Cs and I (GRS-SMM) or Ge
(GRS-HEAO~3) are absent in COMPTEL since they exhibit a ToF value not
used in the analysis (I) or since these elements are not present in
the instrument (Cs, Ge). 

The different detection principles employed in the three instruments
result in significant differences in the relative importance of
specific isotopes, their detectability, and rejectability. COMPTEL
requires coincident interactions in the D1 and D2 detector and
therefore is particularly susceptible to multiple photon events such
as the $\beta^+$-decay of $^{22}$Na or the $\beta^-$-decay of
$^{24}$Na (see Sect.~\ref{line_characteristics}). Consequently,
multiple photon events are relatively more important than single
photon events. On the other hand, individual photons from these decays
produce photopeaks in E$_\mathrm{2}$ that can be used for identifying
individual isotopes and for determining their background
contribution. In addition, the characteristic
E$_\mathrm{1}$-E$_\mathrm{2}$ signature of multiple photon events
provides a wide range of options for their rejection (or
enhancement). In the SMM and HEAO~3 spectrometers events are triggered
by individual photons.
Unlike the COMPTEL veto system, which consists of plastic scintillator
domes, the anti-coincidence systems of the SMM and HEAO~3
spectrometers were made of CsI crystals and therefore sensitive to
photons at $\gamma$-ray line energies. For these two detectors therefore the
probability for a radioactive decay to trigger the detectors is proportional
to the number of emitted photons, as is the probability to trigger the
anti-coincidence systems. To first order, the net probability of a
radioactive decay for generating a background event is therefore
independent of the photon multiplicity for GRS-SMM and GRS-HEAO~3.
As long as the decays occur outside the detectors, individual photons
will give rise to photopeaks in the latter two instruments, and these
backgrounds are harder to suppress by event selections than in
COMPTEL. As far as decays inside the detectors are concerned,
$\beta$-decays are particularly hard to identify for GRS-SMM (e.g.\
activation of Na) and GRS-HEAO~3 (activation of Ge) since the added
$\beta$-continuum will broaden $\gamma$-ray lines beyond recognition. In
COMPTEL, activation in the D2 detector is effectively eliminated by
event (ToF) selection.


Since COMPTEL is the first double-scattering Compton telescope
operated in a near-Earth orbit, the instrumental background
experienced during this mission may provide guidance for the design of
future instruments.  Below, some implications of the COMPTEL
instrumental line background for conducting $\gamma$-ray line studies
with this and future instruments are discussed.

The by far strongest astrophysical $\gamma$-ray line signal detected by
COMPTEL is 1.8~MeV line emission from $^{26}$Al in the
interstellar medium (Diehl et al.\ \cite{diehl_al26}). When observing
along the galactic plane, the average event rate due to this extended
line source is about $7 \times 10^{-4}$~s$^{-1}$ for imaging
selections (see App.~\ref{event_selections_imaging}). Typical event
rates due to point sources observed by COMPTEL in the light of $\gamma$-ray
lines, such as the supernova remnant Cas~A ($^{44}$Ti at 1.12~MeV,
Iyudin et al. \cite{iyudin_cas-a}), or the Type~Ia supernova 1991T
($^{56}$Co at 0.85~MeV and 1.24~MeV, Morris et
al. \cite{morris_sn1991t}), or the Vela region ($^{26}$Al at
1.8~MeV, Diehl et al.\ \cite{diehl_vela}) are about $3 \times
10^{-5}$~s$^{-1}$.  Typical event rates arising from individual
background isotopes are about $10^{-1}$~s$^{-1}$ (see
Fig.~\ref{lines_long-term_plot}).  Below 3~MeV,
instrumental background lines account for 10--50\% of the total
background rate (see Fig.~\ref{stefan_etot-fit}). It follows that
the signal-to-background ratio for astrophysical $\gamma$-ray lines in
general is less than 1\%, which can be enhanced to a few percent by
fine-tuning the imaging event selections described in
App.~\ref{event_selections_imaging} for analyzing a specific $\gamma$-ray
line.

\begin{table}
\small
\begin{tabular}{|cccc|}   \hline
\rule[0ex]{0cm}{3.5ex}\makebox[9ex]{Isotope} & \makebox[12ex]{Efficiency} & 
\makebox[13ex]{Material} & \makebox[12ex]{Activity} \\ 
  & (Imaging Sel.) &  & [g$^{-1}$s$^{-1}$] \\[1ex] \hline 
\rule[0pt]{0ex}{4ex}
$^2$D & $7 \times 10^{-4}$ & D1 scintillator & $3.4 \times 10^{-3}$ 
\\[1.5ex] 
$^{22}$Na & $8 \times 10^{-4}$ & D1 Al structure & $7 \times 10^{-4}$ 
\\[1.5ex]  
$^{24}$Na & $9 \times 10^{-4}$ & D1 Al structure & $9 \times 10^{-4}$ 
\\[1.5ex]    
$^{28}$Al & $2 \times 10^{-4}$ & D1 Al structure & $2 \times 10^{-3}$ 
\\[1.5ex]
$^{40}$K & $2 \times 10^{-4}$ & D1 PMT glass & 0.2 
\\[1.5ex]
$^{52}$Mn & $2 \times 10^{-3}$ & Fe around D1 & $4 \times 10^{-4}$ \\ 
          &                    & Cr around D1 & $3 \times 10^{-4}$ \\
          &                    & Ni around D1 & $2 \times 10^{-4}$ \\
          &                    & Cu around D1 & $7 \times 10^{-5}$ 
\\[1.5ex]
$^{57}$Ni & $5 \times 10^{-4}$ & Ni around D1 & $5 \times 10^{-4}$ \\ 
          &                    & Cu around D1 & $6 \times 10^{-5}$ 
\\[1.5ex] 
$^{208}$Tl & $2 \times 10^{-3}$ & D1 PMT glass & $10^{-2}$  
\\[1.5ex] \hline
\end{tabular}
\normalsize
\caption[]{Background isotope properties relevant for astrophysical
$\gamma$-ray line studies. For each isotope the following quantities are
listed: the efficiency for producing a background event under imaging
event selections, the material(s) in which the isotope is produced,
and the mission-averaged activity. Further details are given in the
text.}
\label{activity_table}
\end{table}

In this respect it is interesting to take a look at the efficiency,
determined by Monte Carlo simulation, for various isotopes to produce
a background event (see Table~\ref{activity_table}). For imaging
selections, this efficiency is about $7 \times 10^{-4}$ and $2 \times
10^{-4}$ for $^{2}$D and $^{40}$K, both of which give rise to type
{\bf A} (single photon) background events. A major source of type {\bf
C} multiple photon events is $^{24}$Na with an efficiency of about $9
\times 10^{-4}$. The efficiency for $^{208}$Tl, a minor source of type
{\bf C} events which is assumed to have the same spatial distribution
as $^{40}$K, is about $2 \times 10^{-3}$. This is an order of
magnitude larger than the efficiency for $^{40}$K and again
illustrates that COMPTEL is more susceptible to multiple photon decays
than to single photon decays. The more stringent CDG event selections
reduce these efficiencies by factors of 4--7 and 2--3 for type {\bf A}
and type {\bf C} background events, respectively. For comparison, the
detection efficiencies for celestial lines around 1--2~MeV are about
$2 \times 10^{-4}$ using imaging event selections.

A goal for future Compton telescopes is to minimize background
production by design. Passive material should be reduced, and
manufactured from materials with low activation.
Based on simulated efficiencies and the COMPTEL mass model,
``mission-averaged'' activities (in decays per g of material per s)
were derived for the eight identified isotopes using data representing
the first 7 years of the mission (June 1991 to April 1998).
These activities, together with the above mentioned simulated
efficiencies and the activated materials, are summarized in
Table~\ref{activity_table}.
The relative yields for $^{52}$Mn and $^{57}$Ni were determined in
hadron simulations (P.~Jean, priv.\ comm., 1997). The numbers for the
average activities are accurate within factors of 2--3. Their major
sources of uncertainty are the simulated efficiencies and the mass
normalization, which are currently based on the assumption that the
activation is homogeneous in a particular material. This may not be
accurate for isotopes produced by thermal neutrons, in particular
$^{28}$Al. As a cross-check, the average activities were compared to
simulations of activation of COMPTEL materials, neglecting the
contributions of secondary particles, for a solar-maximum radiation
environment (P.~Jean, priv.\ comm., 1997). The measured and the
simulated activities are generally consistent within factors of a few
for isotopes mainly produced by SAA protons. Hence these average
activities may be used for estimating the line background rates of
future instruments in a similar orbit, taking into account how the
mass distribution differs from that of COMPTEL.


\section{\label{discussion} Summary and discussion}

We have identified eight different isotopes in the COMPTEL
instrumental line background, namely $^{2}$D, $^{22}$Na, $^{24}$Na,
$^{28}$Al, $^{40}$K, $^{52}$Mn, $^{57}$Ni, and $^{208}$Tl. These
isotopes can account for the major instrumental background lines. Some
minor instrumental lines, however, remain unidentified at this time.
In addition, we have studied the variation of the event rate of these
isotopes with time and incident cosmic-ray intensity, and determined
the average activity of spacecraft materials.

These results provide valuable insight into the physical processes
that give rise to instrumental background lines in $\gamma$-ray detectors in
low-Earth orbits. Because of their importance for MeV astronomy,
these issues have repeatedly been studied (see, e.g., Dean
et al.\ \cite{dean_bgd-review} and references therein). Obviously,
each detector has its own, unique instrumental
background. Nevertheless, background investigations such as those reported
in this work add to a growing pool of background experience to be
found in the literature, which will prove to be important for the
design of future instruments and the understanding of their data.

The successful modelling of the time variation of the background
contributions from the long-lived isotopes $^{22}$Na, $^{24}$Na,
$^{52}$Mn, and $^{57}$Ni in COMPTEL supports the conclusion that
activation occurs predominantly during SAA passages. On average, CGRO
passes through the SAA 6--8 times each day. In 1991 (at solar maximum
and hence at minimum SAA-proton flux) the daily average of the
incident flux of SAA protons with $\mathrm{E}_p > 100$~MeV was about
38 times higher than the corresponding incident galactic cosmic-ray
proton flux (Dyer et al.\ \cite{dyer_cgro-bgd}), illustrating the
importance of SAA passages with respect to the total deposited
radiation dose. The SAA-proton flux varies strongly with altitude and
solar cycle. For example, during its mission the altitude of CGRO
ranged from 350--500~km, which corresponds to a variation of the
SAA-proton flux of a factor $\sim 10$, while the solar activity
results in a variation by a factor $\sim 2$ at these altitudes
(Stassinopoulos \cite{stassinopoulos}). According to our activity
model (Varendorff et al.\ \cite{varendorff_4c}) the anisotropy of the
SAA-proton flux at low satellite altitudes (Watts et al.\
\cite{watts_saa_anisotropy}) has an important effect on the radiation
dose and hence on activation. These findings are consistent with
earlier studies of GRS-SMM instrumental lines, in which activation due
to (anisotropic) radiation-belt protons during SAA transits has been
identified as major source of (long-lived) line background and its
variation with time (see Kurfess et al.\ \cite{kurfess_smm_lines},
Share et al.\ \cite{share_smm_lines}, and references therein).

In contrast, the activities of short-lived isotopes are mostly
determined by the incident primary cosmic-ray and albedo-neutron
fluxes. The enhanced production of these isotopes during SAA transits
is of little importance for their background contribution outside the
SAA, where astrophysical data are recorded, because of their rapid
decay, which precludes build-up.

Two strategies for reducing instrumental line background in future
low-energy $\gamma$-ray detectors due to activation of long-lived isotopes
present themselves. First, care should be taken to avoid detector
materials which are easily activated by SAA protons (such as aluminium) or
which carry primordial radioactivity. Second, instruments should be
placed in orbits that minimize radiation-belt proton dosage, such as 
a low-altitude equatorial orbit below the radiation belts
and outside the SAA, or a high-altitude orbit above the radiation
belts (as scheduled for the INTEGRAL observatory). 
The flux of (low-energy) cosmic rays is significantly 
lower in a low-altitude orbit as compared to a high-altitude orbit
-- at the cost of a much higher albedo-neutron flux. 
Another strategy for limiting the impact of remaining instrumental
lines in future instruments is, e.g., to exploit the expected
improvement in energy resolution to identify (and model) a large
number of instrumental background components. A comprehensive
treatment of the implications of the COMPTEL instrumental line (and
continuum) background for future Compton telescopes is beyond the
scope of this paper and will be given in a separate publication.

Finally, although investigations of background lines are of interest
in their own right, they are ultimately motivated by the struggle to
eliminate background events in astrophysical analyses. The
COMPTEL instrumental background lines, particularly those from
long-lived isotopes, are a concern below about 4~MeV. At these
energies detailed understanding of the dataspace structure and
variability of the instrumental lines is indispensable for optimizing
many astrophysical investigations, such as that of the cosmic diffuse
gamma-ray background (see e.g.\ Kappadath et al.\
\cite{kappadath_cdg}, Weidenspointner et al.\
\cite{weidenspointner_cdg}) or of the galactic 1.8~MeV line emission
from $^{26}$Al (see e.g.\ Oberlack \cite{oberlack_phd},
Pl{\"u}schke et al.\ \cite{plueschke_al-map}). 

Comparing event rates due to $\gamma$-ray line sources and due to
instrumental background it is evident that source fluxes cannot be
derived from global event rates with COMPTEL. This is only possible in
imaging analysis, which exploits the characteristic cone-like
distribution of source events in a three-dimensional data space spanned by
the scatter angle $\bar{\varphi}$ and the direction (zenith and
azimuth) of the scattered photon (see Sch{\"o}nfelder et al.\
\cite{schoenfelder_comptel}). Instrumental background components
exhibit comparatively smooth distributions in this data space,
allowing for their separation from the signal. Further exploitation of
existing knowledge on the dataspace structure of instrumental lines
will remedy some of the difficulties in the search for
astrophysical $\gamma$-ray lines (see e.g.\ Morris et al.\
\cite{morris_sn1991t}). Nevertheless, data
selections that increase the signal-to-background ratio are clearly of
great value.

The importance of instrumental background suppression/rejection can be
illustrated with the original report of nuclear de-excitation lines
from the Orion region (Bloemen et al.\ \cite{bloemen_orion}), which
would have corresponded to an event rate of about $6 \times
10^{-4}$~s$^{-1}$ spread over a rather broad energy range of
3--7~MeV. As was recognized in our recent analyses, the variations in
the $^{24}$Na event rate over the different observation periods of
this galactic region are capable of producing spatial inhomogeneities
in the background of that order, which then can be falsely attributed
to a celestial signal. Only correlations of the supposed Orion signal
with data space domains where $^{24}$Na contamination is strongest
provided the hints to suspect an instrumental background artifact as
the origin of the claimed detection (Bloemen et al.\
\cite{bloemen_revised-orion}).

\begin{acknowledgements}

It is a pleasure to acknowledge laboratory efforts of J.~Macri,
and of R.~Georgii, C.~Wunderer, and W.~Plass in measuring
primordial radioactivity in D1 PMTs.
We are also pleased to acknowledge work on hadron simulations for
estimating activation of spacecraft materials by P.~Jean.
The COMPTEL project is supported by the German government through DARA
grant 50~QV~90968, by NASA under contract NAS5-26645, and by the
Netherlands Organization for Scientific Research NWO.

\end{acknowledgements}

%
%

\appendix


\section{\label{event_selections} Event selections}

The results on the COMPTEL instrumental lines given in the main body
of this paper have been derived using two different sets of event
selections, described below.

\subsection{\label{event_selections_cdg} CDG analysis}

In the COMPTEL analysis of the cosmic diffuse gamma-ray background
(CDG), the intense atmospheric $\gamma$-ray background is eliminated
by selecting data from times when the Earth was sufficiently far
outside a circular field-of-view defined by event selections. Detailed
descriptions of these selections can be found in, e.g., Kappadath
(\cite{kappadath_phd}) and Weidenspointner
(\cite{weidenspointner_phd}).

The standard set of CDG event selections used in this paper are as follows:
\begin{eqnarray}\label{cdg_selections}
70~\mathrm{keV} \le \mathrm{E}_\mathrm{1} \le 20~\mathrm{MeV}
\nonumber \\
730~\mathrm{keV} \le \mathrm{E}_\mathrm{2} \le 30~\mathrm{MeV} \nonumber \\
800~\mathrm{keV} \le \mathrm{E}_\mathrm{tot} \le 30~\mathrm{MeV}
\nonumber \\
110~\mathrm{ch} \le \mathrm{ToF} \le 130~\mathrm{ch} \\
\mathrm{PSD}_\mathrm{low}(\mathrm{E}_\mathrm{1}) \le \mathrm{PSD} \le
\mathrm{PSD}_\mathrm{high}(\mathrm{E}_\mathrm{1}) \nonumber \\
6^\circ \le \bar{\varphi} \le 38^\circ \nonumber \\
\zeta + \bar{\varphi} = \xi \le 40^\circ \nonumber
\end{eqnarray}
where $\bar{\varphi}$ is the photon scatter angle (see
Eq.~\ref{phibar_definition}), and $\zeta$ is the angle between the
direction of the scattered photon and the telescope z-axis. The
selection on $\xi$ defines a field-of-view of half opening angle
40$^\circ$. The values of PSD$_\mathrm{low}$ and PSD$_\mathrm{high}$
are 40~ch and 110~ch at 70~keV, 55~ch and 95~ch at 1~MeV, 55~ch and
92~ch at 10~MeV, and 40~ch and 90~ch at 20~MeV. At intermediate
E$_\mathrm{1}$ energies, these boundaries are linearly interpolated. In
addition, the horizon of the Earth, as seen by the instrument, is
required to be at least 5$^\circ$ outside the field-of-view.

\subsection{\label{event_selections_imaging} Imaging analysis}

In COMPTEL imaging analyses such as that of the $^{26}$Al 1.8~MeV
line emission from the galaxy, $\gamma$-ray albedo photons from the
Earth's atmosphere are rejected by determining for each event the
minimal angular distance between its possible origins in the sky and
the Earth horizon. This angular distance, denoted $\epsilon$ below, is
required to exceed a minimum value (see e.g.\ Sch{\"o}nfelder et
al. \cite{schoenfelder_comptel}). Hence, the effective field-of-view
in imaging analysis changes with time as the instrument orbits the
Earth.

The standard set of imaging event selections used in this paper are as
follows:
\begin{eqnarray}\label{imaging_selections}
70~\mathrm{keV} \le \mathrm{E}_\mathrm{1} \le 20~\mathrm{MeV}
\nonumber \\
650~\mathrm{keV} \le \mathrm{E}_\mathrm{2} \le 30~\mathrm{MeV} \nonumber \\
750~\mathrm{keV} \le \mathrm{E}_\mathrm{tot} \le 30~\mathrm{MeV}
\nonumber \\
115~\mathrm{ch} \le \mathrm{ToF} \le 130~\mathrm{ch} \\
0~\mathrm{ch} \le \mathrm{PSD} \le 110~\mathrm{ch} \nonumber \\
0^\circ \le \bar{\varphi} \le 50^\circ \nonumber \\
\epsilon \ge 5^\circ \nonumber
\end{eqnarray}


\section{\label{line-fitting_procedure} Line fitting procedures in
analyses of CDG and $^{\mathsf{26}}$Al}

Many astrophysical analyses, such as that of the cosmic diffuse
gamma-ray background (CDG), of the galactic 1.8~MeV line emission
from $^{26}$Al, or of $\gamma$-ray line emission from supernovae, can be
optimized by detailed modelling of the instrumental line
background. Below, the (similar) procedures used to determine the
background event rates due to individual isotopes employed by
Weidenspointner et al.\ (\cite{weidenspointner_cdg}), and Oberlack
(\cite{oberlack_phd}) and Pl{\"u}schke et al.\
(\cite{plueschke_al-map}) in investigations of the CDG\footnote{A
similar procedure consisting of four different fits was utilized in
the CDG analysis of Kappadath et al.\ (\cite{kappadath_cdg}) to
determine the background contributions from the five isotopes ($^{2}$D,
$^{22}$Na, $^{24}$Na, $^{28}$Al, and $^{40}$K (fixed))
identified at that time.} and of the $^{26}$Al emission,
respectively, are described.

\subsection{\label{line_fitting_cdg} CDG analysis}

\begin{figure*}
\begin{minipage}{12cm}\makebox[0cm]{}\\
\epsfig{figure=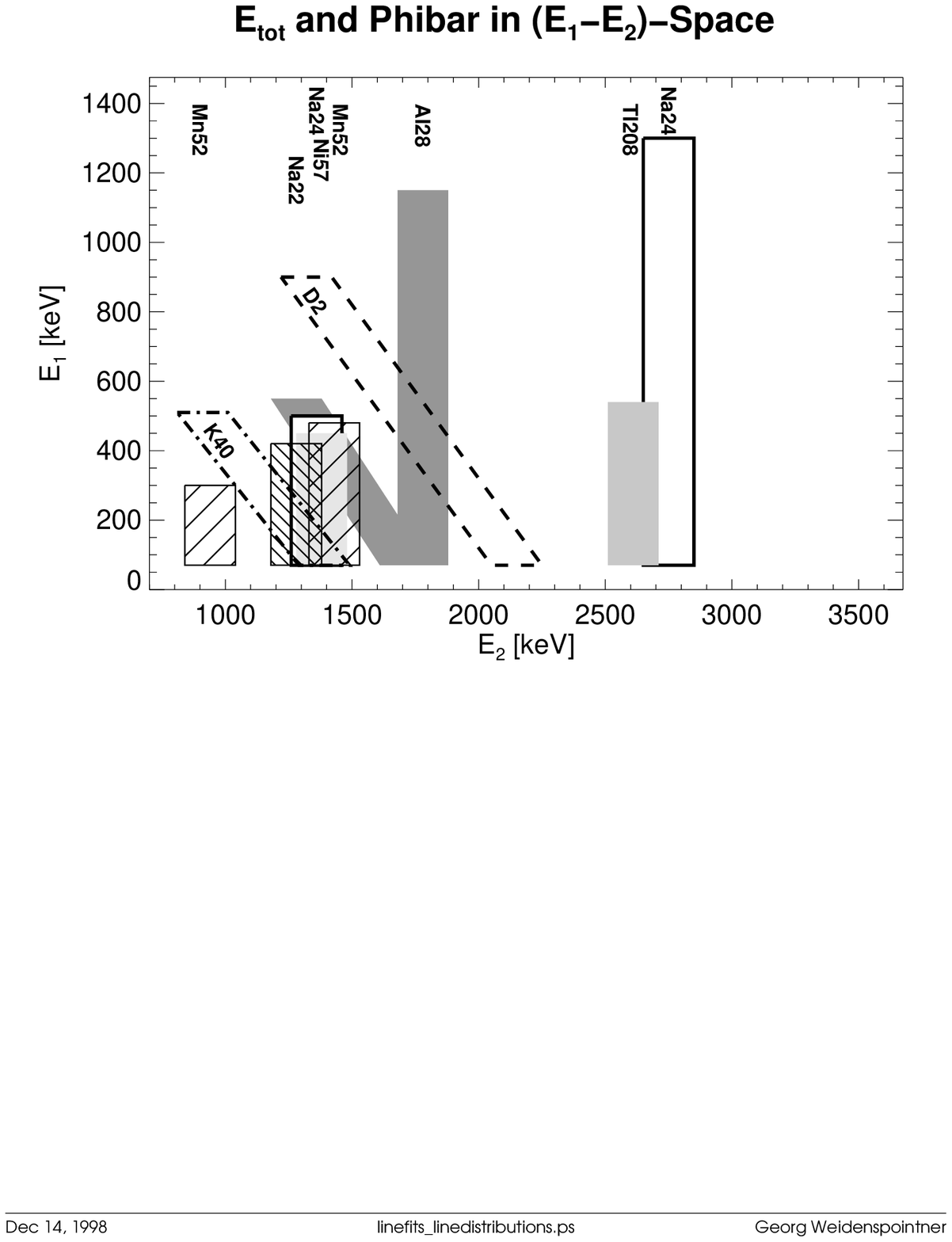,%
bbllx=82pt,bblly=288pt,bburx=468pt,bbury=552pt,width=12cm,clip=}
\end{minipage}
\hfill
\begin{minipage}{5.25cm}
\caption[]{A schematic representation of the simulated
E$_\mathrm{1}$-E$_\mathrm{2}$ distributions of the eight isotopes
identified in the instrumental line background for CDG event
selections. The ``vertical'' and ``diagonal'' bands represent multiple
photon and single photon events, respectively (compare to the detailed
individual diagrams in Sect.~\ref{identified_isotopes}).}
\label{line-fitting_procedure_illustration_1}
\end{minipage}
\begin{minipage}{12cm}\makebox[0cm]{}\\
\epsfig{figure=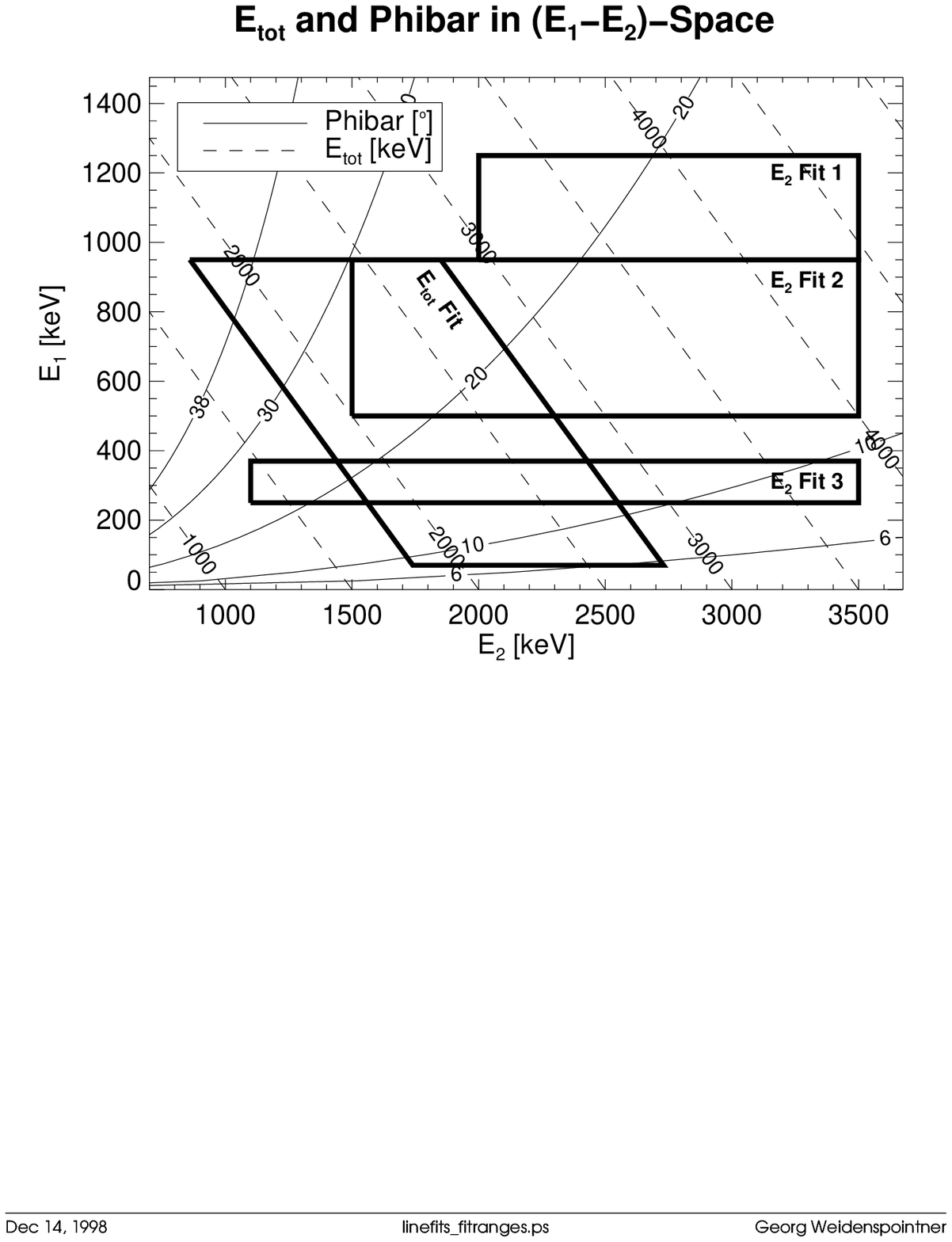,%
bbllx=82pt,bblly=288pt,bburx=468pt,bbury=552pt,width=12cm,clip=}
\end{minipage}
\hfill
\begin{minipage}{5.25cm}
\caption[]{An illustration of the E$_\mathrm{1}$-E$_\mathrm{2}$ ranges
of the three E$_\mathrm{2}$ spectra and the E$_\mathrm{tot}$ spectrum
used to determine the event rates due to the background isotopes. Also
plotted are lines of constant $\bar{\varphi}$ (solid) and
E$_\mathrm{tot}$ (dashed).}
\label{line-fitting_procedure_illustration_2}
\end{minipage}
\end{figure*}

Accounting for the event rates due to the instrumental line
background, in particular its long-lived components, is important for
measuring the CDG at MeV energies (e.g.\ Kappadath et al.\
\cite{kappadath_cdg}, Weidenspointner et al.\
\cite{weidenspointner_cdg}). As explained above, the background
contributions of long-lived isotopes have to be subtracted before
prompt and short-lived backgrounds can be eliminated by veto rate
extrapolation. The event rates due to the eight identified background
isotopes producing the major instrumental lines are determined, as a
function of veto rate, in an iterative procedure by fitting a set of
three E$_\mathrm{2}$ spectra and one E$_\mathrm{tot}$ spectrum for
each veto rate interval\footnote{Ideally, the event rates due to these
isotopes would be determined by fitting the two-dimensional event
distribution in E$_\mathrm{1}$-E$_\mathrm{2}$. However, at present
such two-dimensional fits are not feasible for two reasons. First, in
general the statistics is not sufficient in the CDG analysis, and
second, we do not yet have a reliable model of the
E$_\mathrm{1}$-E$_\mathrm{2}$ distribution of the continuum
background, which has a complex shape as it is a blend of various
components.}.

The rationale of the iterative fitting procedure is illustrated in
Figs.~\ref{line-fitting_procedure_illustration_1} and
\ref{line-fitting_procedure_illustration_2}. The
E$_\mathrm{1}$-E$_\mathrm{2}$ distributions of the eight isotopes as
obtained from Monte Carlo simulations are schematically depicted for
CDG event selections in
Fig.~\ref{line-fitting_procedure_illustration_1}.  In general, there
is considerable overlap in the E$_\mathrm{1}$-E$_\mathrm{2}$
distributions of individual isotopes (in particular around 1.3~MeV in
E$_\mathrm{2}$), which precludes an independent determination of the
isotopes' background contributions.
Therefore, an iterative procedure was introduced, which starts at the
highest energies in E$_\mathrm{1}$ and E$_\mathrm{2}$, where
ambiguities are minimal, and then proceeds down to the increasingly
complex structures at lower E$_\mathrm{1}$ and E$_\mathrm{2}$
energies. The E$_\mathrm{1}$-E$_\mathrm{2}$ ranges of the three
E$_\mathrm{2}$ spectra and the E$_\mathrm{tot}$ spectrum, chosen such
as to enhance or suppress individual lines or spectral features, are
indicated in Fig.~\ref{line-fitting_procedure_illustration_2}. The
E$_\mathrm{1}$-E$_\mathrm{2}$ ranges covered by the second and third
E$_\mathrm{2}$ fit and by the E$_\mathrm{tot}$ fit overlap, hence the
results of these fits are not statistically independent. The overlap
is caused by the E$_\mathrm{tot}$ fit, which was introduced to
properly separate the background events from $^{2}$D and
$^{28}$Al. Inclusion or omission of the E$_\mathrm{tot}$ fit therefore
is a trade-off between systematic and statistical uncertainty. By
iteratively fitting the second and third E$_\mathrm{2}$ spectrum and
the E$_\mathrm{tot}$ spectrum, both the systematic and the statistical
uncertainty in the $^{2}$D and $^{28}$Al event rates are minimized
(other isotopes are hardly affected by the overlap, see below). In
additon, the iterative approach ensures the self-consistency of the
determined isotope background contributions.
Also included in Fig.~\ref{line-fitting_procedure_illustration_2} are
lines of constant $\bar{\varphi}$ and E$_\mathrm{tot}$. Comparison of
Figs.~\ref{line-fitting_procedure_illustration_1} and
\ref{line-fitting_procedure_illustration_2} provides a first
indication of the many options for fitting individual lines, which can
be enhanced or suppressed through the choice of the fit regions, and
in addition through selections on E$_\mathrm{tot}$ and/or $\bar{\varphi}$. In
particular, event selections may be used to suppress unidentified,
long-lived spectral features, which cannot be eliminated by veto rate
extrapolation (see Kappadath et al.\ \cite{kappadath_cdg},
Weidenspointner et al.\ \cite{weidenspointner_cdg}).

The E$_\mathrm{1}$-E$_\mathrm{2}$ ranges of the three E$_\mathrm{2}$
spectra and the E$_\mathrm{tot}$ spectrum indicated in
Fig.~\ref{line-fitting_procedure_illustration_2} were chosen for the
iterative fitting procedure for the following reasons. 
The first E$_\mathrm{2}$ spectrum, covering the E$_\mathrm{1}$-E$_\mathrm{2}$
region of 950--1250~keV in E$_\mathrm{1}$ and 2000--3500~keV in E$_\mathrm{2}$
(see Figs.~\ref{line-fitting_procedure_illustration_1} and
\ref{line-fitting_procedure_illustration_2}), allows us to determine the
event rate from $^{24}$Na. The signal from this isotope is optimized
by selecting E$_\mathrm{1}$ energies around the Compton edge of the 1.37~MeV
photon interacting in D1 and around the photopeak of the 2.75~MeV
photon interacting in D2 (see Fig.~\ref{24_na_dataspace}). 
The $^{2}$D event rate is determined from fitting the E$_\mathrm{tot}$
spectrum (E$_\mathrm{tot}$ 1810--2800~keV, E$_\mathrm{1}$ 70--950~keV,
and E$_\mathrm{2}$ 730--2800~keV, see
Figs.~\ref{line-fitting_procedure_illustration_1} and
\ref{line-fitting_procedure_illustration_2}). The second
E$_\mathrm{2}$ spectrum (E$_\mathrm{1}$ 500--950~keV, E$_\mathrm{2}$
1500--3500~keV, see Figs.~\ref{line-fitting_procedure_illustration_1}
and \ref{line-fitting_procedure_illustration_2}) is used to determine
the event rate from $^{28}$Al. Finally, the third E$_\mathrm{2}$
spectrum is intended for determining the background contributions from
the $\beta^+$-decays of $^{22}$Na, $^{52}$Mn, and $^{57}$Ni, with the
270--350~keV range in E$_\mathrm{1}$ being optimized for the Compton
edge of 511~keV photons, and the E$_\mathrm{2}$ range covering the
energies 1100--3500~keV (see
Figs.~\ref{line-fitting_procedure_illustration_1} and
\ref{line-fitting_procedure_illustration_2}).
To optimize the signal of the instrumental lines, which originate in
the D1 detector material, a ToF range of 2.5--7.5~ns was
selected for the spectra (compare Fig.~\ref{tof_scheme}).

\begin{figure}
\epsfig{figure=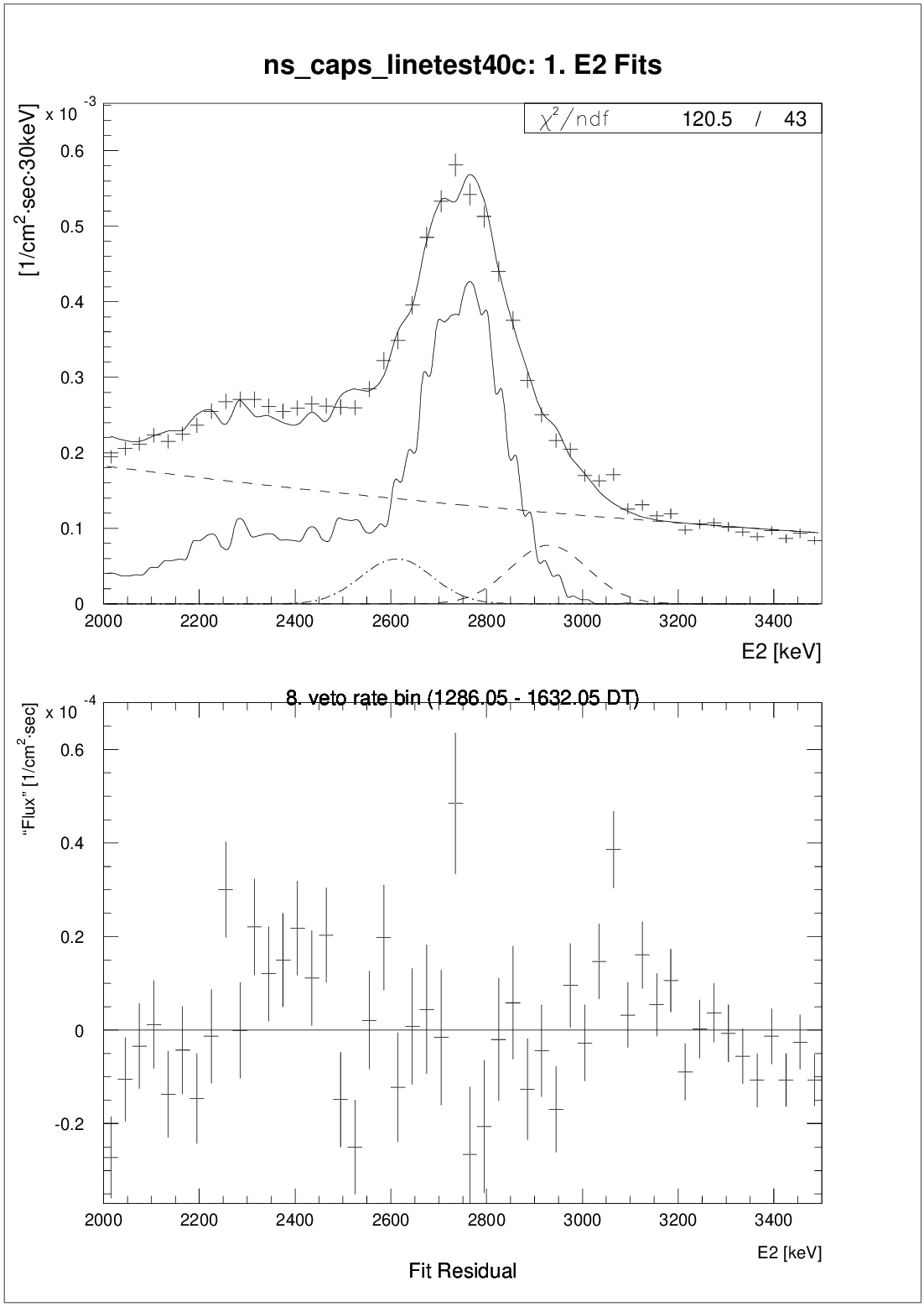,%
bbllx=38pt,bblly=412pt,bburx=508pt,bbury=746pt,width=8.8cm,clip=}
\caption[]{An example of a fit of the first E$_\mathrm{2}$ spectrum, which is
used to determine the event rate due to $^{24}$Na. In addition to
the total fit, the $^{24}$Na template (solid line), the two
unidentified features (dashed and dash-dotted lines) , and the
exponential continuum are indicated.}
\label{e2-fit-1_example} 
\end{figure}

\begin{figure}
\epsfig{figure=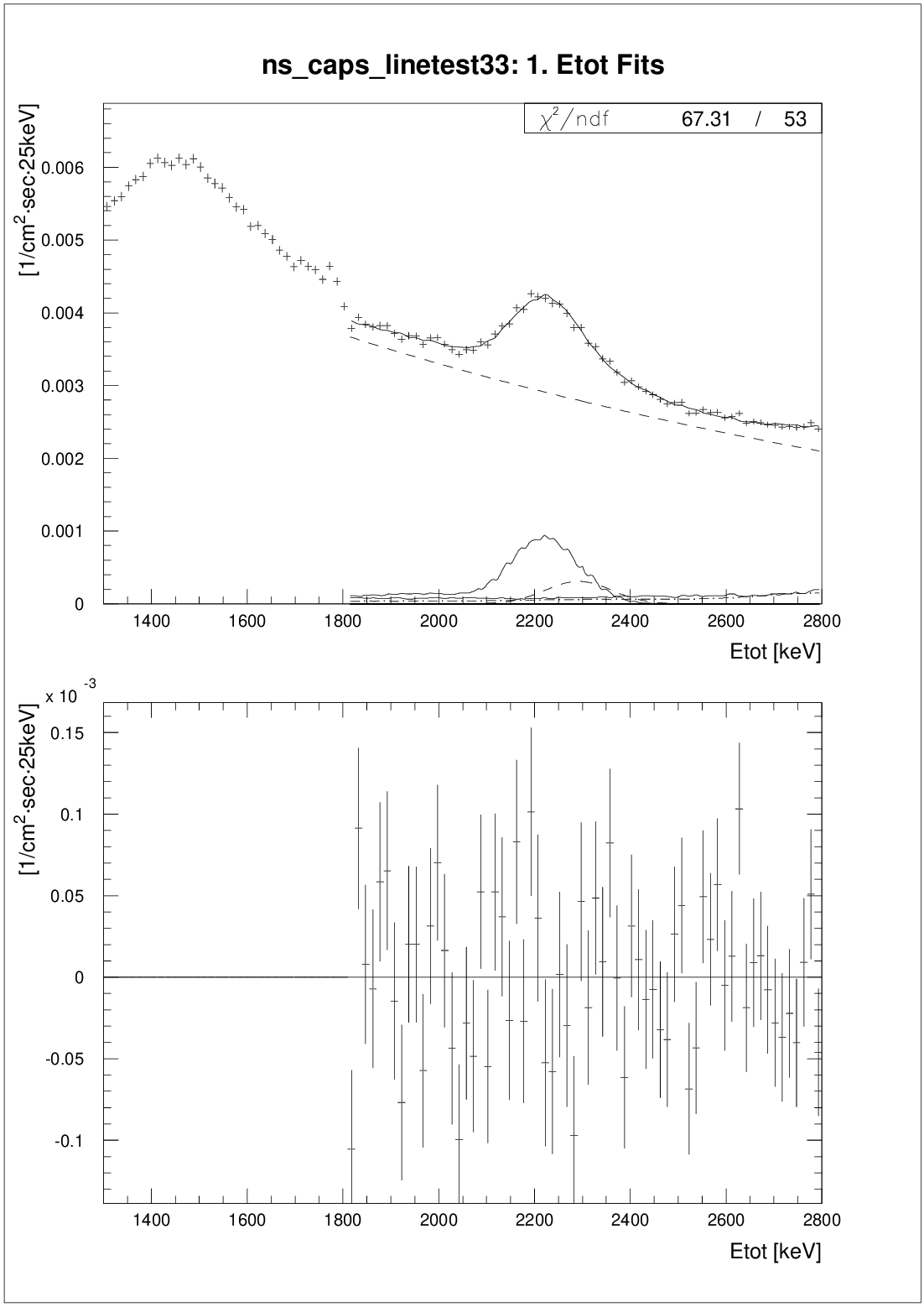,%
bbllx=38pt,bblly=410pt,bburx=516pt,bbury=742pt,width=8.8cm,clip=}
\caption[]{An example of a fit of the E$_\mathrm{tot}$ spectrum for
determining the count rate in the $^{2}$D 2.22~MeV line. In additon to
the total fit, the $^{2}$D template (solid line), an unidentifed
component at about 2.3~MeV (dashed line), the continuum contributions
from $^{24}$Na and $^{208}$Tl (solid and dash-dotted lines), and the
exponential continuum (dashed line) are indicated.}
\label{etot-fit_example} 
\end{figure}

\begin{figure}
\epsfig{figure=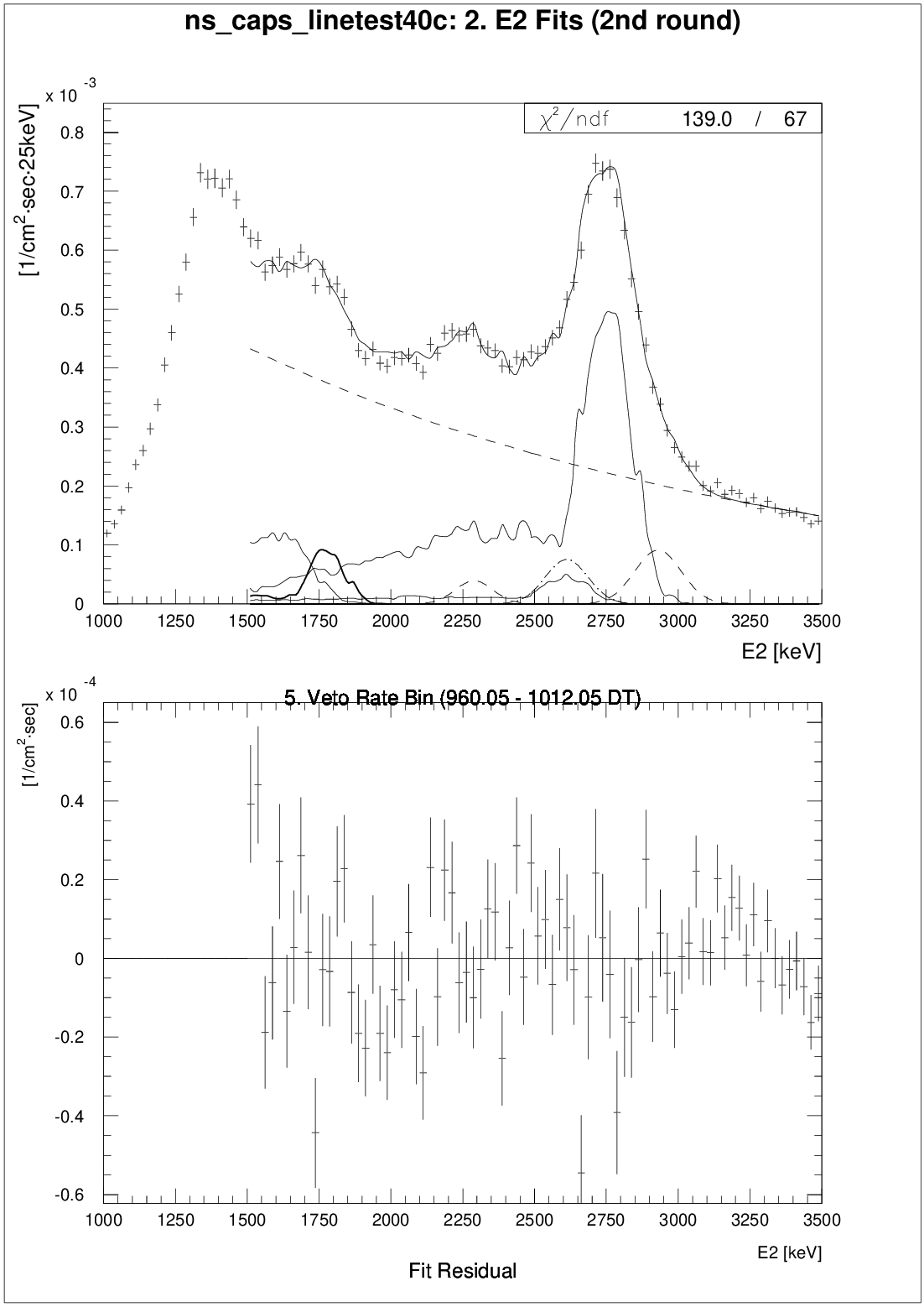,%
bbllx=38pt,bblly=412pt,bburx=516pt,bbury=754pt,width=8.8cm,clip=}
\caption[]{An example for a fit of the second E$_\mathrm{2}$ spectrum,
which is used to determine the event rate from $^{28}$Al. In
addition to the total fit, the templates for $^{2}$D (the
line feature at $\sim$~1.6~MeV, fixed), $^{28}$Al (the 
1.78~MeV line), $^{24}$Na (the strong line at 2.75~MeV, fixed) and 
$^{208}$Tl (the weaker line at 2.61~MeV, fixed), the three
unidentified features (dashed, dashed-dotted, and dashed lines), and the
exponential continuum are indicated.}
\label{e2-fit-2_example} 
\end{figure}

\begin{figure}
\epsfig{figure=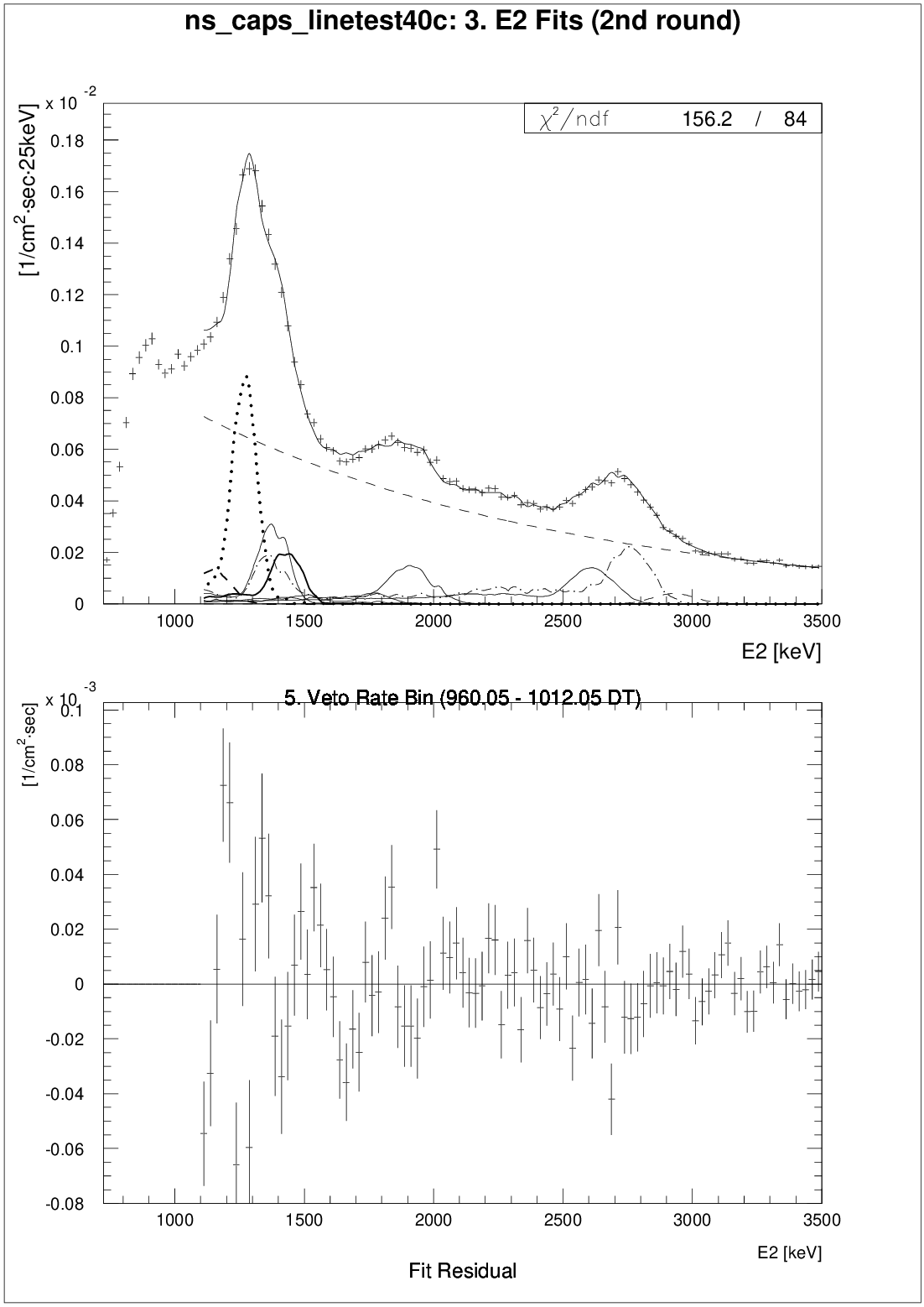,%
bbllx=38pt,bblly=412pt,bburx=516pt,bbury=750pt,width=8.8cm,clip=}
\caption[]{An example for a fit of the third E$_\mathrm{2}$ spectrum,
which is used to determine the event rates from $^{22}$Na (dotted
line), $^{52}$Mn (thick solid line), and $^{57}$Ni (thin solid
line). Also depicted are the fixed components ($^{2}$D: solid line at
1.9~MeV, $^{24}$Na: dash-dotted line, $^{28}$Al: weak solid component
below 1.9~MeV, $^{40}$K: dashed line, $^{208}$Tl: solid line at
2.6~MeV), as well as the total fit, the exponential continuum, and the
unidentified 2.93~MeV feature.}
\label{e2-fit-3_example} 
\end{figure}

The three E$_\mathrm{2}$ spectra and the E$_\mathrm{tot}$ spectrum are
analyzed in an iterative procedure consisting of eight fits, listed
below. The contributions from the primordial radio-nuclides $^{40}$K
and $^{208}$Tl are not determined from the fits, but calculated from
their known ($^{40}$K, see Sect.~\ref{40_k}) or estimated ($^{208}$Tl,
see Sect.~\ref{208_tl}) activities based on Monte Carlo
simulations. Similarly, once the background contribution of an isotope
has been determined from any spectrum, the isotope's contribution to
any other spectrum can be predicted based on Monte Carlo simulations.
The eight identified isotopes can account for the major instrumental
lines, however, some weak lines or spectral features remain
unidentified at this time. In the fits, some of these unidentified
lines were described by gaussians to minimize systematic errors in the
determination of the event rates of the identified components. The
unidentified components are genuinely different from those identified
since their variation with cosmic-ray intensity, as well as their
dependence on event parameter selections, are different (see Kappadath
et al.\ \cite{kappadath_cdg}, Weidenspointner et al.\
\cite{weidenspointner_cdg}).
The identified isotopes are represented by templates obtained from
Monte Carlo simulation. These templates have not been smoothed 
as smoothing inevitably increases the systematic uncertainty due to
distortions of the template shape. The small ``spikes'' in the
templates in Figs.~\ref{e2-fit-1_example}--\ref{e2-fit-3_example} are
an artifact of the plotting software.
The eight fit steps are:
\begin{enumerate}
\item {\sl E$_\mathrm{2}$ Fit 1}: The event rate due to the isotope
$^{24}$Na is determined (see Fig.~\ref{e2-fit-1_example}).
\item {\sl E$_\mathrm{tot}$ Fit}: The $^{2}$D event rate is estimated;
the contributions from $^{24}$Na and $^{208}$Tl are fixed. All other
isotopes are neglected.
\item {\sl E$_\mathrm{2}$ Fit 2}: The $^{2}$D and $^{28}$Al event
rates are estimated; the contributions from $^{24}$Na and $^{208}$Tl
are fixed.
\item {\sl E$_\mathrm{2}$ Fit 3}: The background contributions due to
the $\beta^+$-decays of $^{22}$Na, $^{52}$Mn, and $^{57}$Ni are
estimated; the contributions from $^{2}$D, $^{24}$Na, $^{28}$Al,
$^{40}$K, and $^{208}$Tl are fixed.
\item {\sl E$_\mathrm{tot}$ Fit}: The $^{2}$D event rate is
determined; the contributions from all other isotopes ($^{22}$Na,
$^{24}$Na, $^{28}$Al, $^{40}$K, $^{52}$Mn, $^{57}$Ni, and $^{208}$Tl)
are fixed (see Fig.~\ref{etot-fit_example}).
\item {\sl E$_\mathrm{2}$ Fit 2}: The event rate due to the isotope
$^{28}$Al is determined; the contributions from $^{2}$D, $^{24}$Na,
and $^{208}$Tl are fixed (see Fig.~\ref{e2-fit-2_example}).
\item {\sl E$_\mathrm{2}$ Fit 3}: The event rates due to the
$\beta^+$-decays of $^{22}$Na, $^{52}$Mn, and $^{57}$Ni are
determined; all other isotopes are fixed (see
Fig.~\ref{e2-fit-3_example}).
\item {\sl E$_\mathrm{tot}$ Fit}: The $^{2}$D event rate is
re-determined as in Fit~5.
\end{enumerate}
In this procedure, Fits~5--7 re-iterate Fits~2--4 to ensure
convergence, which is tested by comparing the results of Fits~5 and
8. In general, the two fits yield nearly identical results.
This self-consistent determination of the event rates of long-lived
isotopes inevitably requires the determination of all isotope
contributions, whether long-lived, short-lived, or prompt. In the CDG
analysis, subtraction of the contributions from $^{2}$D,
$^{24}$Na, and $^{28}$Al are based on Fits~8, 1, and 6,
respectively (see Weidenspointer et al.\
\cite{weidenspointner_cdg}). The event rates due to the
$\beta^+$-decays of $^{22}$Na, $^{52}$Mn, and $^{57}$Ni are
derived from the results of Fit~7. As pointed out before, the
contributions from the primordial radio-nuclides $^{40}$K and
$^{208}$Tl need not be determined from the line fits, but can be
computed based on the known or estimated activities.

\subsection{\label{line_fitting_al26} $^{\mathsf{26}}$Al analysis}

\begin{figure}
\epsfig{figure=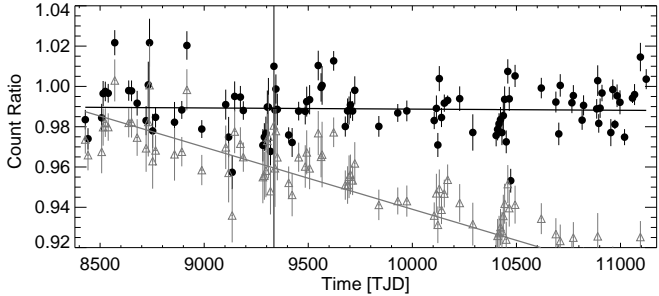,width=8.8cm}
\caption{Count ratios (number of counts in the line interval divided
by the number of counts in the adjacent energy intervals) as
determined for observations at galactic latitude $|b| >
40^{\circ}$. The ratios, and a linear model of their time variation,
are shown before (grey) and after (black) subtraction of the
instrumental line background.}
\label{stefan_al26_bgd_norm}
\end{figure}

The analysis of the galactic 1.8~MeV line emission from $^{26}$Al
is, similar to that of the CDG, affected by the instrumental line
background and its temporal variation. The galactic 1.8~MeV line
emission is determined in the energy band 1.7--1.9~MeV, using
adjacent energy bands for constructing a model for the
background in this so-called line interval (comprehensive
descriptions of this approach can be found in, e.g., Kn\"odlseder
\cite{knoedlseder_phd} and Oberlack \cite{oberlack_phd}). In
particular, the scatter angle $(\bar{\varphi})$ distribution of the
1.7--1.9~MeV background model is derived from an
interpolation of the $\bar{\varphi}$ distributions in narrow adjacent
energy intervals (1.6--1.7~MeV and 1.9--2.0~MeV). Due to the
long-term variation of the instrumental line background (see e.g.\
Fig.~\ref{lines_long-term_plot}), the ratio of the number of counts in
the line interval and in the adjacent energy intervals is decreasing
with time (see Fig.~\ref{stefan_al26_bgd_norm}), mostly due to the
build-up of $^{22}$Na, a major component in the 1.6--1.7~MeV band
in E$_\mathrm{tot}$ (Oberlack \cite{oberlack_phd}). To eliminate the
time dependent contamination of the background reference from adjacent
energies, the number of counts due to each component of the
instrumental line background has to be determined for each individual
observation period, employing a procedure outlined below. After
subtraction of the instrumental line background, the ratio of counts
in the line interval and the adjacent energy intervals to a good
approximation is constant in time (see
Fig.~\ref{stefan_al26_bgd_norm}), and data from individual observation
periods can be summed to analyze the galactic 1.8~MeV line emission.

The procedure used in the $^{26}$Al analysis to determine the
background contributions of the eight identified background isotopes
is a modified version of the CDG procedure described in
App.~\ref{line_fitting_cdg}. The modifications are motivated by
differences in the event selections applied in the two analyses,
particularly differences in the elimination of atmospheric background
(see Apps.~\ref{event_selections_cdg} and
\ref{event_selections_imaging}). One of the consequences is that in
the $^{26}$Al imaging analysis events with larger $\bar{\varphi}$
values are accepted than in the CDG analysis. For some isotopes,
notably $^{24}$Na, this results in significant changes of the
E$_\mathrm{1}$-E$_\mathrm{2}$ distribution as compared to that for CDG
event selections. Consequently, the optimal fit regions in
E$_\mathrm{1}$-E$_\mathrm{2}$ space are somewhat different. Also, in
contrast to the CDG analysis, in the $^{26}$Al analysis the
$\bar{\varphi}$ distribution of the accepted events is different for
each observation period due to the orbit dependent rejection of
atmospheric background. The simulated energy distributions for each
background isotope therefore have to be corrected for the specific
$\bar{\varphi}$ distribution of each individual observation period
(Oberlack \cite{oberlack_phd}).
The correction for selections to reject atmospheric background has
been calculated assuming a homogeneous illumination of the D1
detector. Since there is a certain edge enhancement in the
illumination of the D1 modules for background produced in the D1
structure, slight differences in these corrections result in small
distortions of the templates, and thus add to the observed scatter in
the determined isotope event rates (see
Fig.~\ref{lines_long-term_plot}).
As each observation period is independently analyzed for its
instrumental line background contamination, the background modelling
in the $^{26}$Al analysis yields as a welcome byproduct the
long-term variation of the event rates due to the eight identified
background isotopes depicted in Fig.~\ref{lines_long-term_plot}.

\begin{figure}
\epsfig{figure=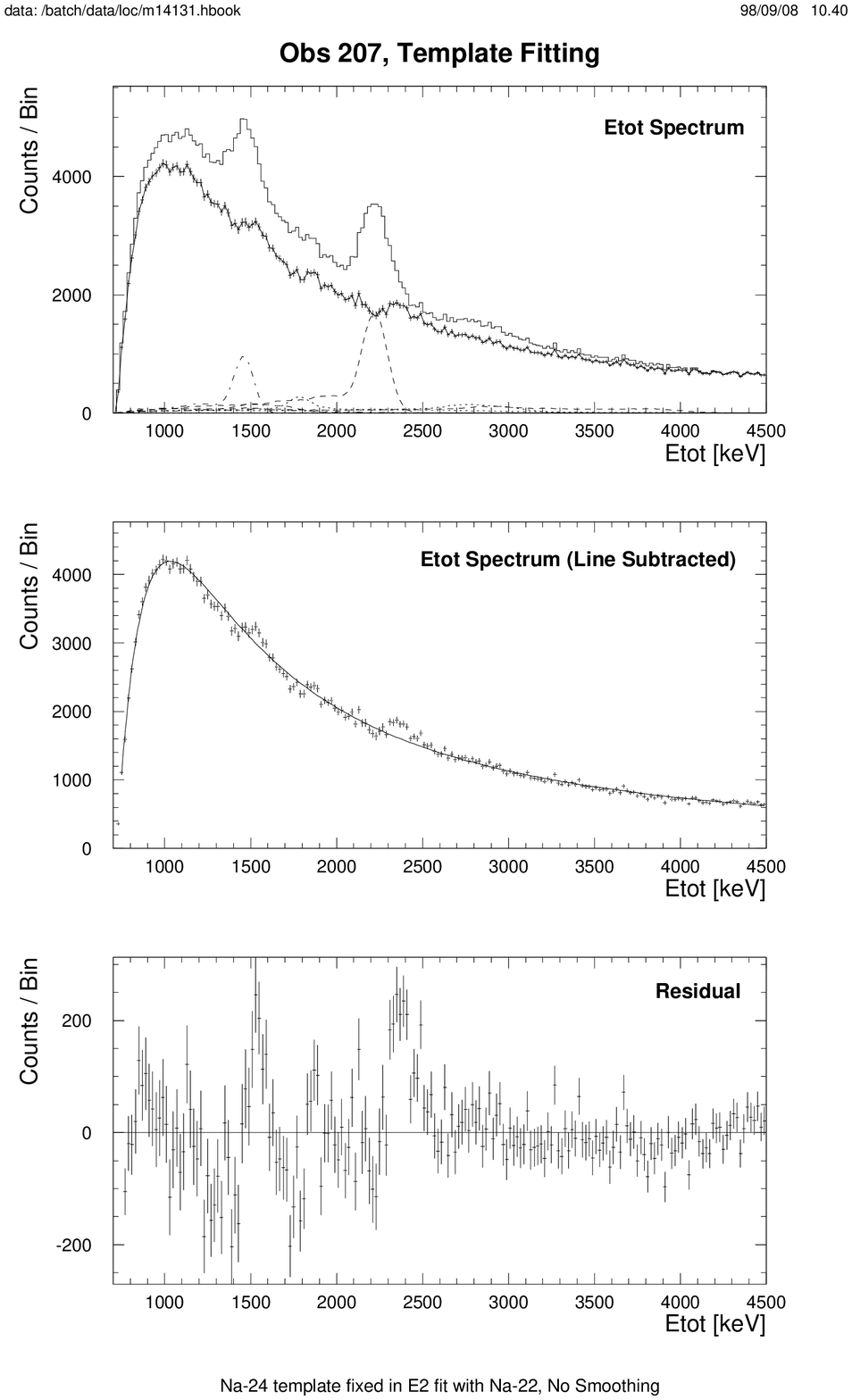,%
bbllx=74pt,bblly=532pt,bburx=482pt,bbury=740pt,width=8.5cm,clip=}
\epsfig{figure=H2362F33.ps,%
bbllx=74pt,bblly=306pt,bburx=482pt,bbury=512pt,width=8.5cm,clip=}
\epsfig{figure=H2362F33.ps,%
bbllx=74pt,bblly=79pt,bburx=482pt,bbury=284pt,width=8.5cm,clip=}
\caption{An example for an E$_\mathrm{tot}$ fit in the $^{26}$Al
analysis. The upper panel depicts the E$_\mathrm{tot}$ spectrum before
(line histogram) and after (data points) subtracting the contributions
from the identified background isotopes (indicated by smooth curves
with different line types). The middle panel shows the
E$_\mathrm{tot}$ spectrum after isotope subtraction, together with the
continuum background fit. The bottom panel gives the residuum of the
fit.}
\label{stefan_etot-fit}
\end{figure}

The four steps of the iterative line fitting procedure in the
$^{26}$Al analysis are similar to those in the CDG analysis, hence we
content ourselves with describing only the differences. In the first
step, two E$_\mathrm{1}$-E$_\mathrm{2}$ regions are fitted
simultaneously to obtain the $^{24}$Na event rate (Oberlack
\cite{oberlack_phd}): the E$_\mathrm{1}$-E$_\mathrm{2}$ region
depicted in Fig.~\ref{line-fitting_procedure_illustration_2}, and an
additional E$_\mathrm{1}$-E$_\mathrm{2}$ region extending from 2000 to
2700~keV in E$_\mathrm{1}$ and from 1100 to 2000~keV in
E$_\mathrm{2}$.  Analogous to the first E$_\mathrm{1}$-E$_\mathrm{2}$
region, the additional E$_\mathrm{1}$-E$_\mathrm{2}$ region optimizes
the $^{24}$Na signal from simultaneous interactions of the 1.37~MeV
and 2.75~MeV photons in D2 and D1, respectively, which are no longer
suppressed due to the larger accepted $\bar{\varphi}$ values for
imaging selections. The E$_\mathrm{1}$-E$_\mathrm{2}$ region for the
second step extends down to 700~keV in E$_\mathrm{1}$, and is used for
determining the $^{28}$Al rate and for obtaining start parameters to
determine the event rates of the remaining isotopes, particularly
$^{22}$Na, in Fit~3. However, unlike in the CDG analysis, the
contributions of the primordial isotopes $^{40}$K and $^{208}$Tl are
not fixed in the $^{26}$Al analysis, but determined in Fit~3 and 1,
respectively. As far as determining the contributions of each
individual isotope is concerned, the procedure ends here, since it has
been demonstrated in the CDG analysis that one fit cycle is sufficient
to obtain self-consistent isotope rates. This advanced background
treatment provided the basis for the latest COMPTEL $^{26}$Al all-sky
maps (e.g.\ Oberlack \cite{oberlack_phd}, Pl\"uschke et al.\
\cite{plueschke_al-map}).

\begin{figure}
\centerline{\hbox{
\epsfig{figure=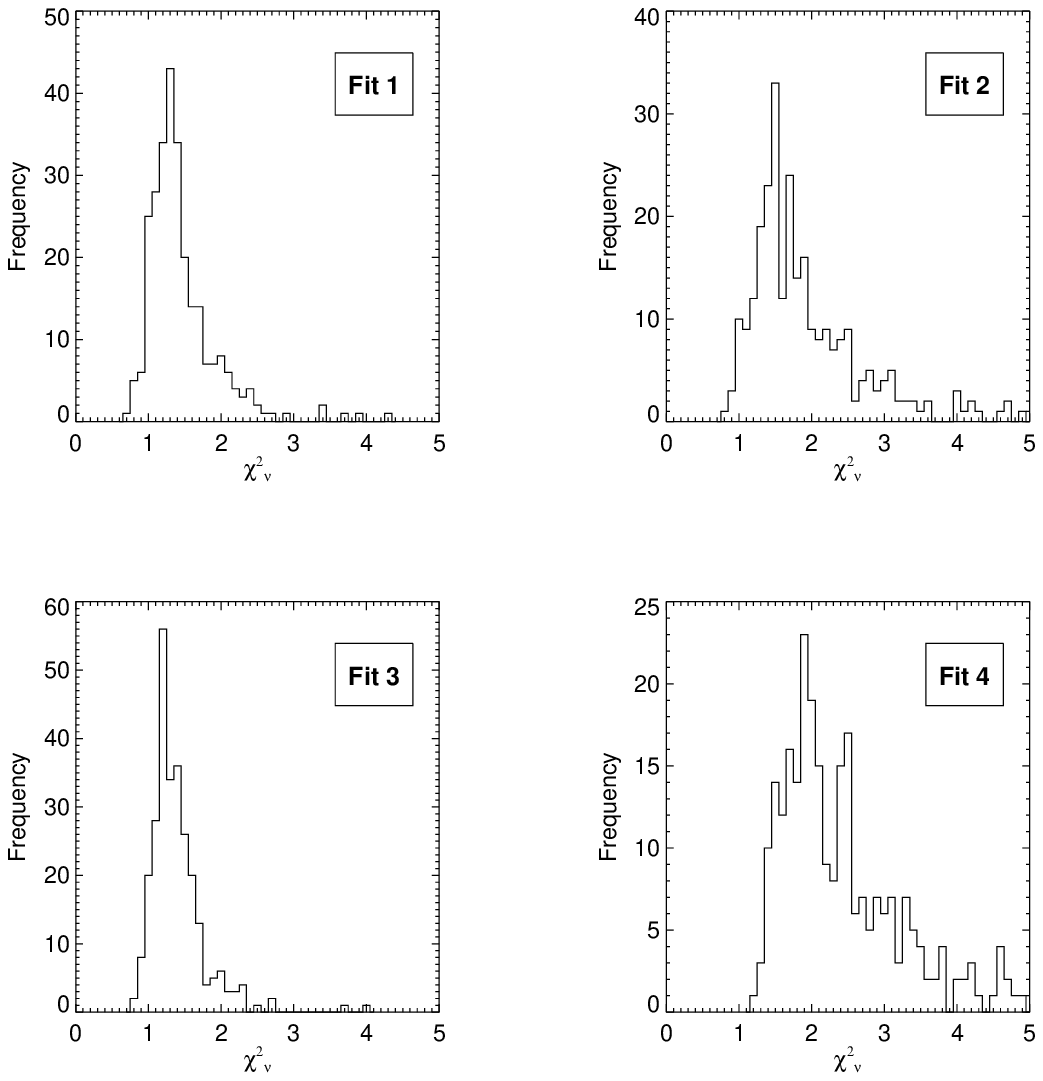,%
bbllx=26pt,bblly=180pt,bburx=162pt,bbury=328pt,width=4cm,clip=}
\epsfig{figure=H2362F34.eps,%
bbllx=196pt,bblly=180pt,bburx=332pt,bbury=328pt,width=4cm,clip=}
}}
\centerline{\hbox{
\epsfig{figure=H2362F34.eps,%
bbllx=26pt,bblly=10pt,bburx=162pt,bbury=156pt,width=4cm,clip=}
\epsfig{figure=H2362F34.eps,%
bbllx=196pt,bblly=10pt,bburx=332pt,bbury=156pt,width=4cm,clip=}
}}
\caption[]{The distribution of the $\chi^2_{\nu}$ values for each of
the four fit steps in determining the instrumental line contamination
in the $^{26}$Al analysis (Pl\"uschke et al.\
\cite{plueschke_al-map}).}
\label{chisquare_stefan}
\end{figure}

To investigate the extent to which the identified isotopes can account
for spectral features in E$_\mathrm{tot}$, an additional fit is
performed (see Fig.~\ref{stefan_etot-fit}). In this E$_\mathrm{tot}$
fit (E$_\mathrm{1} > 70$~keV, E$_\mathrm{2} > 650$~keV), the
contributions from the eight background isotopes are fixed at the
values obtained from the three previous E$_\mathrm{2}$ fits; only the
continuum background, modelled by a power law with an exponential
turn-over at low energies, is varied. As can be seen, the major
spectral features are accounted for by the eight identified isotopes,
however, some minor features remain unidentified at this time.

The distribution of the $\chi^2_{\nu}$ values for each of the four fit
steps, as obtained in the analysis of Pl\"uschke et al.\
(\cite{plueschke_al-map}), is shown in Fig.~\ref{chisquare_stefan}. The
quality of Fits~1 and 3, in which the event rates of most of the
background isotopes are determined, is acceptable. The $\chi^2_{\nu}$
distribtion for the E$_\mathrm{tot}$ fit, however, indicates that further
improvement of the instrumental line background modelling is still
possible.

%
%


\begin{thebibliography}{}

   \bibitem[1994]{bloemen_orion} Bloemen, H., et al., 1994, A\&A {\bf
   281}, L5

   \bibitem[1999]{bloemen_revised-orion} Bloemen, H., et al., 1999,
   ApJ {\bf 512}, L137

   \bibitem[1991]{dean_bgd-review} Dean, A. J., et al., 1991, Space
   Sci.\ Rev.\ {\bf 57}, 109--186

   \bibitem[1995a]{diehl_al26} Diehl, R., et al., 1995a, A\&A {\bf
   298}, 445

   \bibitem[1995b]{diehl_vela} Diehl, R., et al., 1995b, A\&A {\bf 298}, L25

   \bibitem[1994]{dyer_cgro-bgd} Dyer, C. S., et al., 1994, IEEE
   Trans. Nucl. Sci. {\bf 41}, No. 3, 438

   \bibitem[1981]{forrest_smm} Forrest, D. J., et al., 1981, Solar
   Physics {\bf 65}, 15

   \bibitem[1997]{iyudin_cas-a} Iyudin, A. F., et al., 1997, ESA
   SP-382, 37

   \bibitem[1996]{kappadath_3c} Kappadath, S. C., et al., 1996, A\&AS
   {\bf 120}, C619

   \bibitem[1998]{kappadath_phd} Kappadath, S. C., 1998,
      Ph.D. Thesis, University of New Hampshire, USA 

   \bibitem[2000]{kappadath_cdg} Kappadath, S. C., et al., 2000, ApJ,
   in preparation

   \bibitem[1997]{knoedlseder_phd} Kn{\"o}dlseder, J., 1997,
   Th\`ese de l'Universit\'e Paul Sabatier, Toulouse, France

   \bibitem[1989]{kurfess_smm_lines} Kurfess, J. D., et al., 1989, in
   {\sl High-energy radiation background in space} (AIP 186),
   eds. A. Rester, Jr., and J. I. Trombka, 250

   \bibitem[1980]{mahoney_heao3} Mahoney, W. A., et al., 1980, NIM~A
   {\bf 178}, 363

   \bibitem[1995a]{morris_fast_neutrons} Morris, D. J., et al., 1995a,
   Journ. of Geophys. Res. {\bf 100}, 12243

   \bibitem[1997a]{morris_ieee} Morris, D. J., et al., 1997a, in {\sl
   1997 Conference on the High-Energy Background in Space},
   eds. J. I. Trombka, J. S. Schweitzer, and G. P. Lasche, Institute
   of Electrical and Electronic Engineers (IEEE), 26

   \bibitem[1997b]{morris_sn1991t} Morris, D. J., et al., 1997b, in
   {\sl Proc.\ of the Fourth Compton Symposium} (AIP 410),
   eds. Dermer, C. D., Strickman, M. S., and Kurfess, J. D., 1084

   \bibitem[1997]{oberlack_phd} Oberlack, U., 1997,
   Dissertation, Technical University Munich, Germany

   \bibitem[2000]{plueschke_al-map} Pl{\"u}schke, S., et al., 2000, in
   {\sl Proc. of 5$\mathit{th}$ Compton Symposium} (AIP 510),
   eds. M.~McConnell, and J.M.~Ryan, 35

   \bibitem[1997]{ryan_ieee} Ryan, J., et al., 1997, in {\sl 1997
   Conference on the High-Energy Background in Space},
   eds. J. I. Trombka, J. S. Schweitzer, and G. P. Lasche, Institute
   of Electrical and Electronic Engineers (IEEE), 13

   \bibitem[1993]{schoenfelder_comptel} Sch{\"o}nfelder, V., et al.,
   1993, ApJS {\bf 86}, 657

   \bibitem[1989]{share_smm_lines} Share, G. H., et al., 1989, in {\sl
   High-energy radiation background in space} (AIP 186),
   eds. A. Rester, Jr., and J. I. Trombka, 266

   \bibitem[1986]{snelling} Snelling, M., et al., 1986, NIM~A {\bf
   248}, 545

   \bibitem[1989]{stassinopoulos} Stassinopoulos, E. G., 1989, in {\sl
   High-energy radiation background in space} (AIP 186),
   eds. A. Rester, Jr., and J. I. Trombka, 3--63

   \bibitem[1996]{vandijk_phd} van~Dijk, R., 1996, Ph.D. thesis,
   University of Amsterdam, the Netherlands

   \bibitem[1997]{varendorff_4c} Varendorff, M., et al., 1997, in
   {\sl Proc.\ of the Fourth Compton Symposium} (AIP 410),
   eds. Dermer, C. D., Strickman, M. S., and Kurfess, J. D., 1577

   \bibitem[1989]{watts_saa_anisotropy} Watts, J. W., et al., 1989, in
   {\sl High-energy radiation background in space} (AIP 186),
   eds. A. Rester, Jr., and J. I. Trombka, 75

   \bibitem[1996]{weidenspointner_3c} Weidenspointner, G., et al., 1996,
   A\&AS {\bf 120}, C631 

   \bibitem[1999]{weidenspointner_phd} Weidenspointner, G., 1999,
   Dissertation, Technical University Munich, Germany

   \bibitem[2001]{weidenspointner_cdg} Weidenspointner, G., et al.,
   2001, A\&A, in preparation

   \bibitem[1989]{wheaton_heao3} Wheaton, W. A., et al., 1989, in {\sl
   High-energy radiation background in space} (AIP 186),
   eds. A. Rester, Jr., and J. I. Trombka, 304


\end{thebibliography}
\end{document}